\documentclass[floatfix,aps,showpacs,twocolumn,nofootinbib,preprintnumbers]
{revtex4}
\usepackage{graphicx}

\begin{document}

\preprint{FERMILAB-Pub-03/117-A}

\title{WMAPping inflationary physics}
\author{William H.\ Kinney} \email{whkinney@buffalo.edu}
\affiliation{Institute for Strings, Cosmology and Astroparticle Physics,
	Columbia University, 550 W. 120th St., New York, NY 10027\\ and\\
	Dept. of Physics, University at Buffalo,
        the State University of New York, Buffalo, NY 14260-1500}
\author{Edward W. Kolb} \email{rocky@fnal.gov}
\affiliation{Fermilab Astrophysics Center, Fermi
       	National Accelerator Laboratory, Batavia, Illinois \ 60510-0500, USA,\\
       	and Department of Astronomy and Astrophysics, Enrico Fermi Institute,
       	University of Chicago, Chicago, Illinois \ 60637-1433, USA}
\author{Alessandro Melchiorri} \email{melch@astro.ox.ac.uk}
\affiliation{Astrophysics, Denys Wilkinson Building, University of Oxford, 
	Keble road, OX1 3RH, Oxford UK and Dipartimento di Fisica,
Universita' di Roma ``La Sapienza'', Ple Aldo Moro 2, 00185, Italy}
\author{Antonio Riotto} \email{antonio.riotto@pd.infn.it}
\affiliation{INFN, Sezione di Padova, via Marzolo 8, I-35131, Italy}
\date{\today}

\begin{abstract}

We extract parameters relevant for distinguishing among single-field inflation
models from the Wilkinson Microwave Anisotropy Probe (WMAP) data set, and from
a combination of the WMAP data and seven other Cosmic Microwave Background
(CMB) experiments. We use only CMB data and perform a likelihood analysis over
a grid of models including the full error covariance matrix. We find that a
model with a scale-invariant scalar power spectrum ($n=1$), no tensor
contribution, and no running of the spectral index, is within the 1-$\sigma$
contours of both data sets. We then apply the Monte Carlo reconstruction
technique to both data sets to generate an ensemble of inflationary potentials
consistent with observations.  None of the three basic classes of inflation
models (small-field, large-field, and hybrid) are completely ruled out,
although hybrid models are favored by the best-fit region. The reconstruction
process indicates that a wide variety of smooth potentials for the inflaton are
consistent with the data, implying that the first-year WMAP result is still
too crude to constrain significantly either the height or the shape of the
inflaton potential.  In particular, the lack of evidence for tensor
fluctuations makes it impossible to constrain the energy scale of
inflation. Nonetheless, the data rule out a large portion of the available
parameter space for inflation. For instance, we find that potentials of the
form $V = \lambda \phi^4$ are ruled out to $3\sigma$ by the combined data set,
but not by the WMAP data taken alone.

\end{abstract}


\maketitle

\section{Introduction}

One of the fundamental ideas of modern cosmology is that there was an epoch
early in the history of the universe when potential, or vacuum, energy
dominated other forms of energy densities such as matter or radiation. During
such a vacuum-dominated era the scale factor grew exponentially (or nearly
exponentially) in some small time. During this phase, dubbed inflation
\cite{guth81,lrreview}, a small, smooth spatial region of size of order
the Hubble radius grew so large that it easily could encompass the comoving
volume of the entire presently observable universe. If the universe underwent
such a period of rapid expansion, one can understand why the observed universe
is homogeneous and isotropic to such high accuracy.

One of the predictions of the simplest models of inflation is a spatially flat
Universe, {\it i.e.,} $\Omega_{tot}=1$, with great precision.  Inflation has
also become the dominant paradigm for understanding the initial conditions for
structure formation and for Cosmic Microwave Background (CMB) anisotropy. In
the inflationary picture, primordial density and gravity-wave (tensor)
fluctuations are created from quantum fluctuations and ``redshifted'' out of
the horizon during an early period of superluminal expansion of the universe,
where they are ``frozen'' as perturbations in the background metric
\cite{muk81,hawking82,starobinsky82,guth82,bardeen83}. Metric perturbations at
the surface of last scattering are observable as temperature anisotropies in
the CMB.  The first and most impressive confirmation of the inflationary
paradigm came when the CMB anisotropies were detected by the Cosmic
Background Explorer (COBE) satellite in 1992 \cite{smoot92,bennett96,gorski96}.
Subsequently, it became clear that the measurements of the spectrum of the CMB
anisotropy can provide very detailed information about fundamental
cosmological parameters \cite{cp} and other crucial parameters for particle
physics.

In the past few years, a number of balloon-borne and terrestrial experiments
have mapped out the CMB angular anisotropies
\cite{toco,b97,Netterfield,halverson,cbi,vsa,benoit}, revealing a
remarkable agreement between the data and the inflationary predictions of a
flat universe with a nearly scale-invariant spectrum of adiabatic primordial
density perturbations, (see {\it e.g.,}
\cite{debe2001,pryke,stompor,wang,cbit,vsat,saralewis,mesilk,caro}).

Despite the simplicity of the inflationary paradigm, the number of inflation
models that have been proposed in the literature is enormous
\cite{lrreview}. This is true even if we limit ourselves to models with only
one scalar field (the {\it inflaton}).
With the previous data on CMB anisotropies from balloon and terrestrial
experiments it has been possible for the first time to place interesting
constraints on the space of possible inflation models
\cite{dodelson97,kinney98a,probes}. However, the quality of the data were not
good enough to rule out entire classes of models. A boost along these lines has
been very recently provided by the data of the Wilkinson Microwave Anisotropy
Probe (WMAP) mission, which has marked the beginning of the precision era of
the CMB measurements in space \cite{wmap1,kogutt}. The WMAP collaboration has
produced a full-sky map of the angular variations in the microwave flux, in
particular the cosmic microwave background, with unprecedented accuracy.  WMAP
data support the inflationary mechanism for the generation of curvature
superhorizon fluctuations and provide a strong bound on the possible admixture
of isocurvature modes \cite{wmapinf}. Furthermore, consistent with the simplest
single-field models of inflation \cite{nong}, no evidence of nongaussianity is
found \cite{wmapnong}.

The goal of this paper is to use the WMAP data to discriminate among the
various single-field inflationary models. To obtain some indication of the
robustness of our analysis, we also consider a data set consisting of WMAP
augmented with several other CMB experiments.  For single-field inflation
models, the relevant parameter space for distinguishing among models is defined
by the scalar spectral index $n$, the ratio of tensor to scalar fluctuations
$r$, and the running of the scalar spectral index $d n / d\ln{k}$.  We employ
{\em Monte Carlo reconstruction}, a stochastic method for ``inverting''
observational constraints to determine an ensemble of inflationary potentials
compatible with observation \cite{kinney02,easther02}.  In addition to
encompassing a broader set of models than usually considered (large-field,
small-field, hybrid and linear models), Monte Carlo reconstruction makes it
possible easily to incorporate constraints on the running of the spectral index
as well as to include effects to higher order in slow roll.

Since studies on the implications of WMAP data for inflation 
\cite{wmapinf,barger} have already appeared, we briefly mention the different 
elements between our analysis and others. (We will elaborate on these
differences later.) The WMAP collaboration analysis \cite{wmapinf} included
WMAP data, additional CMB data (CBI \cite{cbi} and ACBAR \cite{acbar}),
large-scale structure data (2dFGRS \cite{2df}), as well as Lyman-$\alpha$ power
spectrum data \cite{lyman}.  They used a Markov Chain Monte Carlo technique to
explore the likelihood surface.  Barger {\it et al.}\ \cite{barger}, considered
WMAP data only, but with a top-hat prior on the Hubble constant $h$ ($H_0=$ 100
$h$ km sec$^{-1}$ Mpc$^{-1}$) from the HST key project \cite{hst}.   
Also, Barger {\it et al.}\ did not consider a running of
the scalar spectral index. We only consider CMB data.  We first analyze just
the WMAP results.  We then analyze the WMAP data set in conjunction with other
CMB data sets (BOOMERanG-98~\cite{ruhl}, MAXIMA-1 \cite{lee},
DASI~\cite{halverson}, CBI~\cite{cbi}, ACBAR~\cite{acbar}, VSAE~\cite{grainge},
and Archeops~\cite{benoit}).  We employ a grid of models in the likelihood
analysis, which differs from the method used by the WMAP team.

The paper is organized as follows. In Sec.\ II we discuss single-field
inflation and the relevant observables in more detail. In Sec.\ III we discuss
the inflationary model space, and in Sec.\ IV we describe the Monte Carlo
reconstruction technique. Section V describes the methods used for the CMB
analysis. In Sec.\ VI we present the constraints from the CMB anisotropy data
sets. In Sec.\ VII we present our conclusions.

\section{Single-field inflation and the inflationary observables}

In this section we briefly review scalar field models of inflationary
cosmology, and explain how we relate model parameters to observable quantities.
Inflation, in its most general sense, can be defined to be a period of
accelerating cosmological expansion during which the universe evolves toward
homogeneity and flatness. This acceleration is typically a result of the
universe being dominated by vacuum energy, with an equation of state $p \simeq
-\rho$. Within this broad framework, many specific models for inflation have
been proposed. We limit ourselves here to models with ``normal'' gravity ({\em
i.e.,} general relativity) and a single order parameter for the vacuum,
described by a slowly rolling scalar field $\phi$, the inflaton. These
assumptions are not overly restrictive; the most widely studied inflation
models fall within this category, including Linde's ``chaotic'' inflation
scenario \cite{linde83}, inflation from pseudo Nambu-Goldstone bosons
(``natural'' inflation \cite{freese90}), dilaton-like models involving
exponential potentials (power-law inflation), hybrid inflation
\cite{linde91,linde94,copeland94}, and so forth. Other models, such as
Starobinsky's $R^2$ model \cite{starobinsky80} and versions of extended
inflation, can, through a suitable transformation, be viewed in terms of
equivalent single-field models.  Of course in single-field models of inflation,
the inflaton ``field'' need not be a fundamental field at all. Also, some
``single-field'' models require auxiliary fields. Hybrid inflation models
\cite{linde91,linde94,copeland94}, for example, require a second field to end
inflation. What is significant is that the inflationary epoch be described by a
single dynamical order parameter, the inflaton field.

A scalar field in a cosmological background evolves with an equation of motion
\begin{equation}
\ddot\phi + 3 H \dot\phi + V'\left(\phi\right) = 0.
\label{eqequationofmotion}
\end{equation}
The evolution of the scale factor is given by the scalar field dominated FRW
equation,
\begin{eqnarray}
H^2 & &= {8 \pi \over 3 m_{\rm Pl}^2} \left[{1 \over 2} \dot\phi^2 +
V\left(\phi\right)\right],\cr
\left(\ddot a \over a\right) &&= {8 \pi \over 3 m_{\rm Pl}^2}
\left[V\left(\phi\right) - \dot\phi^2\right].
\label{eqbackground}
\end{eqnarray}
Here $m_{\rm Pl} = G^{-1/2} \simeq 10^{19}\ {\rm GeV}$ is the Planck mass, and
we have assumed a flat Friedmann-Robertson-Walker metric,
\begin{equation}
g_{\mu \nu} =  {\rm diag}(1, -a^2, -a^2 -a^2),
\end{equation}
where $a^2(t)$ is the scale factor of the universe.  {\em Inflation} is defined
to be a period of accelerated expansion, $\ddot a > 0$.  A powerful way of
describing the dynamics of a scalar field-dominated cosmology is to express the
Hubble parameter as a function of the field $\phi$, $H = H(\phi)$, which is
consistent provided $\phi$ is monotonic in time. The equations of motion
become \cite{grishchuk88,muslimov90,salopek90,lidsey95}:
\begin{eqnarray} 
& &\dot\phi =  -{m_{\rm Pl}^2 \over 4 \pi} H'(\phi),\cr
& & \left[H'(\phi)\right]^2 - {12 \pi \over m_{\rm Pl}^2}
H^2(\phi) = - {32 \pi^2 \over m_{\rm Pl}^4}
V(\phi).\label{eqbasichjequations}
\end{eqnarray}
These are completely equivalent to the second-order equation of motion in Eq.\
(\ref{eqequationofmotion}). The second of the above equations is
referred to as the {\it Hamilton-Jacobi} equation, and can be written
in the useful form
\begin{equation} 
H^2(\phi) \left[1 - {1\over 3}
\epsilon(\phi)\right] =  \left({8 \pi \over 3 m_{\rm Pl}^2}\right)
 V(\phi),\label{eqhubblehamiltonjacobi}
\end{equation}
where $\epsilon$ is defined to be
\begin{equation}
\epsilon \equiv {m_{\rm Pl}^2 \over 4 \pi} \left({H'(\phi) \over
 H(\phi)}\right)^2.\label{eqdefofepsilon}
\end{equation}
The physical meaning of $\epsilon$ can be seen by expressing Eq.\
(\ref{eqbackground}) as
\begin{equation}
\left({\ddot a \over a}\right) = H^2 (\phi) \left[1 -
 \epsilon(\phi)\right],
\end{equation}
so that the condition for inflation $(\ddot a / a) > 0$ is given by
$\epsilon < 1$. The scale factor is given by
\begin{equation}
a \propto e^{N} = \exp\left[\int_{t_0}^{t}{H\,dt}\right],
\end{equation}
where the number of e-folds $N$ is
\begin{equation}
N \equiv \int_{t}^{t_e}{H\,dt} = \int_{\phi}^{\phi_e}{{H \over
\dot\phi}\,d\phi} = {2 \sqrt{\pi} \over m_{\rm Pl}}
\int_{\phi_e}^{\phi}{d\phi \over
\sqrt{\epsilon(\phi)}}.\label{eqdefofN}
\end{equation}
To create the observed flatness and homogeneity of the universe, we require
many e-folds of inflation, typically $N \simeq 60$. This figure varies somewhat
with the details of the model. We can relate a comoving scale $k$ in the
universe today to the number of e-folds $N$ before the end of inflation by
\cite{lidsey97}
\begin{equation}
N(k) = 62 - \ln \frac{k}{a_0 H_0} - \ln \frac{10^{16}
        {\rm GeV}}{V_k^{1/4}}
        + \ln \frac{V_k^{1/4}} {V_e^{1/4}} - \frac{1}{3} \ln
        \frac{{V_e}^{1/4}}{\rho_{{\rm RH}}^{1/4}} \, .
\end{equation}
Here $V_k$ is the potential when the mode leaves the horizon, $V_e$ is the
potential at the end of inflation, and $\rho_{{\rm RH}}$ is the energy density
after reheating. Scales of order the current horizon size exited the horizon at
$N\left(k\right) \sim 60$. Since this number depends, for example, on the
details of reheating, we will allow $N$ to vary within the range $40 \leq N
\leq 70$ for any given model in order to consider the most general case. 
(Dodelson and Hui have recently argued that the value of $N$ corresponding
to the current horizon size can be no larger than 60 \cite{dodelson03}, so
in this sense we are being more general than is necessary.)

We will frequently work within the context of the {\em slow roll} approximation
\cite{linde82,albrecht82}, which is the assumption that the evolution of the
field is dominated by drag from the cosmological expansion, so that $\ddot\phi
\simeq 0$ and
\begin{equation}
\dot \phi \simeq -{V' \over 3 H}.
\end{equation}
The equation of state of the scalar field is dominated by the potential,
so that $p \simeq -\rho$, and the expansion rate is approximately
\begin{equation}
H \simeq \sqrt{{8 \pi \over 3 m_{\rm Pl}^2} V\left(\phi\right)}.
\label{eqhslowroll}
\end{equation}
The slow roll approximation is consistent if both the slope and curvature of
the potential are small, $V',\ V'' \ll V$. In this case the parameter
$\epsilon$ can be expressed in terms of the potential as
\begin{equation}
\epsilon \equiv {m_{\rm Pl}^2 \over 4 \pi} \left({H'\left(\phi\right) \over
H\left(\phi\right)}\right)^2 \simeq {m_{\rm Pl}^2 \over 16 \pi}
\left({V'\left(\phi\right) \over V\left(\phi\right)}\right)^2.
\end{equation}
We will also define a second ``slow roll parameter'' $\eta$ by:
\begin{eqnarray}
\eta\left(\phi\right) &\equiv& {m_{\rm Pl}^2 \over 4 \pi} 
\left({H''\left(\phi\right)
\over H\left(\phi\right)}\right)\cr
&\simeq& {m_{\rm Pl}^2 \over 8 \pi}
\left[{V''\left(\phi\right) \over V\left(\phi\right)} - {1 \over 2}
\left({V'\left(\phi\right) \over V\left(\phi\right)}\right)^2\right].
\end{eqnarray}
Slow roll is then a consistent approximation for $\epsilon,\ \eta \ll 1$. 

Inflation models not only explain the large-scale homogeneity of the universe,
but also provide a mechanism for explaining the observed level of {\em
inhomogeneity} as well. During inflation, quantum fluctuations on small scales
are quickly redshifted to scales much larger than the horizon size, where they
are ``frozen'' as perturbations in the background metric
\cite{muk81,hawking82,starobinsky82,guth82,bardeen83}. The metric perturbations
created during inflation are of two types: scalar, or {\it curvature}
perturbations, which couple to the stress-energy of matter in the universe and
form the ``seeds'' for structure formation, and tensor, or gravitational wave
perturbations, which do not couple to matter.  Both scalar and tensor
perturbations contribute to CMB anisotropy. Scalar fluctuations can also be
interpreted as fluctuations in the density of the matter in the
universe. Scalar fluctuations can be quantitatively characterized by
perturbations $P_{\cal R}$ in the intrinsic curvature scalar. As long as the
equation of state $\epsilon$ is slowly varying,\footnote{This assumption is
{\em not} identical to the assumption of slow roll (see, e.g., Ref.\
\cite{kinney97a}), although in most cases it is equivalent.} the curvature
perturbation can be shown to be 
\cite{mukhanov85,mukhanov88,mukhanov92,stewart93}
\begin{equation}
P_{\cal R}^{1/2}\left(k\right) =  \frac{1}{2\pi} \left [{H \over m_{\rm Pl} }
{1 \over \sqrt{\epsilon}}\right]_{k = a H}.
\end{equation}
The fluctuation power spectrum is in general a function of wavenumber $k$, and
is evaluated when a given mode crosses outside the horizon during inflation, $k
= a H$. Outside the horizon, modes do not evolve, so the amplitude of the mode
when it crosses back {\em inside} the horizon during a later radiation- or
matter-dominated epoch is just its value when it left the horizon during
inflation.

The {\em spectral index} $n$ for $P_{\cal R}$ is defined by
\begin{equation}
n - 1 \equiv {d\ln P_{\cal R} \over d\ln k},
\end{equation}
so that a scale-invariant spectrum, in which modes have constant amplitude at
horizon crossing, is characterized by $n = 1$. Some inflation models predict
running of the spectral index with scale
\cite{stewart97,linde97,stewart97a,copeland97,kinney97,covi98,kinney98,covi99,covi00} 
or even sharp features in the power spectrum \cite{featuresinpowerspectrum}. 
We will consider the running of the spectral index in more detail on
Sec.\ \ref{secmontecarlorecon}.

Instead of specifying the fluctuation amplitude directly as a function of $k$,
it is often convenient to specify it as a function of the number of e-folds $N$
before the end of inflation at which a mode crossed outside the horizon. Scales
of interest for current measurements of CMB anisotropy crossed outside the
horizon at $N \sim 60$, so that $P_{\cal R}$ is conventionally
evaluated at $P_{\cal R}\left({N \sim 60}\right)$. 

The power spectrum of tensor fluctuation modes is given
by \cite{starobinsky79,rubakov82,fabbri83,abbot84,starobinsky85}
\begin{equation}
P_{T}^{1/2}\left(k_N\right) = {1 \over 2 \pi} \left[\frac{H}{m_{\rm Pl}}
\right]_{N}.
\end{equation}
The ratio of tensor to scalar modes is then
\begin{equation}
{P_{T} \over P_{\cal R}} = \epsilon,
\end{equation}
so that tensor modes are negligible for $\epsilon \ll 1$. Tensor and scalar
modes both contribute to CMB temperature anisotropy. If the contribution of
tensor modes to the CMB anisotropy can be neglected, normalization to the COBE
four-year data gives \cite{bunn96,lyth96} $P_{\cal R}^{1/2} = 4.8 \times
10^{-5}$. In the next section, we will describe the
predictions of various models in this parameter space.

\section{The inflationary model space}
\label{seczoology}

To summarize the results of the previous section, inflation generates scalar
(density) and tensor (gravity wave) fluctuations which are generally well
approximated by power laws:
\begin{equation}
P_{\cal R}\left(k\right) \propto k^{n - 1}; \qquad
P_{T}\left(k\right) \propto k^{n_{T}}.
\end{equation}
In the limit of slow roll, the spectral indices $n$ and $n_{T}$ vary slowly 
or not at all with scale.
We can write the spectral indices $n$ and $n_{T}$ to lowest order in terms
of the slow roll parameters $\epsilon$ and $\eta$ as \cite{stewart93}:
\begin{eqnarray}
n \simeq&& 1 - 4 \epsilon + 2 \eta,\cr
n_{T} \simeq&& - 2 \epsilon.
\end{eqnarray}
The tensor spectral index is {\em not} an independent parameter, but
is proportional to the tensor/scalar ratio, given to lowest order in
slow roll by
\begin{equation}
n_{T} \simeq - 2 \epsilon = - 2 {P_{T} \over P_{\cal R}}.
\end{equation}
This is known as the {\it consistency relation} for inflation. (This relation
holds only for single-field inflation, and weakens to an inequality for
inflation involving multiple degrees of freedom
\cite{polarski95,bellido95,sasaki96}.) A given inflation model can therefore be
described to lowest order in slow roll by three independent parameters,
$P_{\cal R}$, $P_{T}$, and $n$. If we wish to include higher-order effects, we
have a fourth parameter describing the running of the scalar spectral index, $d
n / d\ln{k}$.

The tensor/scalar ratio is frequently expressed as a ratio of their
contributions to the CMB quadrupole,
\begin{equation}
r \equiv {C^{\rm Tensor}_2 \over C^{\rm Scalar}_2}.
\end{equation}
The relation between $r$ and ratio of amplitudes in the primordial power
spectra $P_{T} / P_{\cal R}$ depends on the background cosmology, in particular
the densities of matter ($\Omega_{\rm m}$) and cosmological constant
($\Omega_{\Lambda}$). For the currently favored values of $\Omega_{\rm m}
\simeq 0.3$ and $\Omega_{\Lambda} \simeq 0.7$, the relation is approximately
\begin{equation}
r \simeq 10 \epsilon,
\end{equation}
to lowest order in slow roll.  Conventions for the normalization of this
parameter vary widely in the literature. In particular, Peiris et
al.\ \cite{wmapinf} use $r \simeq 16 \epsilon$.

Calculating the CMB fluctuations from a particular inflationary model reduces
to the following basic steps: (1) from the potential, calculate $\epsilon$ and
$\eta$. (2) From $\epsilon$, calculate $N$ as a function of the field $\phi$.
(3) Invert $N\left(\phi\right)$ to find $\phi_N$. (4) Calculate $P_{\cal R}$,
$n$, and $P_T$ as functions of $\phi$, and evaluate them at $\phi =
\phi_N$. For the remainder of the paper, all parameters are assumed to be
evaluated at $\phi = \phi_N$.  

Even restricting ourselves to a simple single-field inflation scenario, the
number of models available to choose from is large \cite{lrreview}.  It is
convenient to define a general classification scheme, or ``zoology'' for models
of inflation. We divide models into three general types: {\it large-field},
{\it small-field}, and {\it hybrid}, with a fourth classification, {\it linear}
models, serving as a boundary between large- and small-field. A generic
single-field potential can be characterized by two independent mass scales: a
``height'' $\Lambda^4$, corresponding to the vacuum energy density during
inflation, and a ``width'' $\mu$, corresponding to the change in the field
value $\Delta \phi$ during inflation:
\begin{equation}
V\left(\phi\right) = \Lambda^4 f\left({\phi \over \mu}\right).
\end{equation}
Different models have different forms for the function $f$. The height
$\Lambda$ is fixed by normalization, so the only free parameter is the width
$\mu$.

With the normalization fixed, the relevant parameter space for distinguishing
between inflation models to lowest order in slow roll is then the $r\,-\,n$
plane.  (To next order in slow-roll parameters, one must introduce the running
of $n$.)  Different classes of models are distinguished by the value of the
second derivative of the potential, or, equivalently, by the relationship
between the values of the slow-roll parameters $\epsilon$ and
$\eta$.\footnote{The designations ``small-field'' and ``large-field'' can
sometimes be misleading. For instance, both the $R^2$ model
\cite{starobinsky80} and the ``dual inflation'' model \cite{bellido98} are
characterized by $\Delta \phi \sim m_{\rm Pl}$, but are ``small-field'' in the
sense that $\eta < 0 < \epsilon$, with $n < 1$ and negligible tensor modes.} 
Each class of models has a different relationship between $r$ and $n$. For a
more detailed discussion of these relations, the reader is referred to Refs.\
\cite{dodelson97,kinney98a}.  

First order in $\epsilon$ and $\eta$ is sufficiently accurate for the purposes
of this Section, and for the remainder of this Section we will only work to
first order. The generalization to higher order in slow roll will be discussed 
in Sec.\ \ref{secmontecarlorecon}.

\subsection{Large-field models: $-\epsilon < \eta \leq \epsilon$}

Large-field models have inflaton potentials typical of ``chaotic'' inflation
scenarios \cite{linde83}, in which the scalar field is displaced from the
minimum of the potential by an amount usually of order the Planck mass. Such
models are characterized by $V''\left(\phi\right) > 0$, and $-\epsilon < \eta
\leq \epsilon$. The generic large-field potentials we consider are polynomial
potentials $V\left(\phi\right) = \Lambda^4 \left({\phi / \mu}\right)^p$,
and exponential potentials, $V\left(\phi\right) = \Lambda^4 \exp\left({\phi /
\mu}\right)$. For the case of an exponential potential, $V\left(\phi\right)
\propto \exp\left({\phi / \mu}\right)$, the tensor/scalar ratio $r$ is simply
related to the spectral index as
\begin{equation}
r = 5 \left(1 - n\right).
\end{equation}
This result is often incorrectly generalized to all slow-roll models, but is in
fact characteristic {\it only} of power-law inflation. For inflation with a
polynomial potential, $V\left(\phi\right) \propto \phi^p$,  we again have $r
\propto 1 - n$,
\begin{equation}
r = 5 \left({p \over p + 2}\right) \left(1 - n\right),
\end{equation}
so that tensor modes are large for significantly tilted spectra. We will be
particularly interested in models with $p = 4$ as a test case for our ability
to rule out models. For $p = 4$, the observables are given in terms of the
number of e-folds $N$ by
\begin{eqnarray}
r &=& {10 \over N + 1},\cr
1 - n &=& {3 \over N + 1}.
\end{eqnarray}

\subsection{Small-field models: $\eta < -\epsilon$}

Small-field models are the type of potentials that arise naturally from
spontaneous symmetry breaking (such as the original models of ``new'' inflation
\cite{linde82,albrecht82}) and from pseudo Nambu-Goldstone modes (natural
inflation \cite{freese90}). The field starts from near an unstable equilibrium
(taken to be at the origin) and rolls down the potential to a stable
minimum. Small-field models are characterized by $V''\left(\phi\right) < 0$ and
$\eta < -\epsilon$. Typically $\epsilon$ (and hence the tensor amplitude) is
close to zero in small-field models. The generic small-field potentials we
consider are of the form $V\left(\phi\right) = \Lambda^4 \left[1 - \left({\phi
/ \mu}\right)^p\right]$, which can be viewed as a lowest-order Taylor expansion
of an arbitrary potential about the origin. The cases $p = 2$ and $p > 2$ have
very different behavior. For $p = 2$,
\begin{equation}
r = 5 (1 - n) \exp\left[- 1 - N\left(1 - n\right)\right],
\end{equation}
where $N$ is the number of e-folds of inflation. For $p > 2$, the scalar
spectral index is
\begin{equation}
n \simeq 1 - {2 \over N} \left({p - 1 \over p - 2}\right),
\end{equation}
{\it independent} of $r$. Assuming $\mu < m_{\rm Pl}$ results in an upper bound
on $r$ of
\begin{equation}
r < 5 {p \over N \left(p - 2\right)} \left({8 \pi \over N p \left(p -
2\right)}\right)^{p / \left(p - 2\right)}.
\end{equation}

\subsection{Hybrid models: $0 < \epsilon < \eta$}

The hybrid scenario \cite{linde91,linde94,copeland94} frequently appears in
models which incorporate inflation into supersymmetry. In a typical hybrid
inflation model, the scalar field responsible for inflation evolves toward a
minimum with nonzero vacuum energy. The end of inflation arises as a result of
instability in a second field. Such models are characterized by
$V''\left(\phi\right) > 0$ and $0 < \epsilon < \eta$. We consider generic
potentials for hybrid inflation of the form $V\left(\phi\right) = \Lambda^4
\left[1 + \left({\phi / \mu}\right)^p\right].$ The field value at the end of
inflation is determined by some other physics, so there is a second free
parameter characterizing the models. Because of this extra freedom, hybrid
models fill a broad region in the $r\,-\,n$ plane (see Fig.\ \ref{figregions}).
There is, however, no overlap in the $r\,-\,n$ plane between hybrid inflation
and other models. The distinguishing feature of many hybrid models is a {\it
blue} scalar spectral index, $n > 1$. This corresponds to the case $\eta > 2
\epsilon$. Hybrid models can also in principle have a red spectrum, $n < 1$.

\subsection{Linear models: $\eta = - \epsilon$}

Linear models, $V\left(\phi\right) \propto \phi$, live on the boundary between
large-field and small-field models, with $V''\left(\phi\right) = 0$ and $\eta =
- \epsilon$. The spectral index and tensor/scalar ratio are related as:
\begin{equation}
r = {5 \over 3} \left(1 - n\right).
\end{equation}

This enumeration of models is certainly not exhaustive. There are a number of
single-field models that do not fit well into this scheme, for example
logarithmic potentials $V\left(\phi\right) \propto
\ln\left(\phi\right)$ typical of supersymmetry
\cite{lrreview}. Another example is potentials
with negative powers of the scalar field $V\left(\phi\right) \propto
\phi^{-p}$ used in intermediate inflation \cite{barrow93} and dynamical
supersymmetric inflation \cite{kinney97,kinney98}. Both of these cases require
an auxiliary field to end inflation and are more properly categorized as
hybrid models, but fall into the small-field region of the $r\,-\,n$ plane.
However, the three classes categorized by the relationship between the
slow-roll parameters as $-\epsilon <
\eta \leq \epsilon$ (large-field), $\eta \leq -\epsilon$ (small-field, linear),
and $0 < \epsilon < \eta$ (hybrid), cover the entire $r\,-\,n$ plane and are in
that sense complete.\footnote{Ref.\ \cite{kinney98a} incorrectly specified $0 <
\eta \leq \epsilon$ for large-field and $\eta < 0$ for small-field.} Figure 1
\cite{dodelson97} shows the $r\,-\,n$ plane divided into regions representing
the large field, small-field and hybrid cases. Figure 2 shows a ``zoo plot'' of
the particular potentials considered here plotted on the $r\,-\,n$ plane.

\section{Monte Carlo reconstruction}
\label{secmontecarlorecon}

In this section we describe {\em Monte Carlo reconstruction}, a stochastic
method for ``inverting'' observational constraints to determine an ensemble of
inflationary potentials compatible with observation. The method is described in
more detail in Refs.\ \cite{kinney02,easther02}.  In addition to encompassing a
broader set of models than we considered in Sec.\ \ref{seczoology}, Monte Carlo
reconstruction allows us easily to incorporate constraints on the running of
the spectral index $d n / d \ln{k}$ as well as to include effects to higher
order in slow roll.

We have defined the slow roll parameters $\epsilon$ and $\eta$ in terms of
the Hubble parameter $H\left(\phi\right)$ as
\begin{eqnarray}
\epsilon &\equiv& {m_{\rm Pl}^2 \over 4 \pi} \left({H'(\phi) \over
 H(\phi)}\right)^2,\cr
\eta\left(\phi\right) &\equiv& {m_{\rm Pl}^2 \over 4 \pi} 
\left({H''\left(\phi\right)
\over H\left(\phi\right)}\right).
\end{eqnarray}
These parameters are simply related to observables $r \simeq 10 \epsilon$, and
$n - 1 \simeq 4 \epsilon - 2 \eta$ to first order in slow roll. (We discuss
higher order expressions for the observables below.) Taking higher derivatives
of $H$ with respect to the field, we can define an infinite hierarchy of slow
roll parameters \cite{liddle94}:
\begin{eqnarray}
\sigma &\equiv& {m_{\rm Pl} \over \pi} \left[{1 \over 2} \left({H'' \over
 H}\right) -
\left({H' \over H}\right)^2\right],\cr
{}^\ell\lambda_{\rm H} &\equiv& \left({m_{\rm Pl}^2 \over 4 \pi}\right)^\ell
{\left(H'\right)^{\ell-1} \over H^\ell} {d^{(\ell+1)} H \over d\phi^{(\ell +
1)}}.
\end{eqnarray}
Here we have chosen the parameter $\sigma \equiv 2 \eta - 4 \epsilon \simeq n
-1 $ to make comparison with observation convenient.

It is convenient to use $N$ as the measure of time during inflation. As above,
we take $t_e$ and $\phi_e$ to be the time and field value at end of
inflation. Therefore, $N$ is defined as the number of e-folds before the end of
inflation, and increases as one goes {\em backward} in time ($d t > 0
\Rightarrow d N < 0$):
\begin{equation}
{d \over d N} = {d \over d\ln a} = { m_{\rm Pl} \over 2 \sqrt{\pi}}
\sqrt{\epsilon} {d \over d\phi},
\end{equation}
where we have chosen the sign convention that $\sqrt{\epsilon}$ has the same
sign as $H'\left(\phi\right)$:
\begin{equation}
\sqrt{\epsilon} \equiv + {m_{\rm PL} \over 2 \sqrt{\pi}} {H' \over H}.
\end{equation}
Then $\epsilon$ itself can be expressed in terms of $H$ and $N$ simply as,
\begin{equation}
\label{eqepsilonfromN}
{1 \over H} {d H \over d N} = \epsilon.
\end{equation}
Similarly, the evolution of the higher order parameters during inflation is
determined by a set of ``flow'' equations \cite{hoffman00,schwarz01,kinney02},
\begin{eqnarray}
{d \epsilon \over d N} &=& \epsilon \left(\sigma + 2
\epsilon\right),\cr {d \sigma \over d N} &=& - 5 \epsilon \sigma - 12
\epsilon^2 + 2 \left({}^2\lambda_{\rm H}\right),\cr {d
\left({}^\ell\lambda_{\rm H}\right) \over d N} &=& \left[
\frac{\ell - 1}{2} \sigma + \left(\ell - 2\right) \epsilon\right]
\left({}^\ell\lambda_{\rm H}\right) + {}^{\ell+1}\lambda_{\rm
H}.\label{eqfullflowequations}
\end{eqnarray}
The derivative of a slow roll parameter at a given order is higher order in
slow roll. A boundary condition can be specified at any point in the
inflationary evolution by selecting a set of parameters
$\epsilon,\sigma,{}^2\lambda_{\rm H},\ldots$ for a given value of $N$. This is
sufficient to specify a ``path'' in the inflationary parameter space that
specifies the background evolution of the spacetime. Taken to infinite order,
this set of equations completely specifies the cosmological evolution, up to
the normalization of the Hubble parameter $H$. Furthermore, such a
specification is exact, with no assumption of slow roll necessary. In practice,
we must truncate the expansion at finite order by assuming that the
${}^\ell\lambda_{\rm H}$ are all zero above some fixed value of $\ell$.  We
choose initial values for the parameters at random from the following ranges:
\begin{eqnarray}
N &=& [40,70]\cr
\epsilon &=& \left[0,0.8\right]\cr
\sigma &=& \left[-0.5,0.5\right]\cr
{}^2\lambda_{\rm H} &=& \left[-0.05,0.05\right]\cr
{}^3\lambda_{\rm H} &=& \left[-0.025,0.025\right],\cr
&\cdots&\cr
{}^{M+1}\lambda_{\rm H} &=& 0.\label{eqinitialconditions}
\end{eqnarray}
Here the expansion is truncated to order $M$ by setting ${}^{M+1}\lambda_{\rm
H} = 0$. In this case, we still generate an exact solution of the background
equations, albeit one chosen from a subset of the complete space of
models. This is equivalent to placing constraints on the form of the potential
$V\left(\phi\right)$, but the constraints can be made arbitrarily weak by
evaluating the expansion to higher order. For the purposes of this analysis, we
choose $M = 5$.  The results are not sensitive to either the choice of order
$M$ (as long as it is large enough) or to the specific ranges from which the
initial parameters are chosen.

Once we obtain a solution to the flow equations
$[\epsilon(N),\sigma(N),{}^\ell\lambda_{\rm H}(N)]$, we can calculate the
predicted values of the tensor/scalar ratio $r$, the spectral index $n$, and
the ``running'' of the spectral index $d n / d\ln k$.  To lowest order, the
relationship between the slow roll parameters and the observables is especially
simple: $r = 10 \epsilon$, $n - 1 = \sigma$, and $d n / d \ln k = 0$. To
second order in slow roll, the observables are given by
\cite{liddle94,stewart93},
\begin{equation}
r = 10 \epsilon \left[1 - C \left(\sigma + 2
 \epsilon\right)\right],\label{eqrsecondorder}
\end{equation}
for the tensor/scalar ratio, and 
\begin{equation}
n - 1 = \sigma - \left(5 - 3 C\right) \epsilon^2 - {1 \over 4} \left(3
- 5 C\right) \sigma \epsilon + {1 \over 2}\left(3 - C\right)
\left({}^2\lambda_{\rm H}\right)\label{eqnsecondorder}
\end{equation}
for the spectral index. The constant $C \equiv 4 (\ln{2} +
\gamma) - 5 = 0.0814514$, where $\gamma \simeq 0.577$ is Euler's
constant.\footnote{Some earlier papers \cite{kinney02,easther02}, due to a long
unnoticed typographic error in Ref.\ \cite{liddle94}, used an incorrect value
for the constant $C$, given by $C \equiv 4 (\ln{2} + \gamma) = 5.0184514$. The
effect of this error is significant at second order in slow roll.}  Derivatives
with respect to wavenumber $k$ can be expressed in terms of derivatives with
respect to $N$ as \cite{liddle95}
\begin{equation}
{d \over d N} = - \left(1 - \epsilon\right) {d \over d \ln k},
\end{equation}
The scale dependence of $n$ is then  given by the simple expression
\begin{equation}
{d n \over d \ln k} = - \left({1 \over 1 - \epsilon}\right) {d n \over d N},
\end{equation}
which can be evaluated by using Eq.~(\ref{eqnsecondorder}) and the flow
equations.  For example, for the case of $V \propto \phi^4$, 
the observables to lowest order are
\begin{eqnarray}
\label{eqphi4obs}
r &\simeq& {10 \over N + 1},\cr
n - 1 &\simeq& - {3 \over N + 1},\cr
{dn \over d\ln k} &\simeq& - {3 \over N \left(N + 1\right)}.
\end{eqnarray}
The final result following the evaluation of a particular path in the
$M$-dimensional ``slow roll space'' is a point in ``observable parameter
space,'' i.e., $(r,n,dn/d\ln k)$, corresponding to the observational prediction
for that particular model.  This process can be repeated for a large number of
models, and used to study the attractor behavior of the inflationary
dynamics. In fact, the models cluster strongly in the observable parameter
space~\cite{kinney02}. Figure \ref{figzooplot2} shows an ensemble of models
generated stochastically on the $(r,n)$ plane, along with the predictions of
the specific models considered in Sec. \ref{seczoology}.

Figure \ref{nnk} shows an ensemble of models generated stochastically on the
$(n,dn/d\ln k)$ plane. As one can see, and contrary to what commonly believed,
there are single-field models of inflation which predict a significant running
of the spectral index.  The same can be appreciated in Fig.\ \ref{figrnk},
where we plot an ensemble of models generated stochastically on the $(r,dn/d\ln
k)$ plane.

The reconstruction method works as follows:
\begin{enumerate}
\item Specify a ``window'' of parameter space: {\em e.g.,} central values for
$n-1$, $r$, or $d n /d \ln{k}$ and their associated error bars.
\item Select a random point in slow roll space, 
$[\epsilon,\eta,{}^\ell\lambda_{\rm H}]$, truncated at order $M$ in
the slow roll expansion.
\item Evolve forward in time ($d N < 0$) until either (a) inflation ends
 ($\epsilon > 1$), or (b) the evolution reaches a late-time fixed
 point ($\epsilon = {}^\ell\lambda_{\rm H} = 0,\ \sigma = {\rm
 const.}$).
\item If the evolution reaches a late-time fixed point, calculate the
 observables $r$, $n - 1$, and $d n / d \ln k$ at this point.
\item If inflation ends, evaluate the flow equations backward $N$ e-folds from
 the end of inflation. Calculate the observable parameters at that
 point.
\item If the observable parameters lie within the specified window of
parameter  space, compute the potential and add this model to the ensemble 
of ``reconstructed'' potentials.
\item Repeat steps 2 through 6 until the desired number of models
have been found.
\end{enumerate}

The condition for the end of inflation is that $\epsilon = 1$. Integrating the
flow equations forward in time will yield two possible outcomes. One
possibility is that the condition $\epsilon = 1$ may be satisfied for some
finite value of $N$, which defines the end of inflation. We identify this point
as $N=0$ so that the primordial fluctuations are actually generated when $N
\sim 60$. Alternatively, the solution can evolve toward an inflationary
attractor with $r = 0$ and $n > 1$, in which case inflation never
stops.\footnote{See Ref.\ \cite{kinney02} for a detailed discussion of the
fixed-point structure of the slow roll space.}  In reality, inflation must stop
at some point, presumably via some sort of instability, such as the
``hybrid'' inflation mechanism \cite{linde91,linde94,copeland94}. Here we make
the simplifying assumption that the observables for such models are the values
at the late-time attractor.

Given a path in the slow roll parameter space, the form of the potential
is fixed, up to normalization \cite{hodges90,copeland93,beato00,easther02}.
The starting point is the Hamilton-Jacobi equation,
\begin{equation}
V(\phi) = \left({3 m_{\rm Pl}^2} \over 8 \pi\right)
H^2(\phi) \left[1 - {1\over 3}
\epsilon(\phi)\right].\label{eqHJpotential}
\end{equation}
We have $\epsilon(N)$ trivially from the flow equations. In order to calculate
the potential, we need to determine $H(N)$ and $\phi(N)$. With $\epsilon$
known, $H(N)$ can be determined by inverting the definition of $\epsilon$, Eq.\
(\ref{eqepsilonfromN}).  Similarly, $\phi(N)$ follows from the first
Hamilton-Jacobi equation (\ref{eqbasichjequations}):
\begin{equation}
{d \phi \over d N} = {m_{\rm PL} \over 2 \sqrt{\pi}}
\sqrt{\epsilon}.
\end{equation}
Using these equations and Eq.~(\ref{eqHJpotential}), the form of the potential
can then be fully reconstructed from the numerical solution for $\epsilon(N)$.
The only necessary observational input is the normalization of the Hubble
parameter $H$, which enters the above equations as an integration constant.
Here we use the simple condition that the density fluctuation amplitude (as
determined by a first-order slow roll expression) be of order $10^{-5}$,
\begin{equation}
{\delta \rho \over \rho} \simeq \frac{1}{2\pi} {H \over m_{\rm Pl}} 
\frac{1}{\sqrt{\epsilon}} = 10^{-5}.
\end{equation}
A more sophisticated treatment would perform a full normalization to
the COBE CMB data \cite{bunn94,stompor95}.  The value of the field,
$\phi$, also contains an arbitrary, additive constant.

\section{CMB Analysis}
\label{secCMBanalysis}

Our analysis method is based on the computation of a likelihood
distribution over a fixed grid of pre-computed theoretical models.
We restrict our analysis to a flat, adiabatic, $\Lambda$-CDM model template
computed with CMBFAST (\cite{sz}), sampling the parameters as follows:
$\Omega_{cdm}h^2\equiv \omega_{cdm}= 0.01,...0.25$, in steps of $0.01$;
$\Omega_{b}h^2\equiv\omega_{b} = 0.009, ...,0.028$, in steps of $0.001$ and
$\Omega_{\Lambda}=0.5, ..., 0.95$, in steps of $0.05$. The value of the Hubble
constant is not an independent parameter, since:
\begin{equation}
h=\sqrt{{\omega_{cdm}+\omega_b} \over {1-\Omega_{\Lambda}}} ,
\end{equation}
and we use the further prior: $h=0.72\pm0.15$.  We allow for a reionization of
the intergalactic medium by varying the Compton optical depth parameter
$\tau_c$ in the range $\tau_c=0.05,...,0.30$ in steps of $0.05$.  Our choice of
the above parameters is motivated by Big Bang Nucleosynthesis bounds on
$\omega_b$ (both from D \cite{burles} and $^4\textrm{He}+^7\textrm{Li}$
\cite{cyburt}), from supernovae (\cite{super1}) and galaxy clustering
observations (see {\em e.g.,} \cite{thx}), and by the WMAP
temperature-polarization cross-correlation data, which indicate an optical
depth $\tau = 0.17\pm 0.04$ (\cite{kogutt}).  Our choice for an upper limit of
$\tau_c < 0.30$ is very conservative respect to the maximum values expected in
numerical simulations (see {\rm e.g.,} \cite{ciardi}) even in the case of
non-standard reionization processes.  From the grid above, we only consider
models with an age of the universe in excess of 11 Gyr.  Variations in the
inflationary parameters $n$, $r$ and $dn /d\ln{k}$ are not computationally
relevant, and for the range of values we considered they can be assumed as free
parameters.  The tensor spectral index $n_t$ is determined by the consistency
relation.

For the WMAP data we use the recent temperature and cross polarization results
from Ref.\ \cite{wmap1} and compute the likelihood ${\cal L}^{WMAP}$ for each
theoretical model as explained in Ref.\ \cite{Verde:2003ey}, using the publicly
available code on the LAMBDA web site ({http://lambda.gsfc.nasa.gov/}).

We further include the results from seven other experiments:
BOOMERanG-98~\cite{ruhl}, MAXIMA-1
\cite{lee}, DASI~\cite{halverson}, CBI~\cite{cbi}, ACBAR~\cite{acbar},
VSAE~\cite{grainge}, and Archeops~\cite{benoit}. The expected
theoretical Gaussian signal inside the bin $C_B^{th}$ is computed by using the
publicly available window functions and lognormal prefactors as in Ref.\
\cite{BJK}.  The likelihood ${\cal L}^{pre-WMAP}$ from this dataset and for a
given theoretical model is defined by
\begin{equation}
-2\ln{\cal L}^{pre-WMAP}=(C_B^{th}-C_B^{ex})M_{BB'}(C_{B'}^{th}-C_{B'}^{ex}),
\end{equation}
where $M_{BB'}$ is the Gaussian curvature of the likelihood matrix at the peak
and $C_B^{ex}$ is the experimental signal in the bin.  We consider $7 \%$, $10
\%$, $4 \%$, $5 \%$, $5 \%$, $5 \%$ and $5 \%$ Gaussian distributed calibration
errors for the Archeops, BOOMERanG-98, DASI, MAXIMA-1, VSAE, ACBAR, and CBI,
experiments respectively and include the beam uncertainties using the
analytical marginalization method presented in \cite{bridle}.  We use as
combined likelihood just the normalized product of the two likelihood
distributions: ${\cal L}\sim{\cal L}^{WMAP}\times{\cal
L}^{pre-WMAP}$.\footnote{We do not take into account correlations in the
variance of the experimental data introduced by observations of the same
portions of the sky (such as in the case of WMAP and Archeops, for example).
We have verified that such correlations have negligible effect by removing one
experiment at a time and testing for the stability of our results.}

In order to constrain a set of parameters $\vec{x}$ we marginalize over the
values of the remaining ``nuisance'' parameters $\vec{y}$. This yields the
marginalized likelihood distribution
\begin{equation}
\mathcal{L}(\vec{x}) \equiv
        P(\vec{x}|{\cal C}_B) =
        \int {\cal L}(\vec{x},\vec{y})  d\vec{y}.
\label{likelihood1}
\end{equation}

In the next section we will present constraints in several two-dimensional
planes.  To construct plots in two dimensions, we project (not marginalize)
over the third, ``nuisance'' parameter. For example, likelihoods in the $(r,n)$
plane are calculated for given choice of $r$ and $n$ by using the value of
$dn/d\ln{k}$ which maximizes the likelihood function. The error contours are
then plotted relative to likelihood falloffs of $0.17$, $0.018$ and $0.0035$ as
appropriate for $1\sigma$, $2\sigma$, and $3\sigma$ contours of a {\em
three-dimensional} Gaussian. In effect we are taking the shadow of the
three-dimensional error contours rather than a slice, which makes clear the
relationship between the error contours and the points generated by Monte
Carlo. All likelihoods used to constrain models are calculated relative the the
full three-dimensional likelihood function.

\section{Results}

We will plot the likelihood contours obtained from our analysis on three
different planes: $dn/d\ln{k}$ vs.\ $n$, $r$ vs.\ $n$ and $r$ vs.\
$dn/d\ln{k}$.  Presenting our results on these planes is useful for
understanding the effects of theoretical assumptions and/or external priors.

We do this in Fig.\ \ref{fig_like} for two cases: the WMAP dataset alone (left
column), and WMAP plus the additional CMB experiments BOOMERanG-98, MAXIMA-1,
DASI, CBI, ACBAR, VSAE, and Archeops (right column).  By analyzing these
different datasets we can check the consistency of the previous experiments
with WMAP. The dots superimposed on the likelihood contours show the models
sampled by the Monte Carlo reconstruction.

In the top row of Fig.\ \ref{fig_like}, we show the 68\%, 95\%, and 99\%
likelihood contours on the $n$ vs.\ $dn/d\ln{k}$ plane (refer to the end of
Sec.\ \ref{secCMBanalysis} for a discussion of the method used to plot the
contours.) The pivot scale $k_0$ is $k_0=0.002 h$ Mpc$^{-1}$.  As we can see,
both datasets are consistent with a scale invariant $n=1$ power law spectrum
with no further scale dependence ($dn/d\ln{k}=0$). A degeneracy is also
evident: an increase in the spectral index $n$ is equivalent to a negative
scale dependence ($dn/d\ln{k} < 0$). We emphasize, however, that this beahavior
depends strictly on the position of the pivot scale $k_0$: choosing $k_0=0.05h$
Mpc$^{-1}$ would change the direction of the degeneracy.  Models with $n \sim
1.1$ need a negative running at about the $3\sigma$ level.  It is interesting
also to note that models with lower spectral index, $n \sim 0.9$, are in better
agreement with the data with a zero or positive running.  For $n<1$, the
running is bounded by $0.005 \gtrsim dn/d\ln{k} \gtrsim -0.025$ at $1\sigma$.

In the center row of Fig.\ \ref{fig_like} we plot the 68\%, 95\%, and 99\%
likelihood contours on the $r$ vs.\ $n$ plane.  As we can see, the present data
only weakly constrain the presence of tensor modes, although a gravity wave
component is not preferred. Models with $n<0.9$ must have a negligible tensor
component, while models with $n>1$ can have $r$ larger than $0.4$ ($2\sigma$
C.L.). However, as we can see from the bottom row of Fig.\ \ref{fig_like},
there is no correlation between the tensor component and the running of the
scalar index.

Figures \ref{fig_potentials}, \ref{fig_xsigma}, and \ref{fig_xsigma_rescaled}
show a subset of 300 reconstructed potentials selected from the sampled
set. Note in particular the wide range of inflationary energy scales compatible
with the observational constraint. This is to be expected, since there was no
detection of tensor modes in the WMAP data, which would be seen here as a
detection of a nonzero tensor/scalar ratio $r$. In addition, the {\em shape} of
the inflationary potential is also not well constrained by WMAP.

Figure \ref{fig_like_zoo} shows the models sampled by the Monte Carlo
categorized by their ``zoology'', {\em i.e.,} whether they fit into the
category of small-field, large-field, or hybrid. We see that all three types of
potential are compatible with the data, although hybrid class models are
preferred by the best-fit region. 

Figures \ref{fig_V_zoo}, \ref{fig_V_zoo_type}, and 
\ref{fig_V_zoo_type_rescaled} show the reconstructed potentials divided by 
type. Perhaps the only conclusion to be drawn here is that the WMAP data places
no significant constraint on the shape of the inflationary potential; many
``reasonable'' potentials are consistent with the data. However, significant
portions of the observable parameter space are ruled out by WMAP, and future
observations can be expected to significantly tighten these constraints
\cite{dodelson97,kinney98a}. 

For the particular example of $V \propto \phi^4$, using Eq.\ \ref{eqphi4obs} we
find that this choice of potential ruled out to $3\sigma$ only for $N < 40$ for
the WMAP data set. This constraint is even weaker than that claimed by Barger
{\it et al.}\ \cite{barger}, most likely because we allow for a running of the
spectral index in our constraint.

When augmenting the WMAP data set with the data from seven other CMB
experiments, the most noticeable improvement in the constraints is a better
upper limit to the tensor/scalar ratio $r$, which results in a slightly
improved upper limit on the height of the potential. Also, the width of the
reconstructed potentials in Planck units is somewhat less than in the case of
the WMAP-only constraint, showing that the additional data more strongly limit
the form of the inflationary potential. 

The combined data rules out a potential with $V \propto \phi^4$ for $N < 66$ to
$3\sigma$, effectively killing such models as observationally viable candidates
for the inflaton potential.

\section{Conclusions}

In this paper, we presented an analysis of the WMAP data set with an emphasis
on parameters relevant for distinguishing among the various possible models for
inflation. In contrast to previous analyses, we confined ourselves to CMB data
only and performed a likelihood analysis over a grid of models including the
full error covariance data from the WMAP satellite alone, and in conjunction
with measurements from BOOMERanG-98, MAXIMA-1, DASI, CBI, ACBAR, VSAE, and
Archeops.

We found that the WMAP data alone are consistent with a scale-invariant power
spectrum, $n = 1$, with no running of the spectral index, $dn/d\ln{k} =
0$. However, a great number of models indicates a compatibility of the data
with a blue spectral index and a substantial negative running. This is
consistent with the result published by the WMAP team including data from
large-scale structure measurements and the Lyman-$\alpha$ forest. The WMAP
result is also consistent with previous CMB experiments. The inclusion of
previous datasets in the analysis has the effect of reducing the error bars and
give a better determination of the inflationary parameters. Still, no clear
evidence for the running is present in the combined analysis.  This result
differs from the result obtained in Peiris {\em et al.}\ in the case of
combined (WMAP+CBI+ACBAR) analysis, where a mild (about $1.5\sigma$) evidence
for running was reported.  The different and more conservative method of
analysis adopted here, and the larger CMB dataset used in our paper can explain
this difference. 

In addition, we applied the Monte Carlo reconstruction technique to generate an
ensemble of inflationary potentials consistent with observation. Of the three
basic classes of inflation model, small-field, large-field, and hybrid, none
are conclusively ruled out, although hybrid models are favored by the best fit
region. The reconstruction process indicates that a wide variety of smooth
potentials for the inflaton are consistent with the data, indicating that the
WMAP result is too crude to significantly constrain either the height or the
shape of the inflaton potential.  In particular, the lack of evidence for
tensor fluctuations makes it impossible to constrain the energy scale at which
inflation takes place. Nonetheless, WMAP rules out a large portion of the
available parameter space for inflation, itself a significant improvement over
previous measurements.  For the particular case of a potential of the form
$V\left(\phi\right) = \lambda \phi^4$, WMAP rules out all such potentials for
$N < 40$ at the $3\sigma$ level, which means that $\phi^4$ potentials are not
conclusively ruled out by WMAP alone. The combined data set, however, rules out
$\phi^4$ models for $N < 66$, which kills such potentials as viable candidates
for inflation.

After this paper first appeared, Leach and Liddle also released a reanalysis of
the WMAP data \cite{leach03}, which is in general agreement with our results
here.

\section*{Acknowledgments}

W.H.K.\ is supported by ISCAP and the Columbia University Academic Quality
Fund.  ISCAP gratefully acknowledges the generous support of the Ohrstrom
Foundation.  E.W.K.\ is supported in part by NASA grant NAG5-10842.  We thank
Richard Easther for helpful conversations and for the use of computer code.



\newpage

\begin{widetext}

\begin{figure}[b]
\centerline{\includegraphics[width=4.0in]{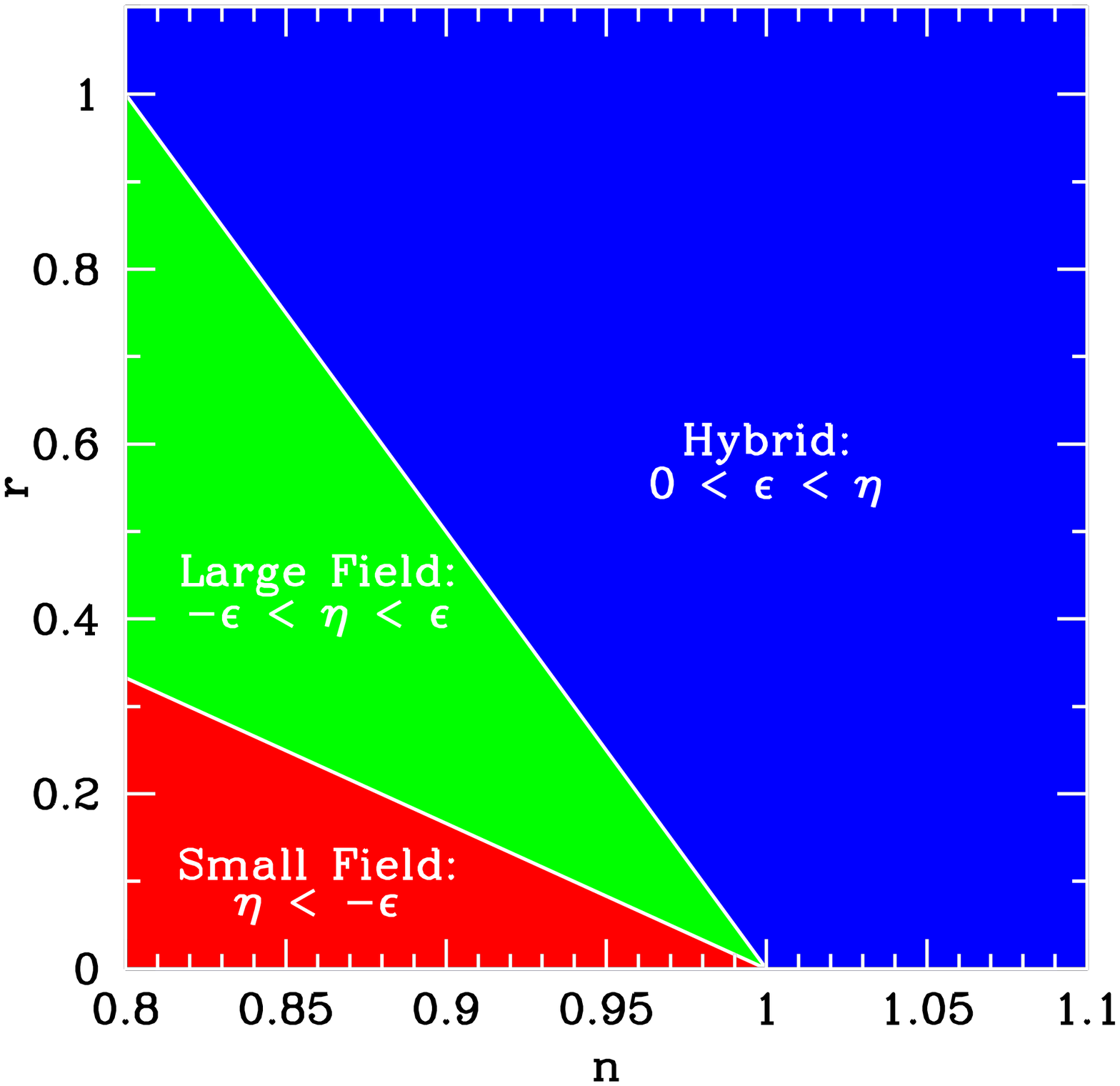}}  
\caption{\label{figregions} Regions on the $r\,-\,n$ plane. The different 
types of potentials, small field, large field, and hybrid, occupy different
regions of the observable parameter space.}
\end{figure}

\begin{figure}
\centerline{\includegraphics[width=4.0in]{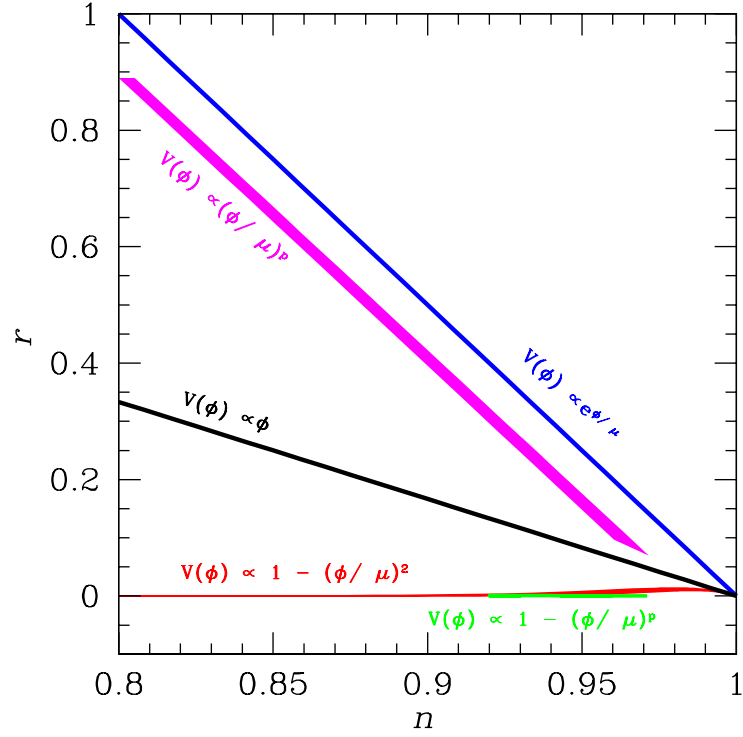}}  
\caption{\label{figzooplot1} A ``zoo plot'' of models in the $r\,-\,n$ plane,
plotted to first order in slow roll.}
\end{figure}

\begin{figure}
\centerline{\includegraphics[width=4.0in]{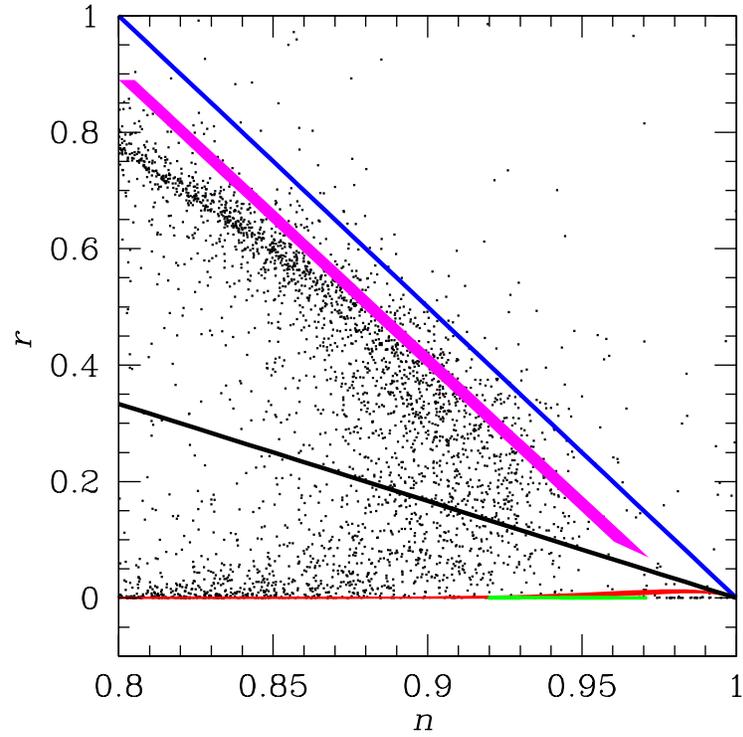}}  
\caption{\label{figzooplot2} Models generated by Monte Carlo plotted 
on the $(r,n)$ plane (black dots). The colored lines are the same models as in
Fig. 2. For comparison with the models, points are plotted to first order in 
slow roll.}
\end{figure}

\begin{figure}
\centerline{\includegraphics[width=4.0in]{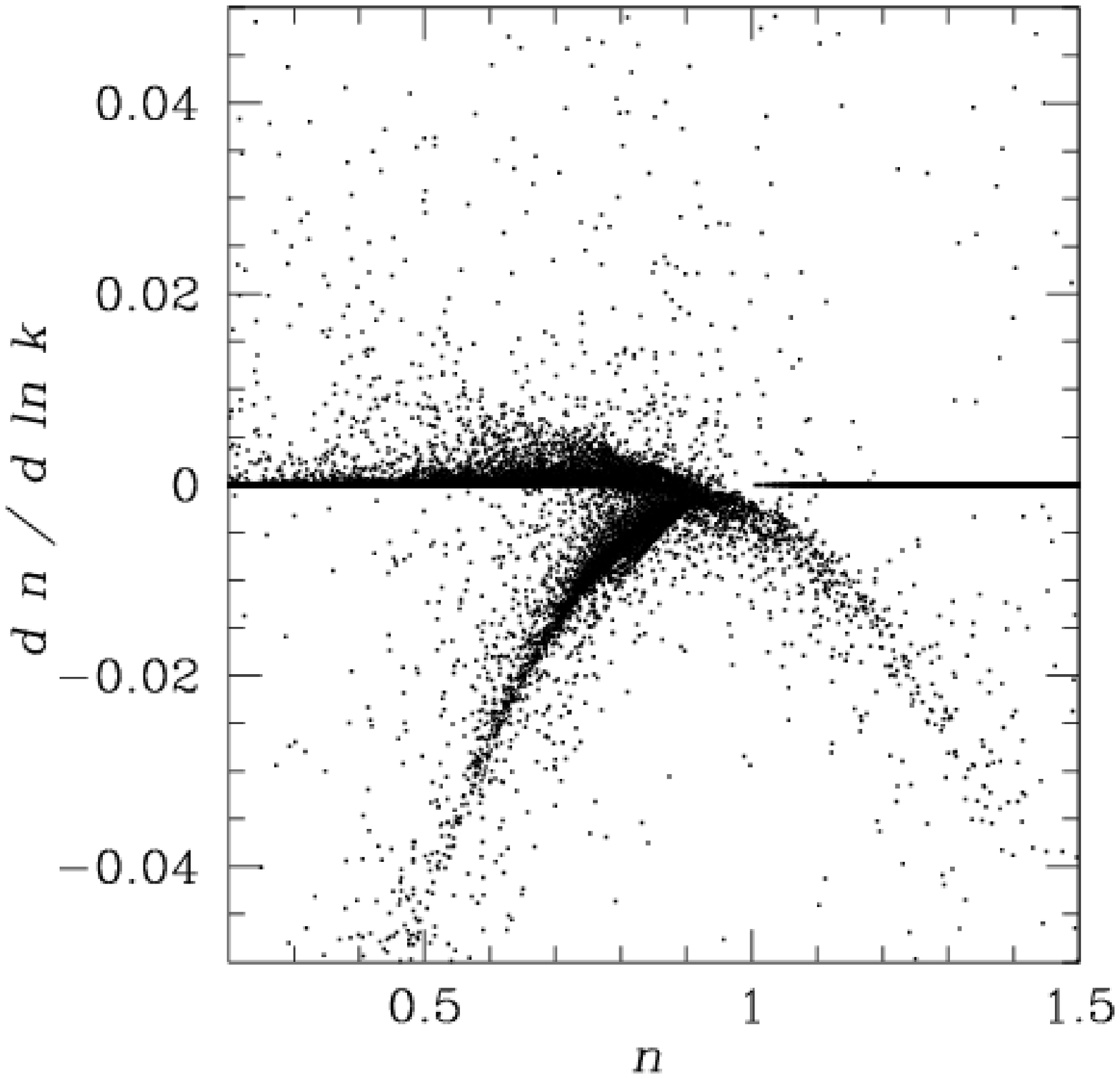}}
\caption{\label{nnk} Models generated by Monte Carlo plotted
on the $(n,dn/d\ln k)$ plane. Points are plotted to second order
in slow roll.}
\end{figure}

\begin{figure}
\centerline{\includegraphics[width=4.0in]{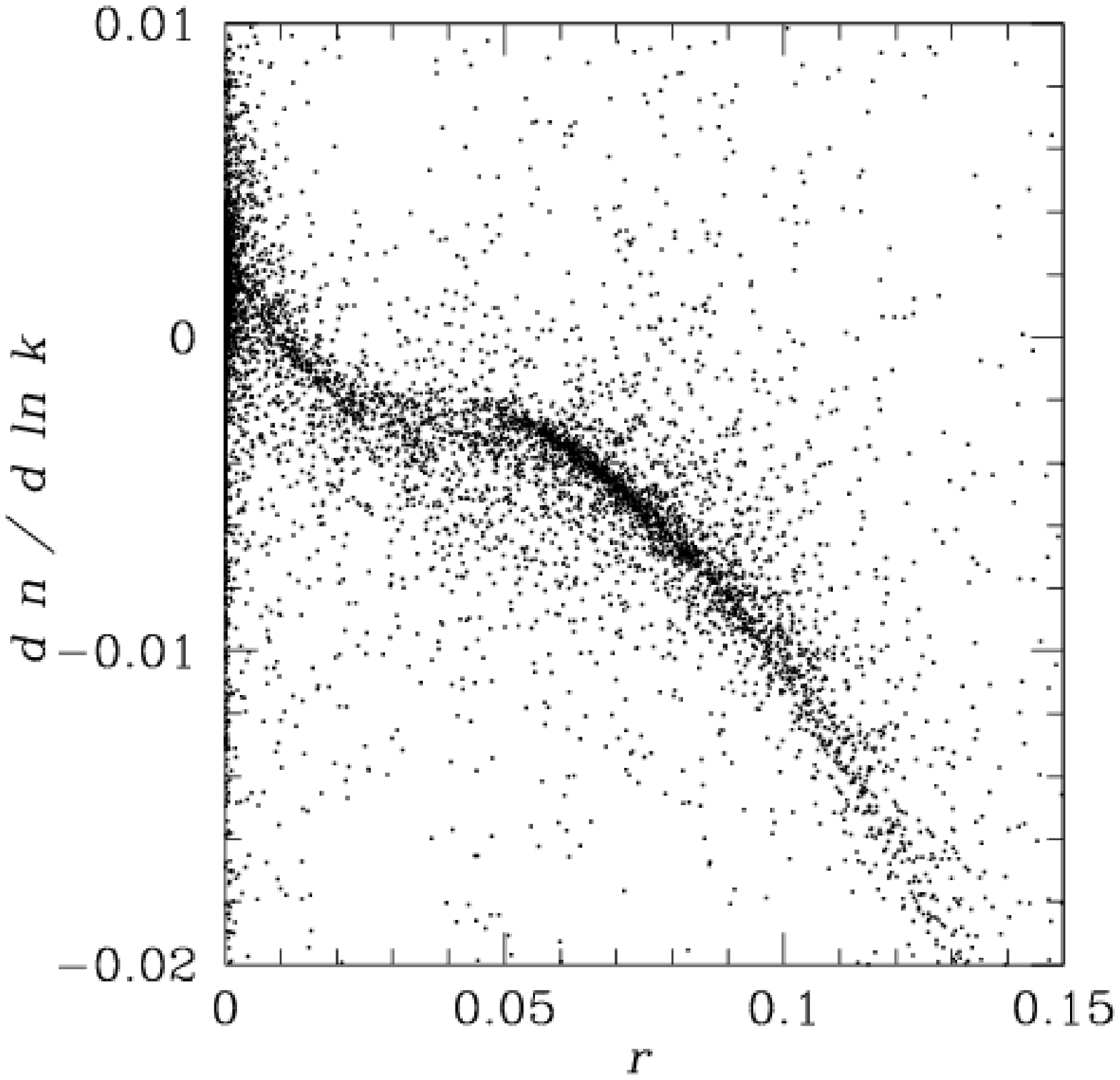}}
\caption{\label{figrnk} Models generated by Monte Carlo plotted 
on the $(r,dn/d\ln k)$ plane. Points are plotted to second order in slow roll.}
\end{figure}

\begin{figure}
\hspace*{1.0in} {\bf {\large WMAP only\ \ \ \ }} \hspace*{1.0in} 
{\bf {\large WMAP plus seven other experiments}}
\centerline{\includegraphics[width=2.7in]{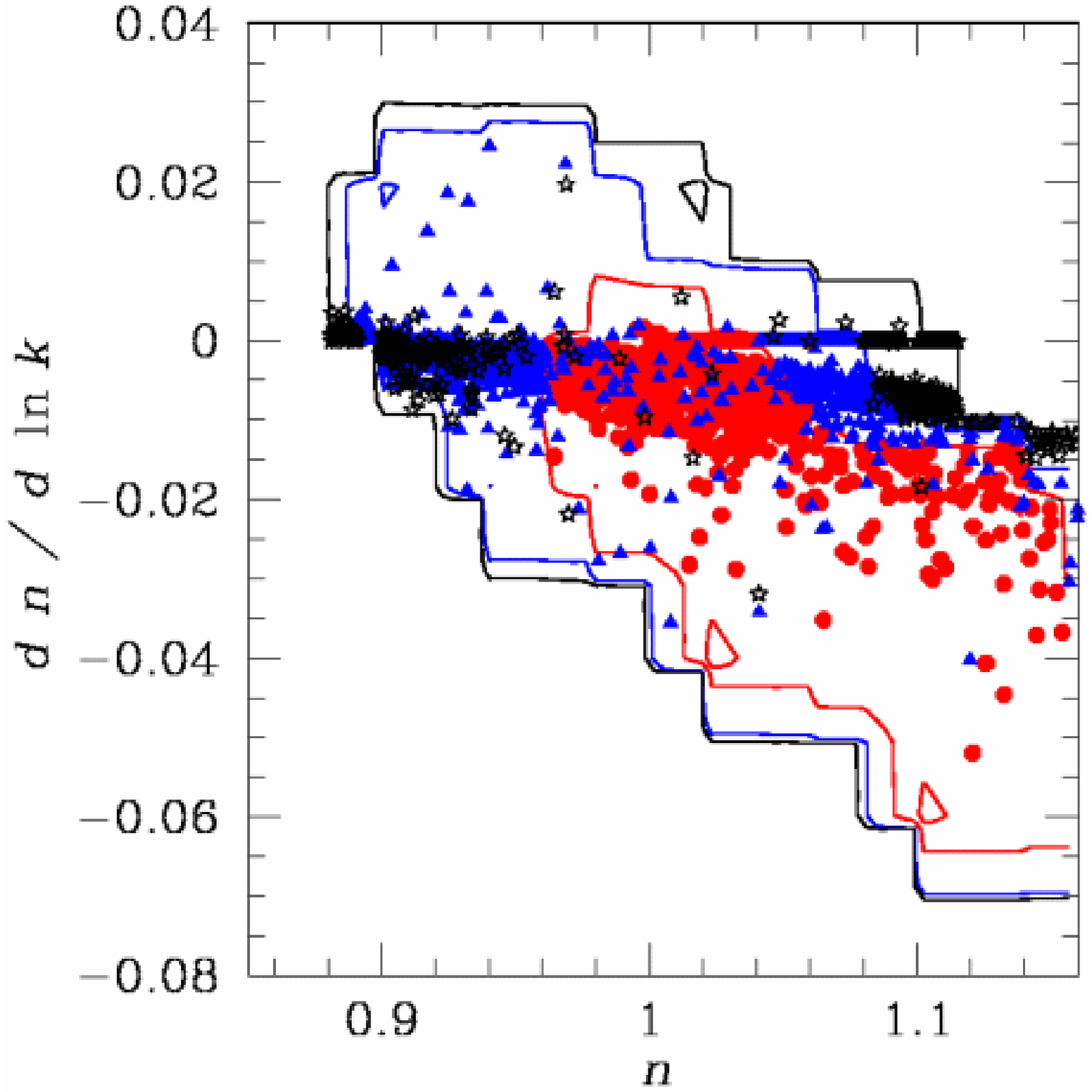} 
\hspace*{24pt}
\includegraphics[width=2.7in]{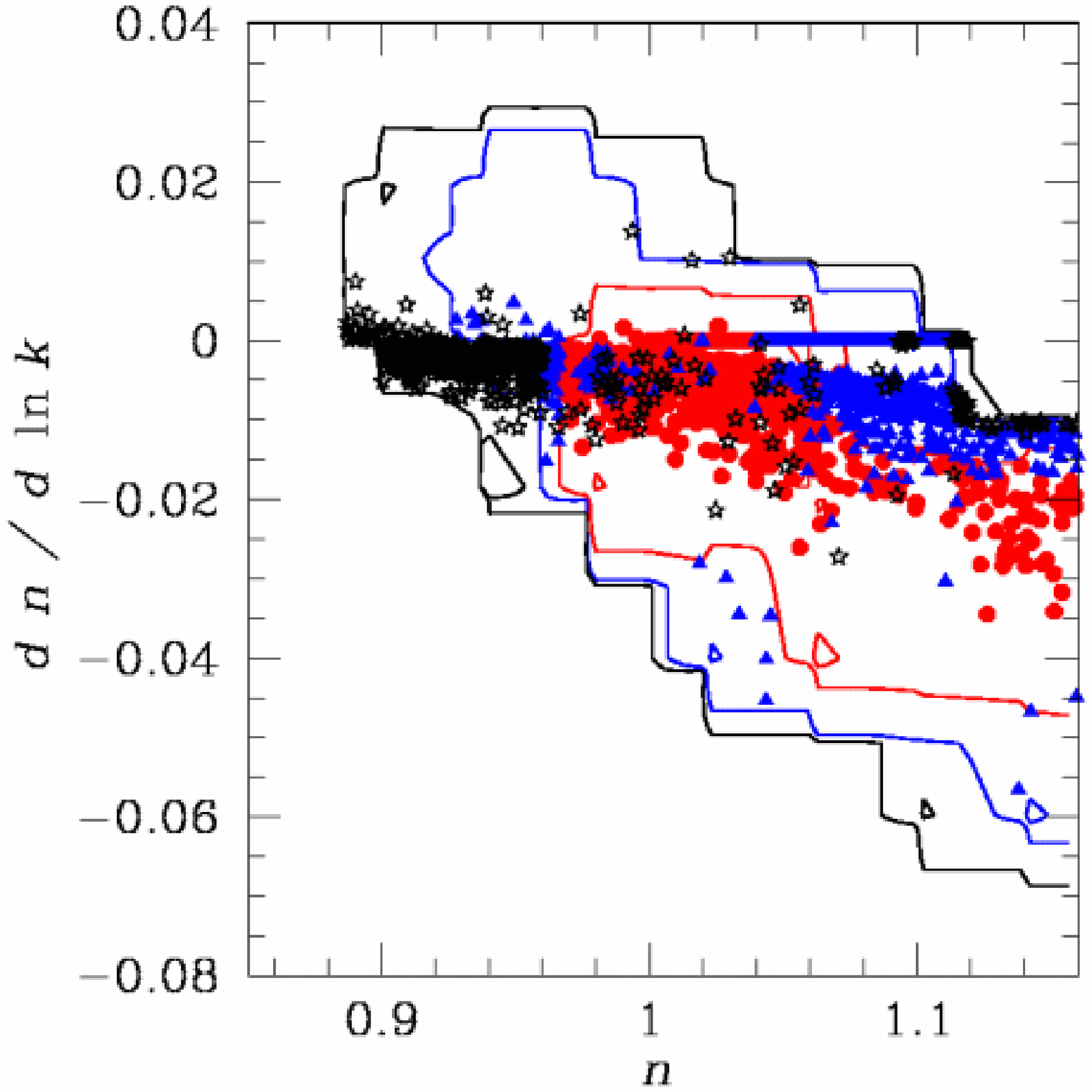}}
\centerline{\includegraphics[width=2.7in]{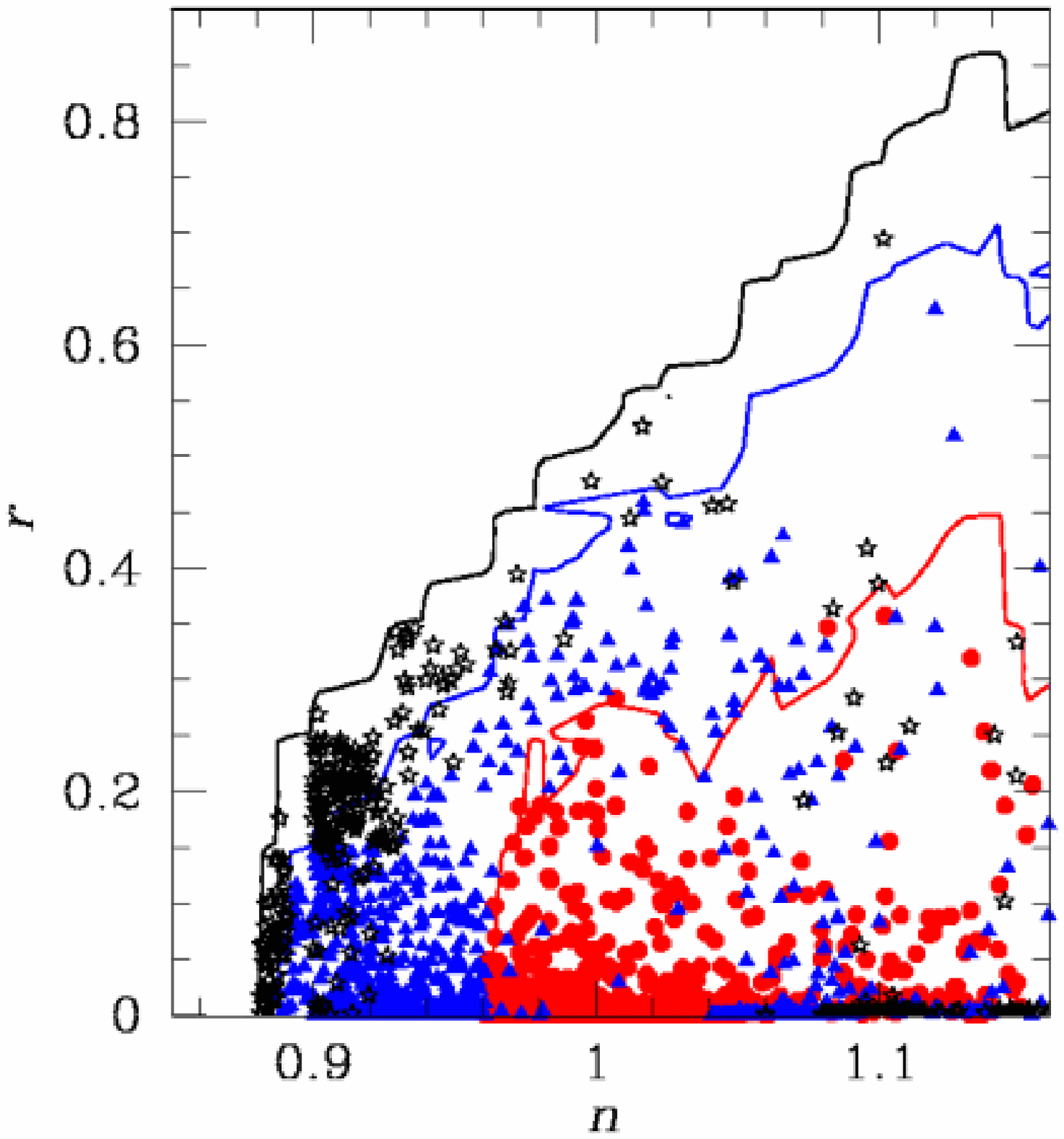}  
\hspace*{24pt}
\includegraphics[width=2.7in]{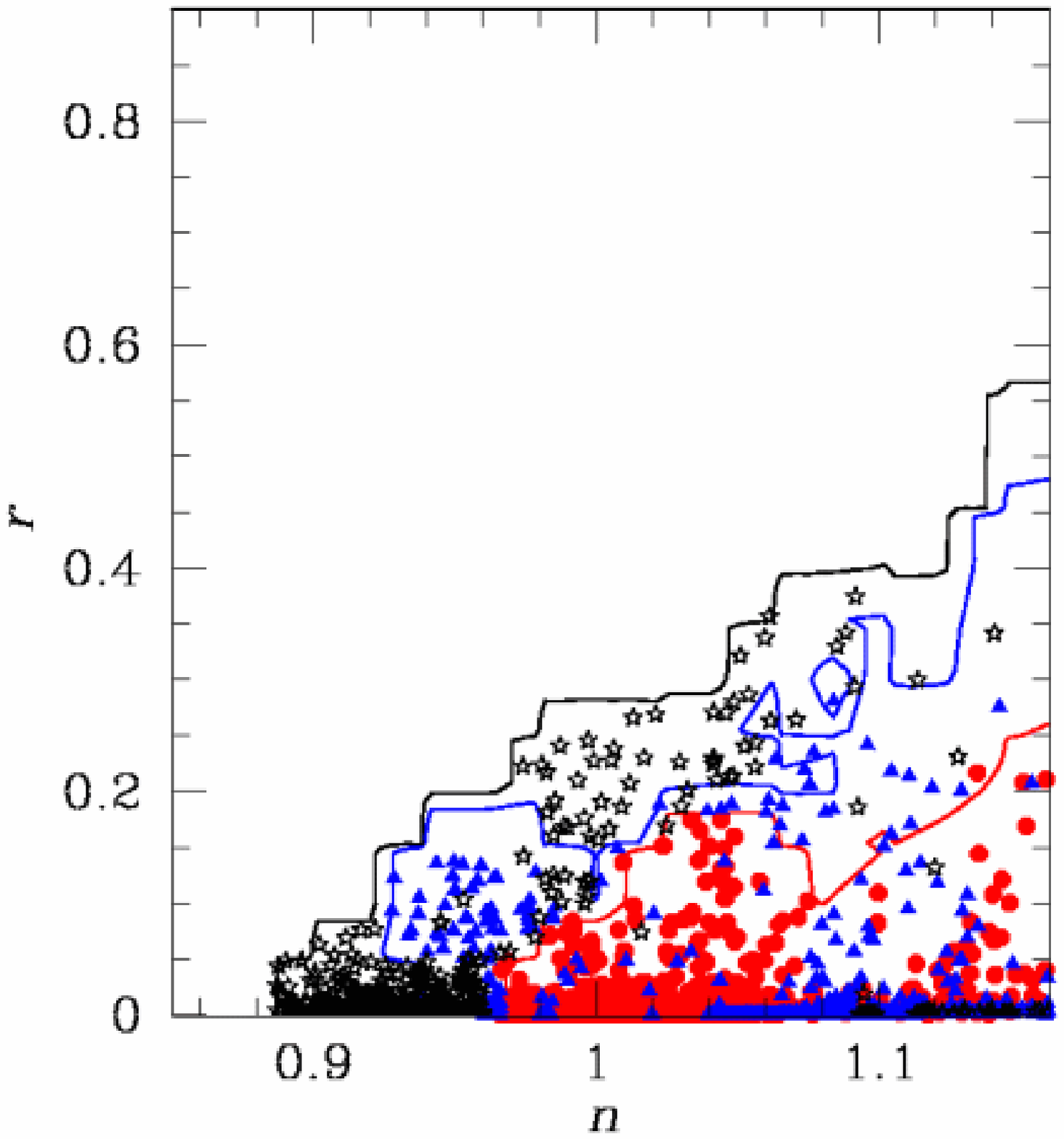}}
\centerline{\includegraphics[width=2.7in]{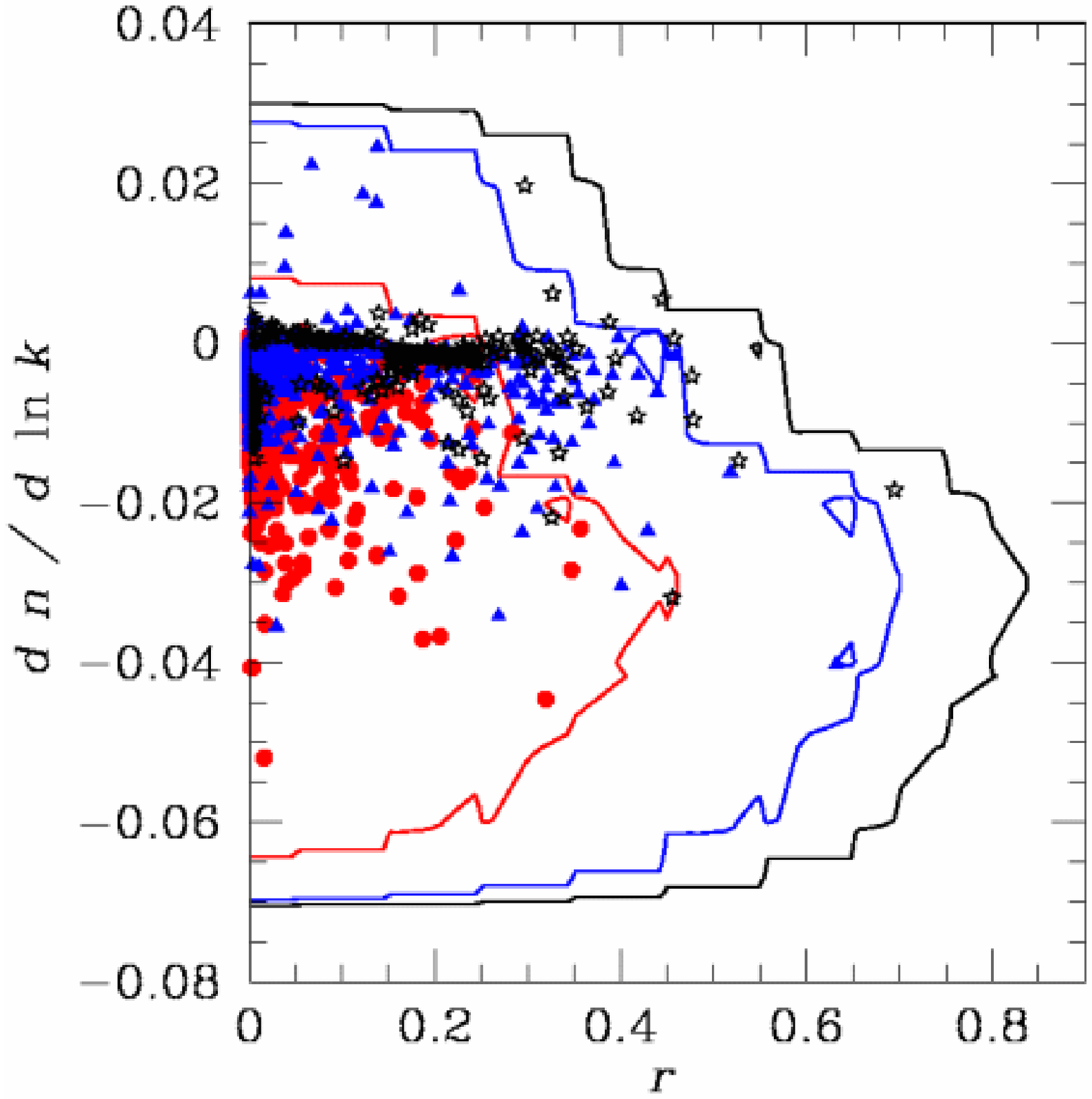}  
\hspace*{24pt}
\includegraphics[width=2.7in]{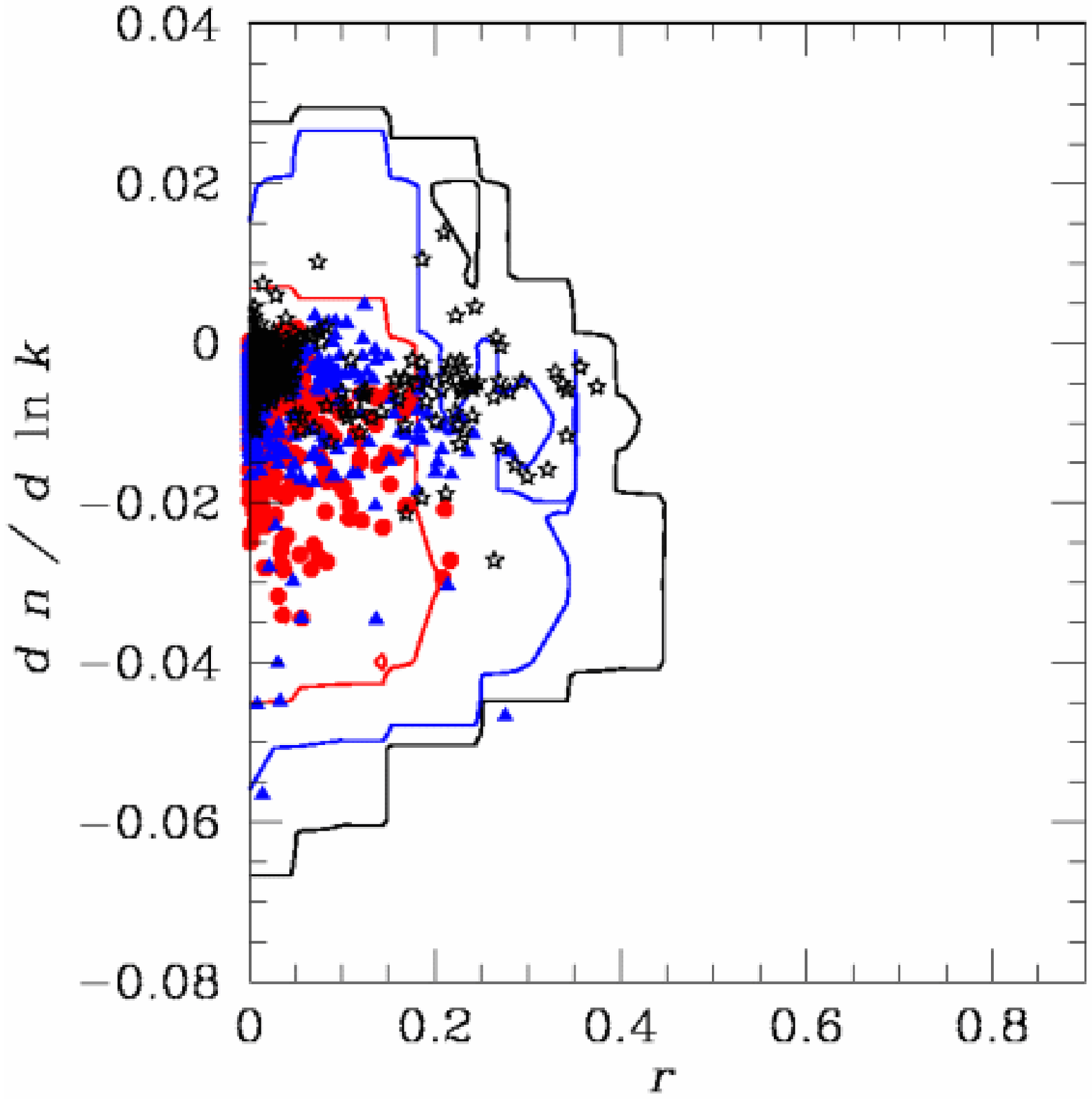}}
\caption{\label{fig_like} Likelihood contours for WMAP alone (left column)
and WMAP plus seven other experiments (right column), plotted in the
($n,dn/d\ln{k}$) plane (top), the ($n,r$) plane (center), and the
($r,dn/d\ln{k}$) plane (bottom).  The points represent the results of the Monte
Carlo sampling of inflationary models consistent with $1\sigma$ (red, dots),
$2\sigma$ (blue, triangles), and $3\sigma$ (black, stars) contours. All points
in this and subsequent figures are plotted to second order in slow roll.}
\end{figure}

\begin{figure}
\hspace*{1.0in} {\bf {\large WMAP only\ \ \ \ }} \hspace*{1.0in} 
{\bf {\large WMAP plus seven other experiments}}
\centerline{\includegraphics[width=3.0in]{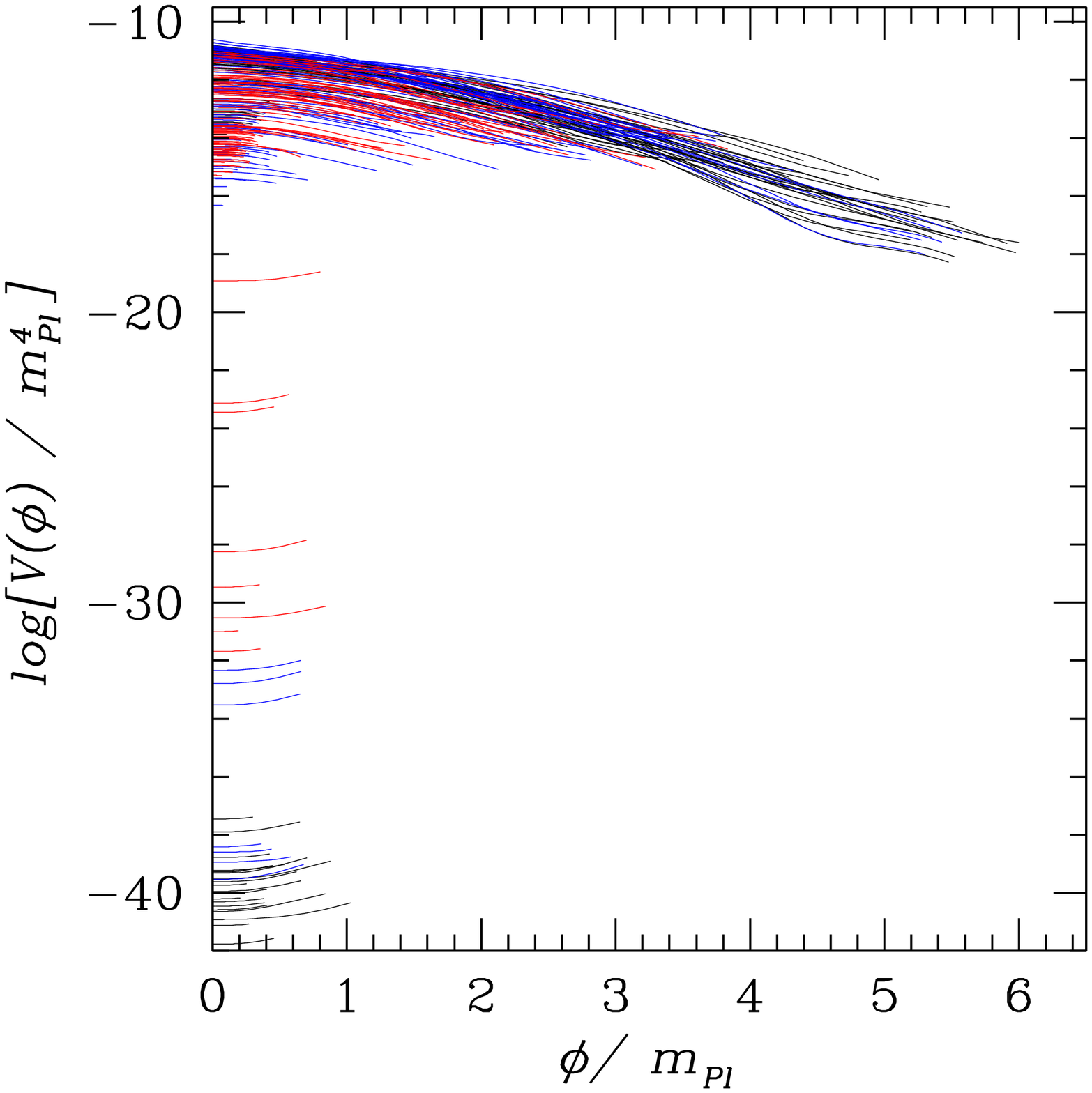} 
\hspace*{24pt}
\includegraphics[width=3.0in]{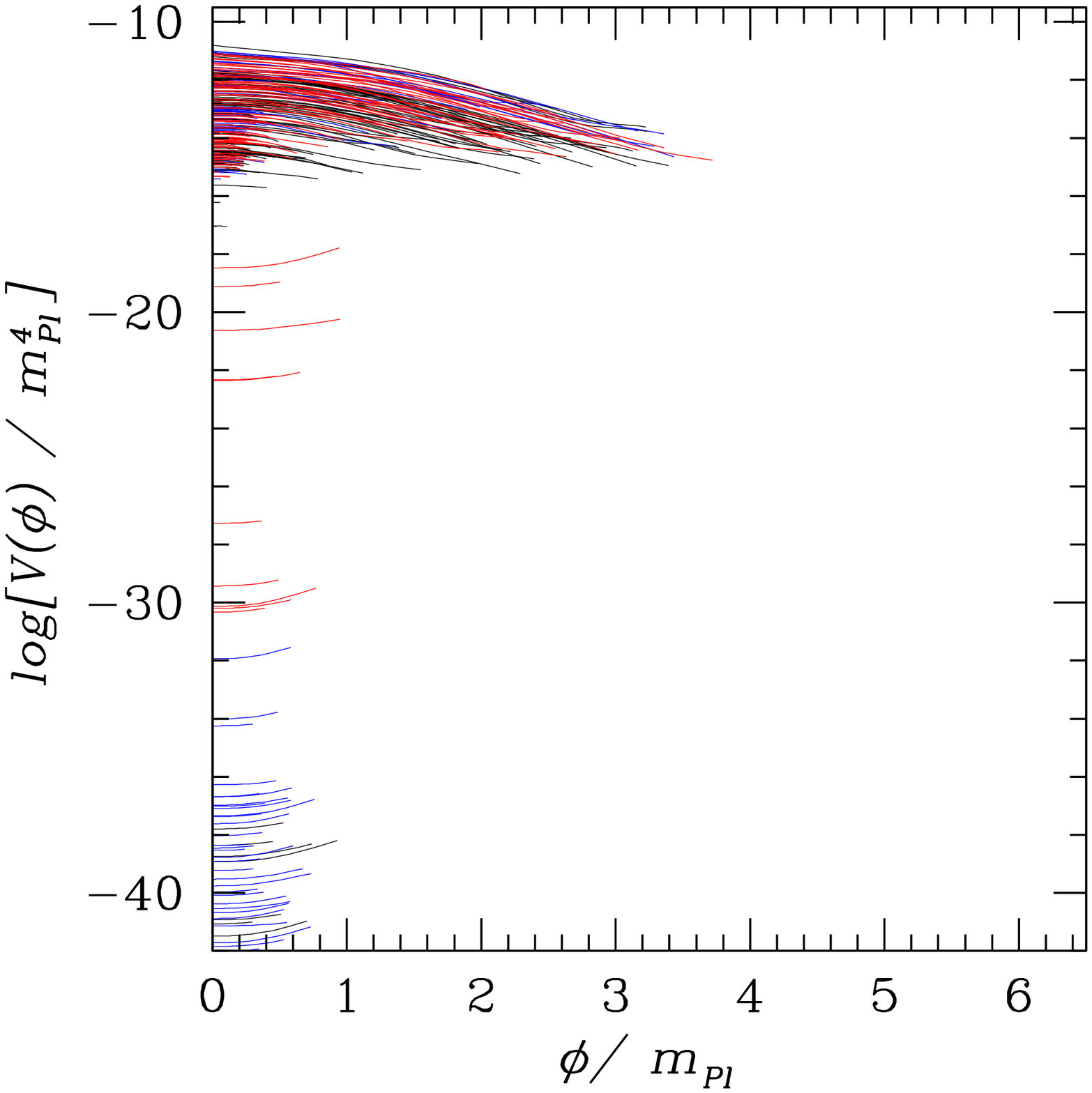}}
\centerline{\includegraphics[width=3.0in]{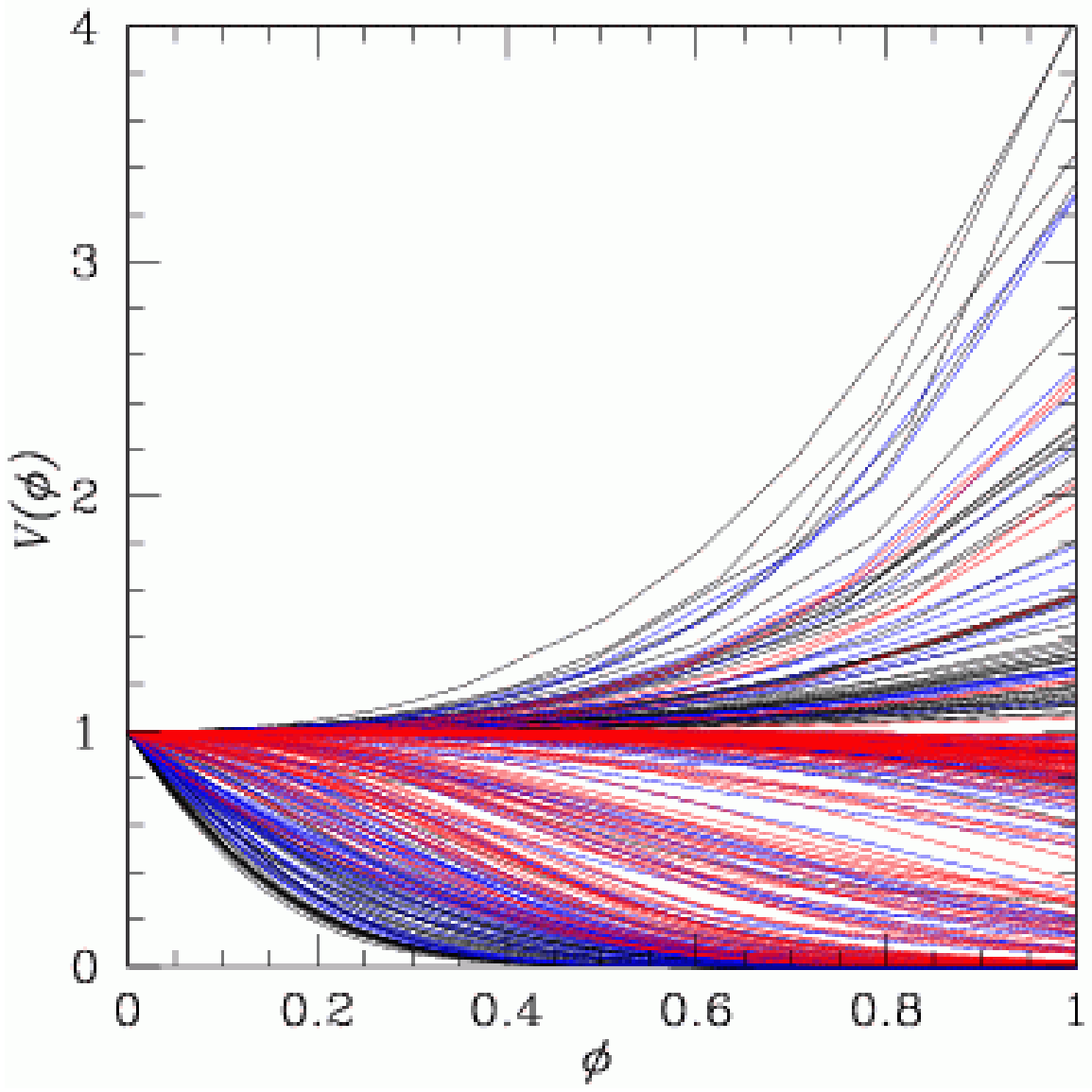} 
\hspace*{24pt}
\includegraphics[width=3.0in]{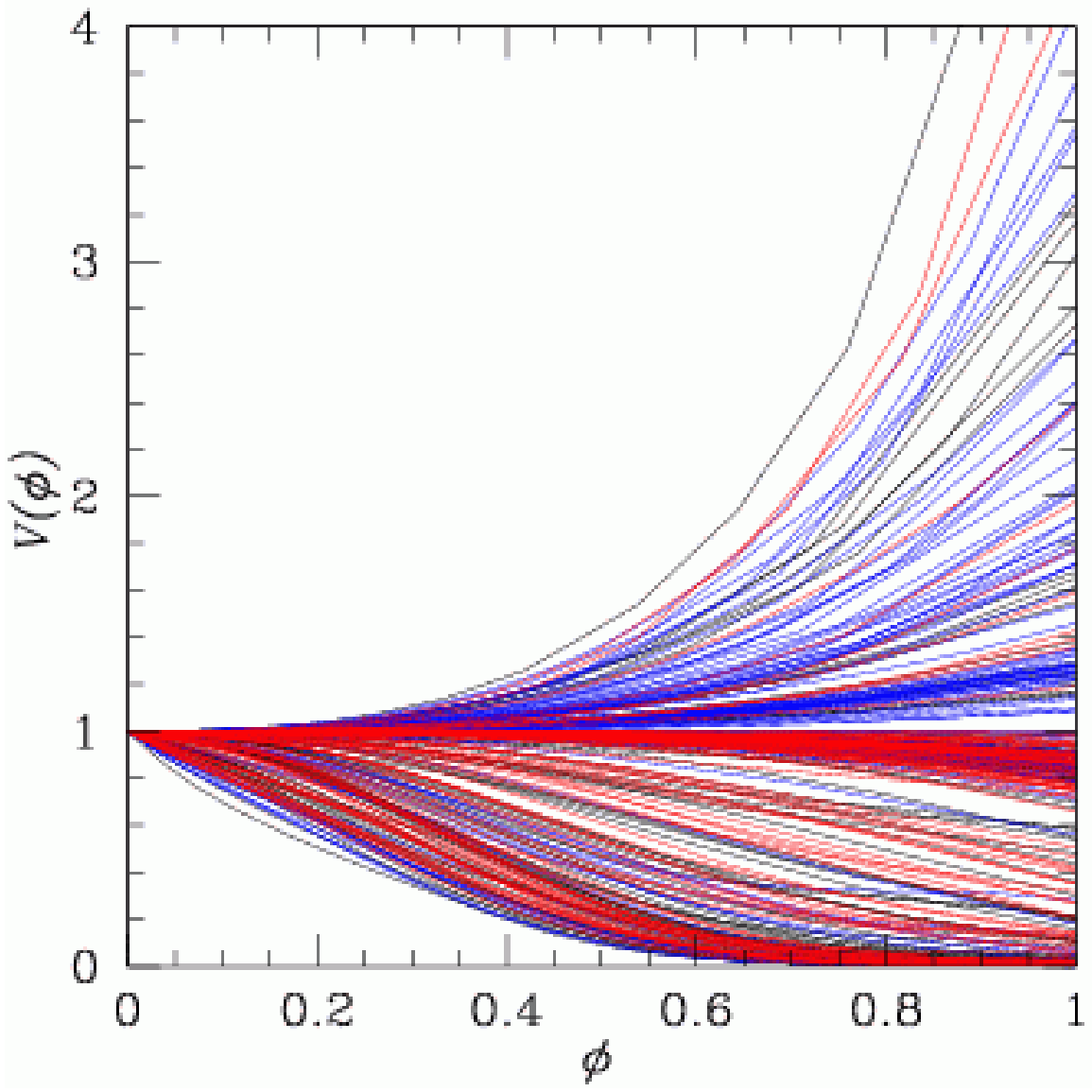}}
\caption{\label{fig_potentials}
Three hundred reconstructed potentials chosen from the sampling shown in
Fig.\ \ref{fig_like} for the WMAP dataset (left column) and the WMAP data set
plus seven other experiments (right column). The potentials are color-coded
according to their likelihoods: $1\sigma$ (red) $2\sigma$ (blue), and $3\sigma$
(black). The top figure shows the potentials with height and width plotted in
units of $m_{\rm Pl}$, and the bottom figure shows the same potentials rescaled
to all have the same height and width.}
\end{figure}

\begin{figure}
\hspace*{1.0in} {\bf {\large WMAP only\ \ \ \ }} \hspace*{1.0in} 
{\bf {\large WMAP plus seven other experiments}}
\centerline{\includegraphics[width=2.7in]{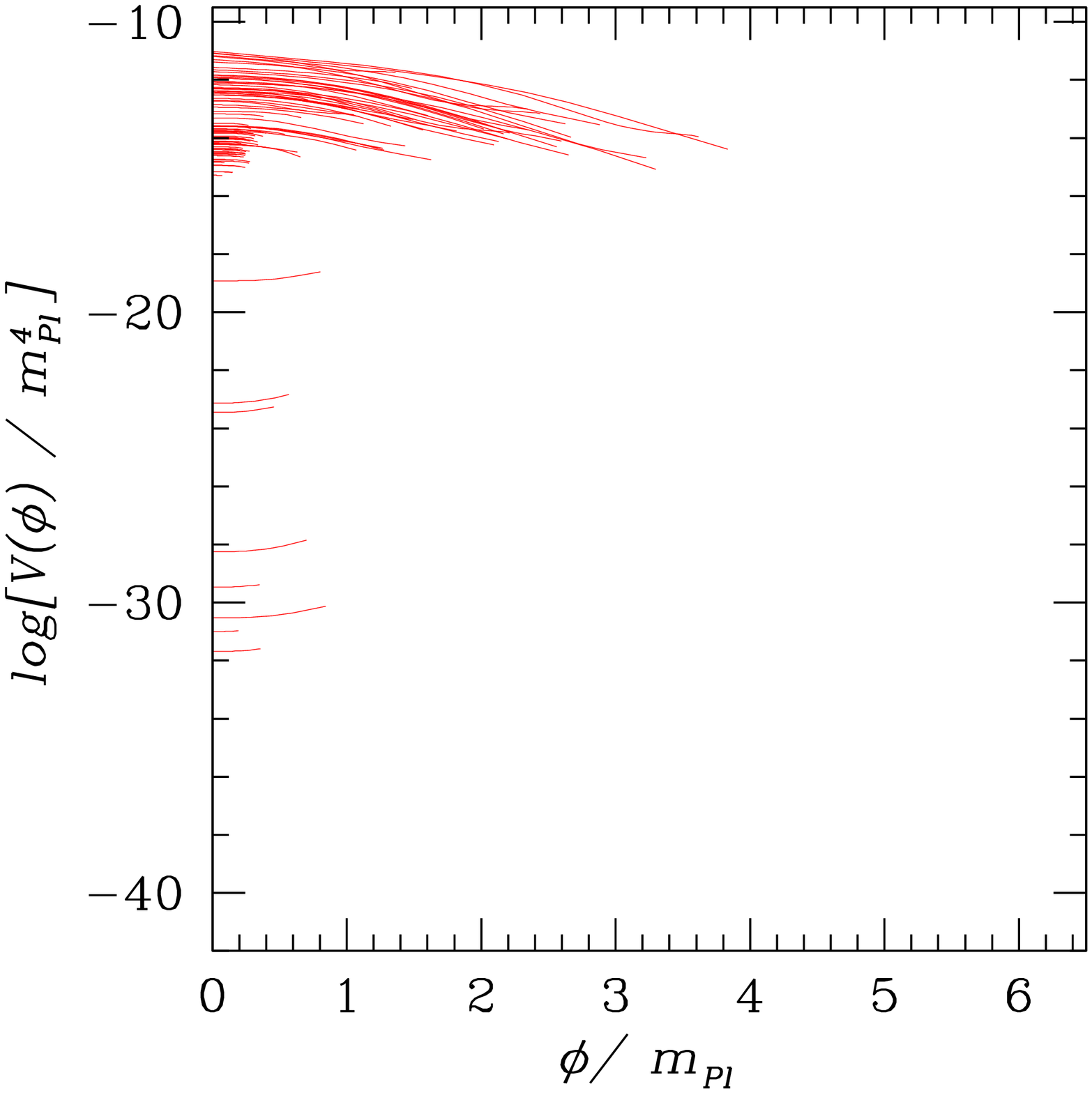}
\hspace*{24pt}
\includegraphics[width=2.7in]{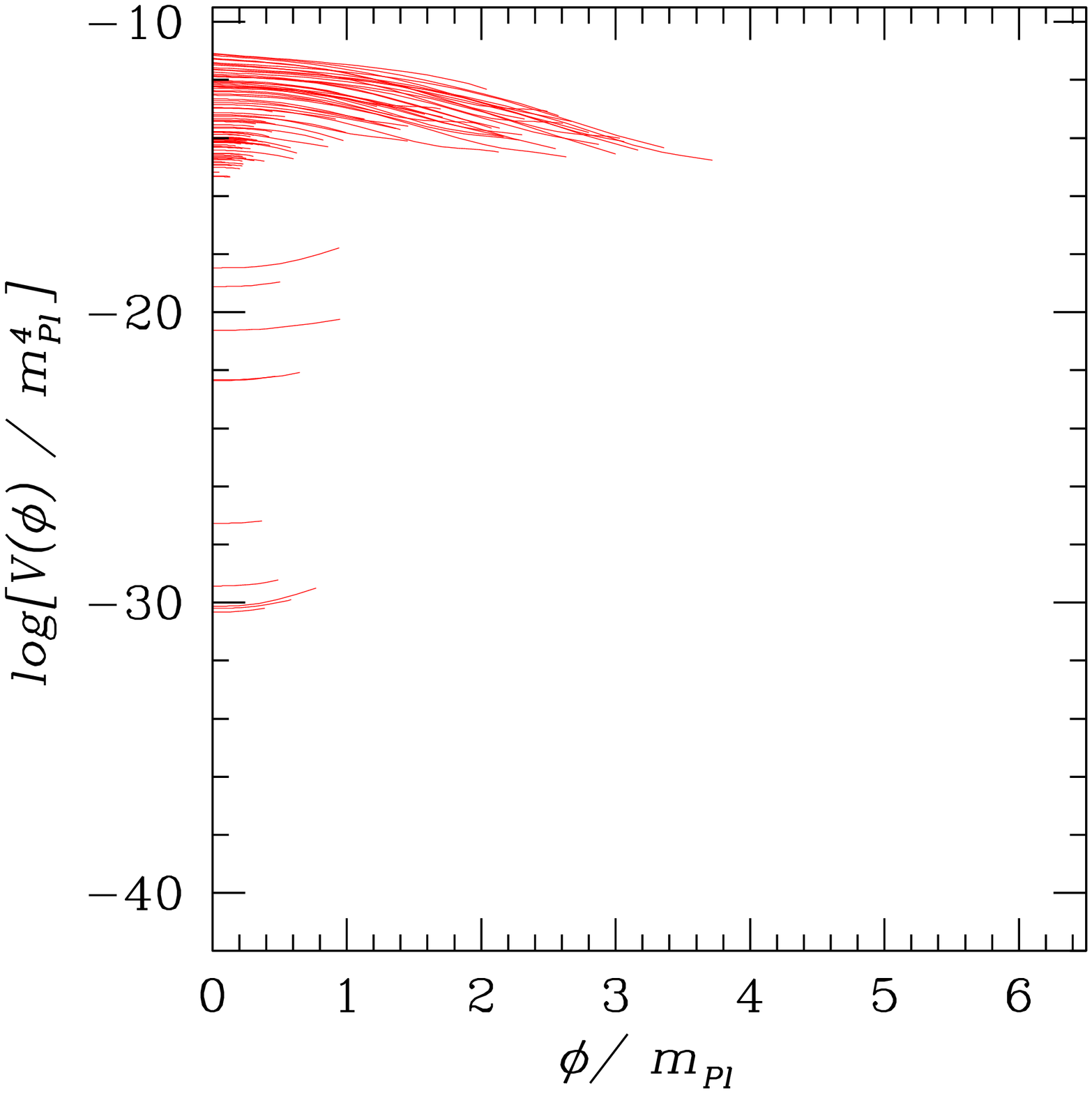}}
\centerline{\includegraphics[width=2.7in]{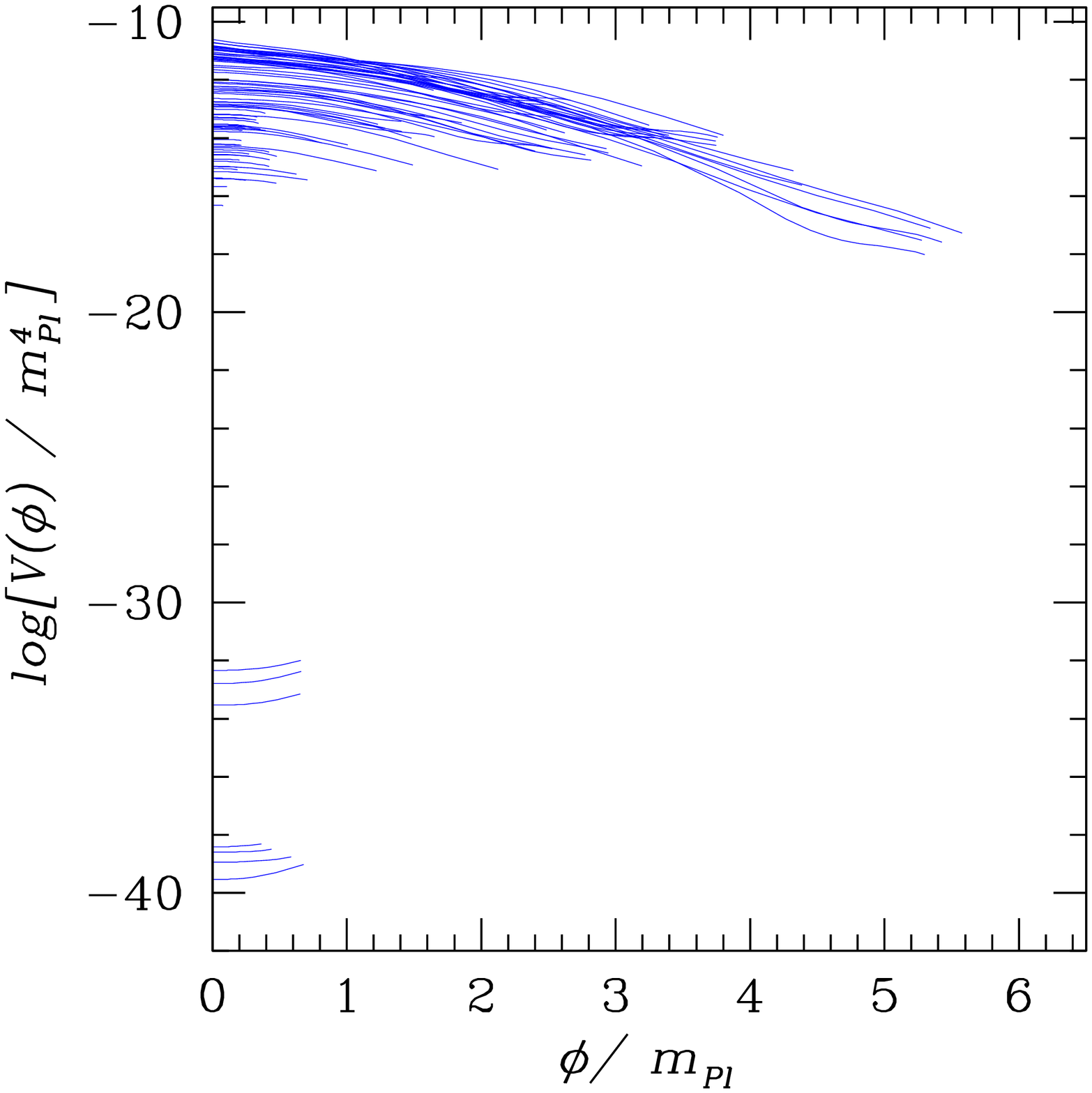}
\hspace*{24pt}
\includegraphics[width=2.7in]{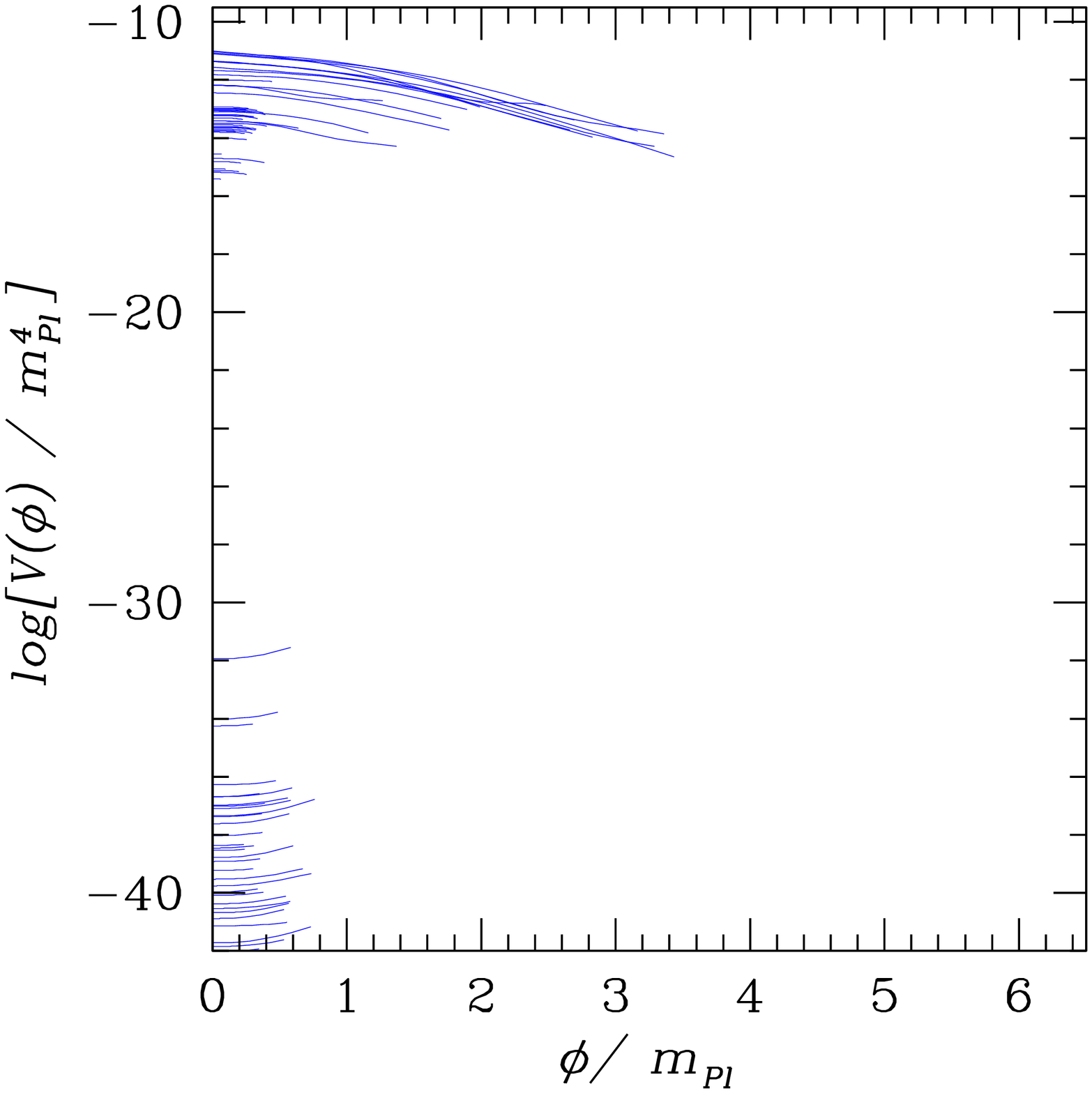}}
\centerline{\includegraphics[width=2.7in]{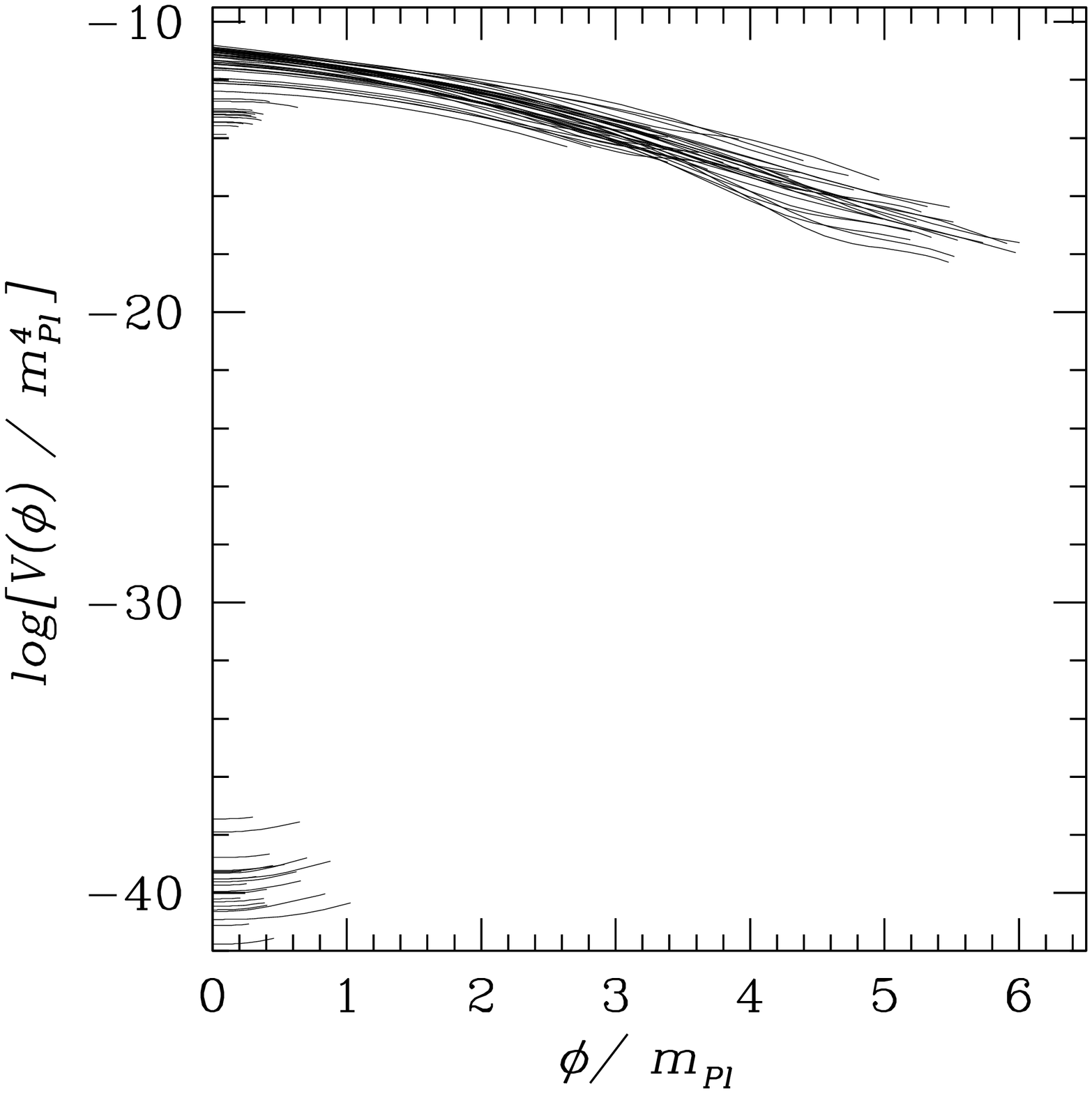}
\hspace*{24pt}
\includegraphics[width=2.7in]{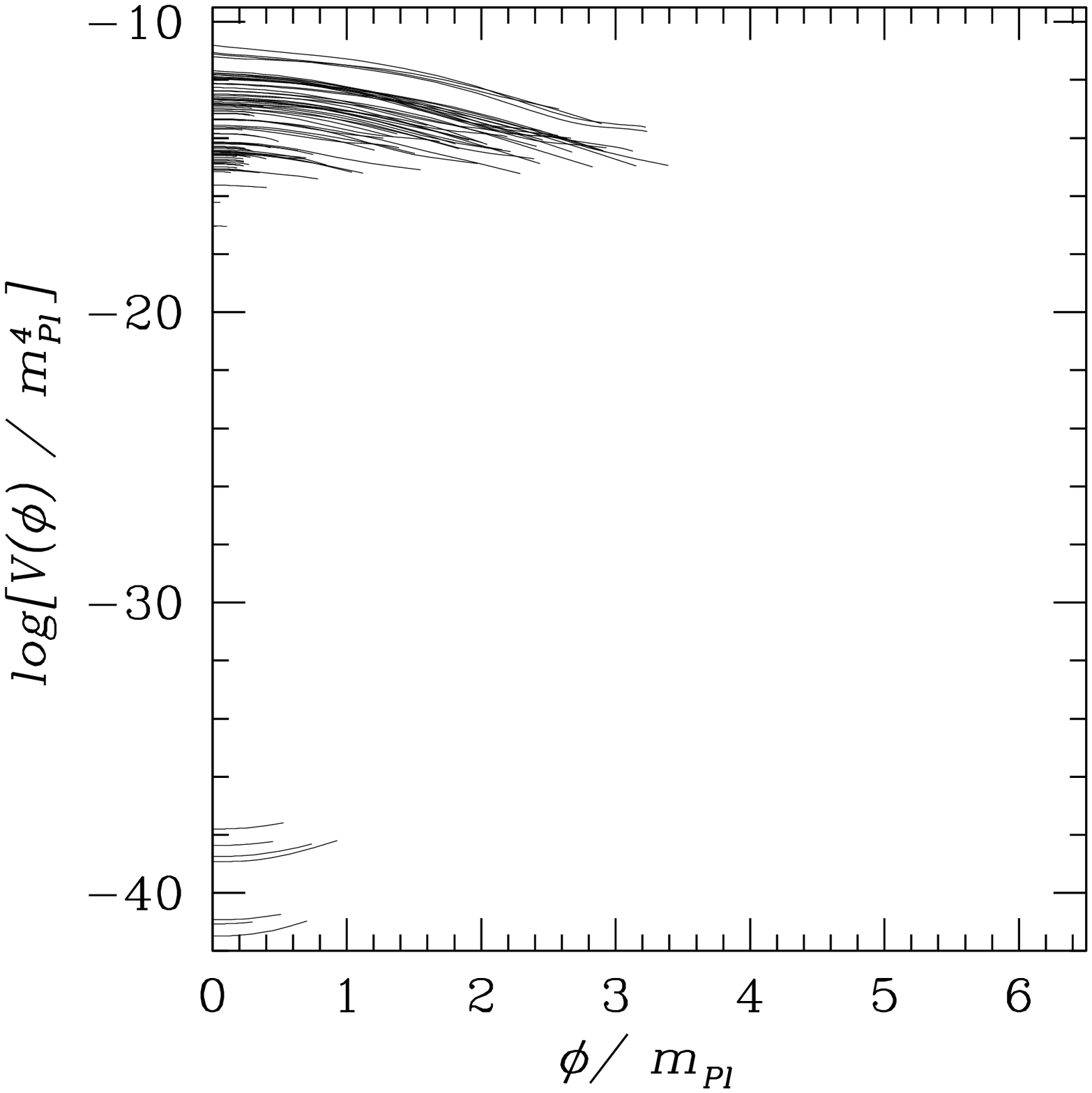}}
\caption{\label{fig_xsigma}
Reconstructed potentials from Fig.\ \ref{fig_potentials} plotted
separately by likelihood: $1\sigma$ (red, top), $2\sigma$ (blue, center), and
$3\sigma$ (black, bottom).  The WMAP results are in the left column, and WMAP
plus seven other experiments are in the right column.}
\end{figure}

\begin{figure}
\hspace*{1.0in} {\bf {\large WMAP only\ \ \ \ }} \hspace*{1.0in} 
{\bf {\large WMAP plus seven other experiments}}
\centerline{\includegraphics[width=2.7in]{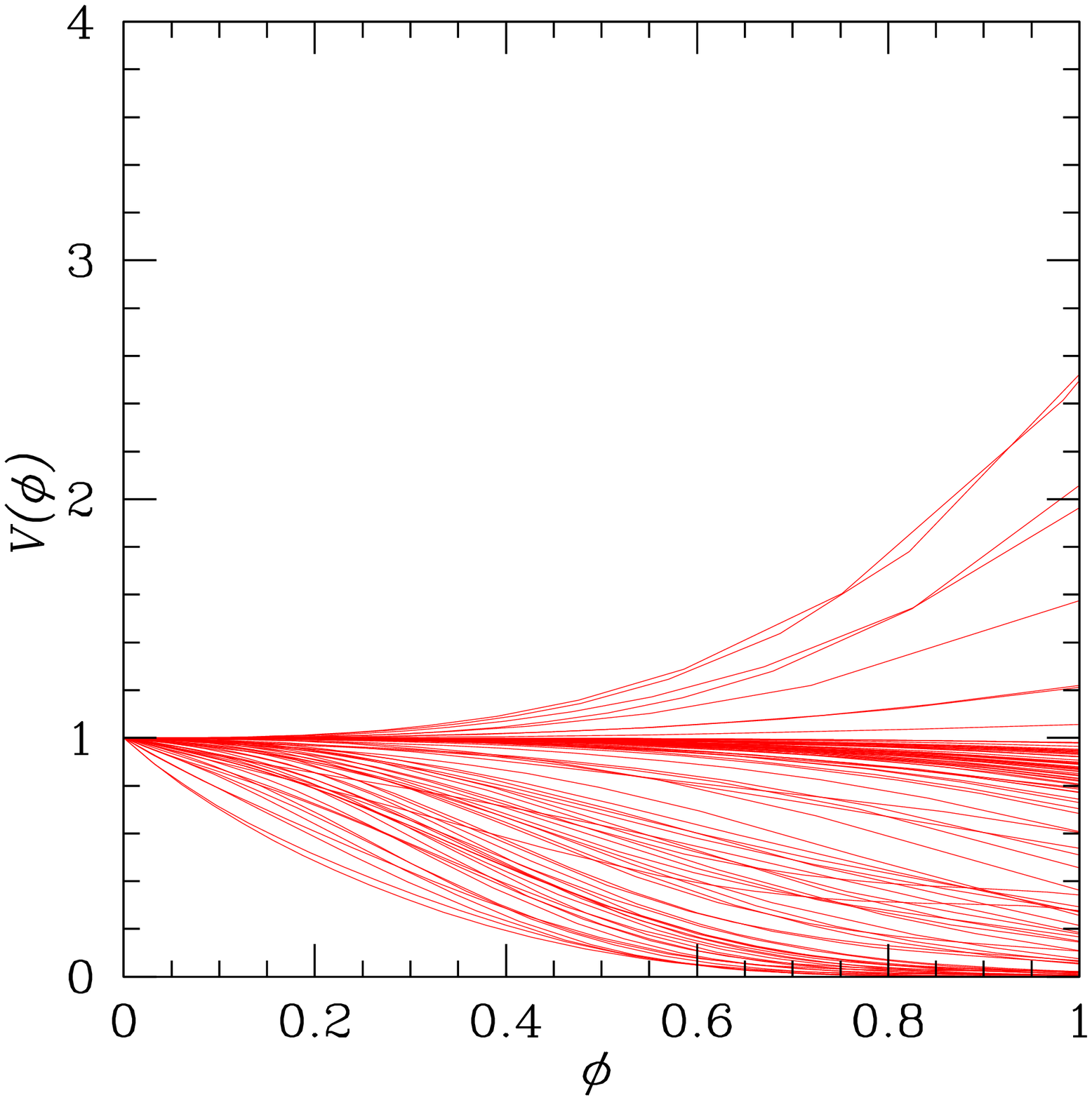}
\hspace*{24pt}
\includegraphics[width=2.7in]{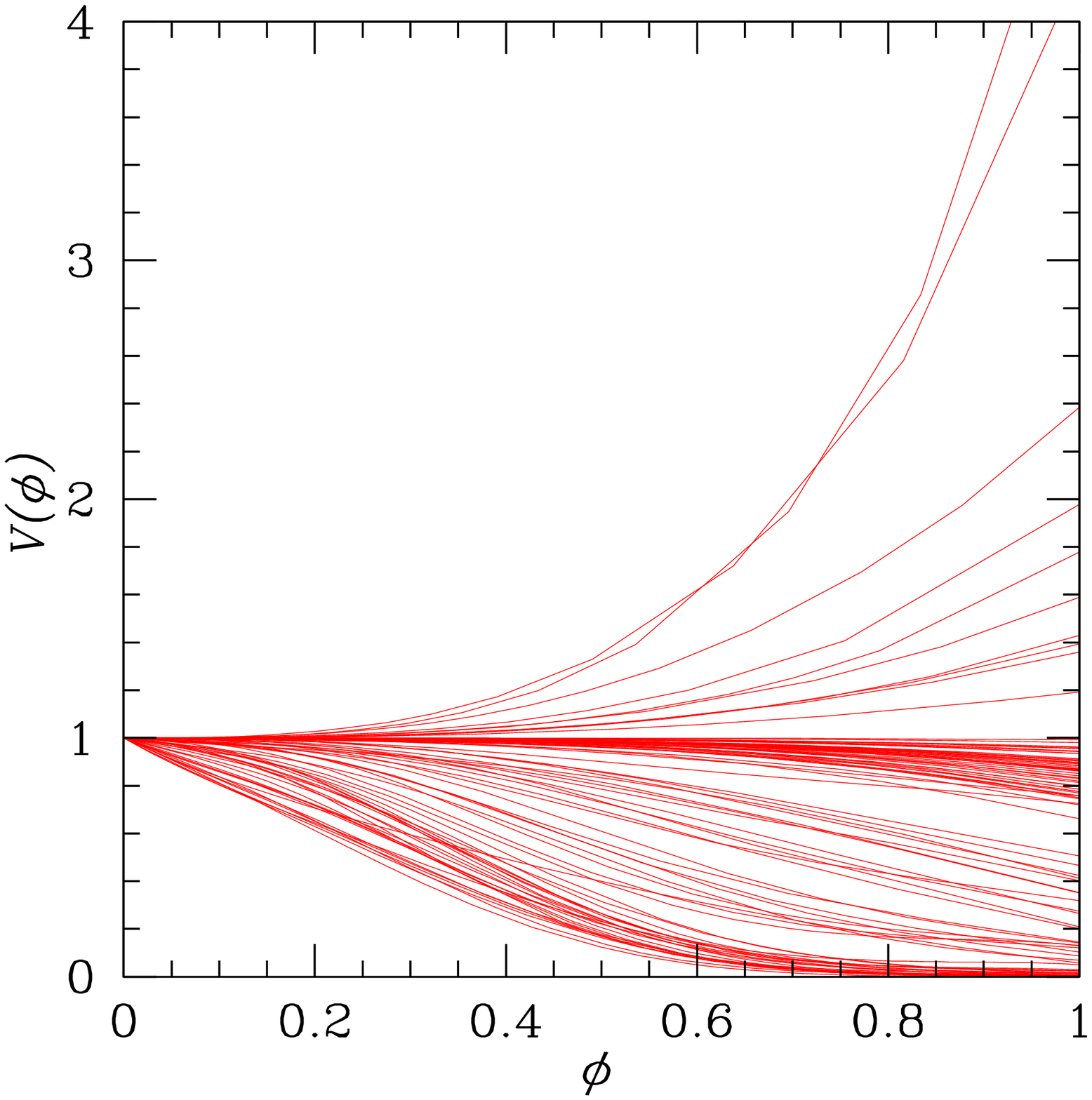}}
\centerline{\includegraphics[width=2.7in]{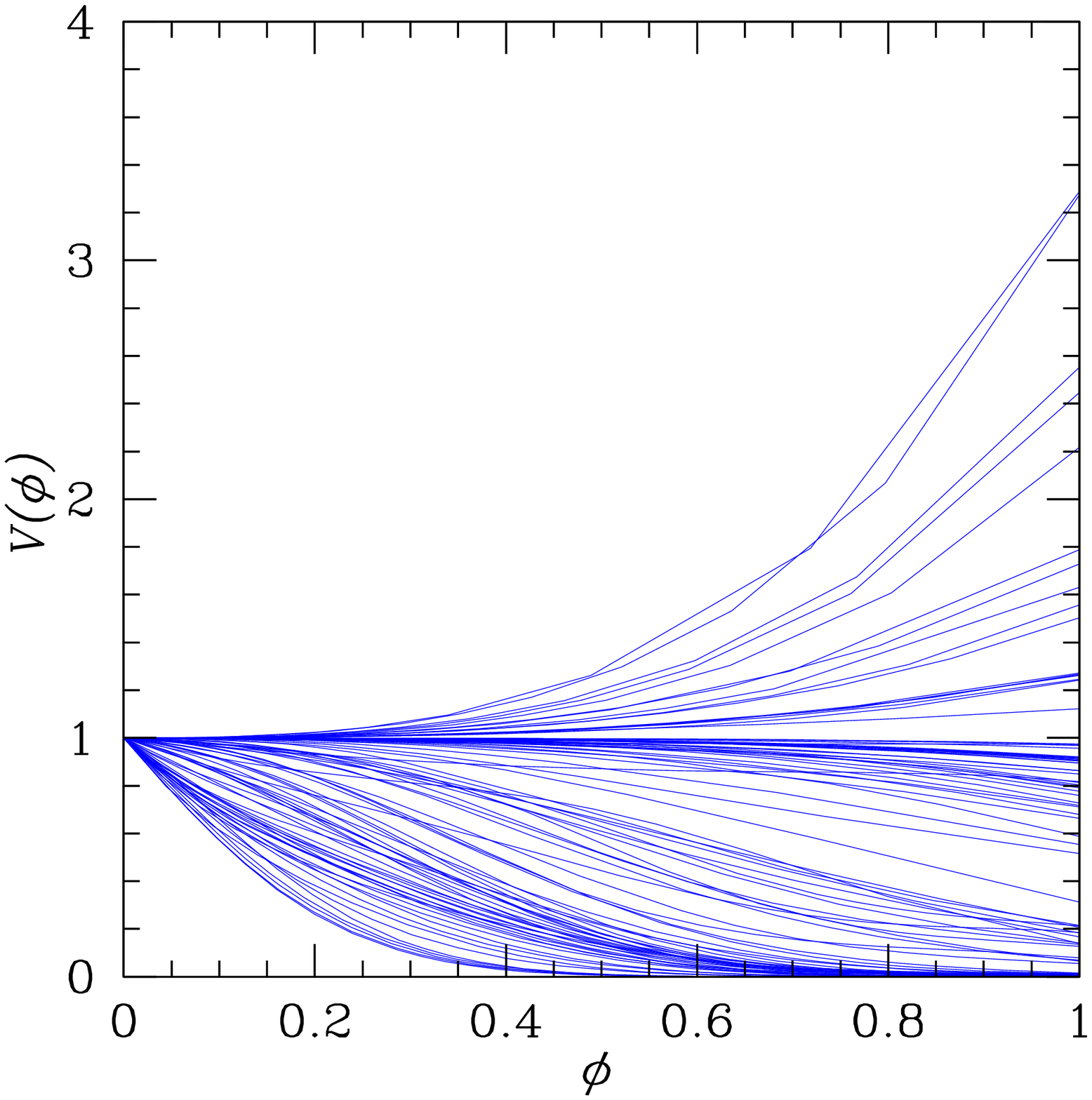}
\hspace*{24pt}
\includegraphics[width=2.7in]{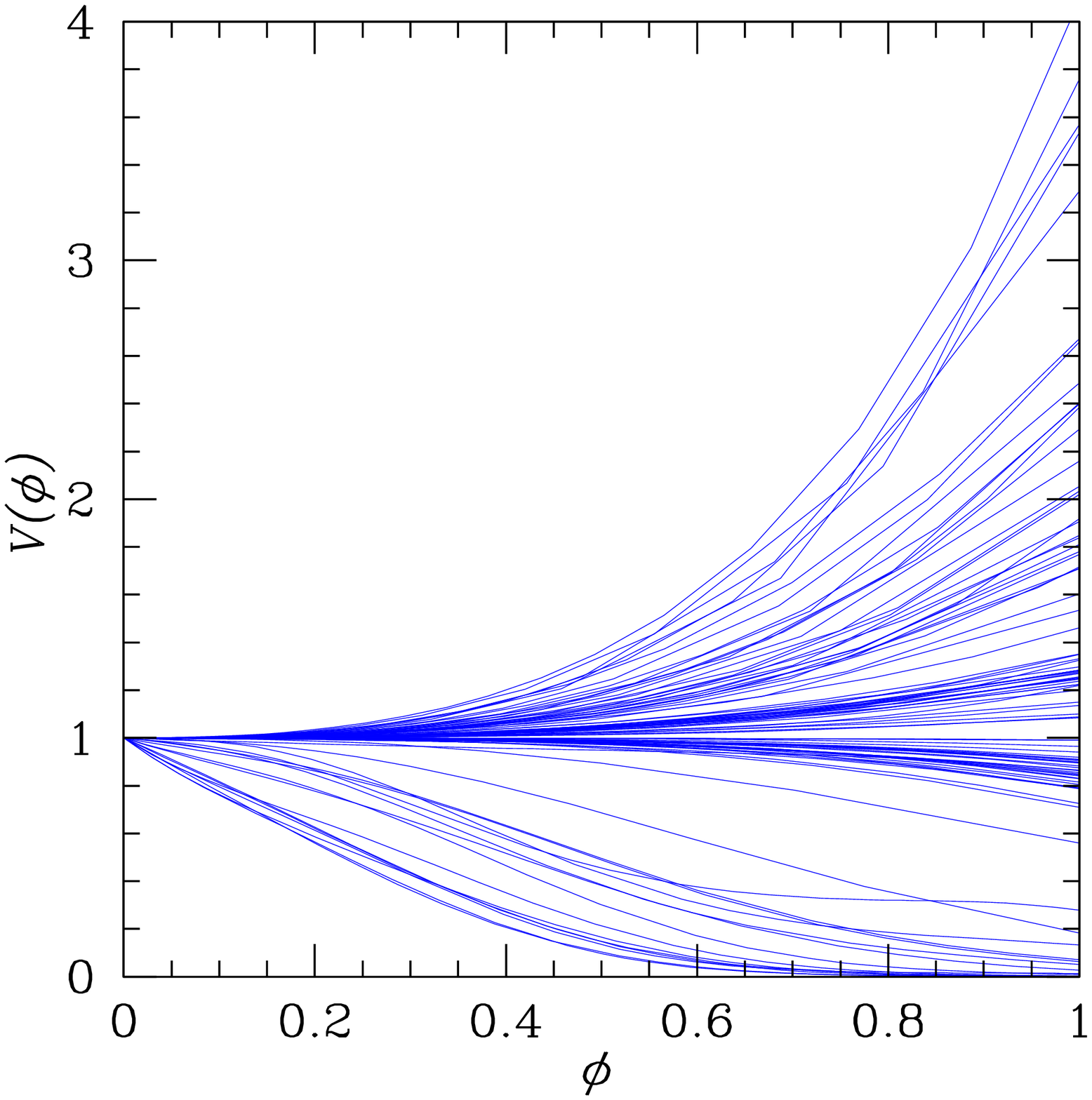}}
\centerline{\includegraphics[width=2.7in]{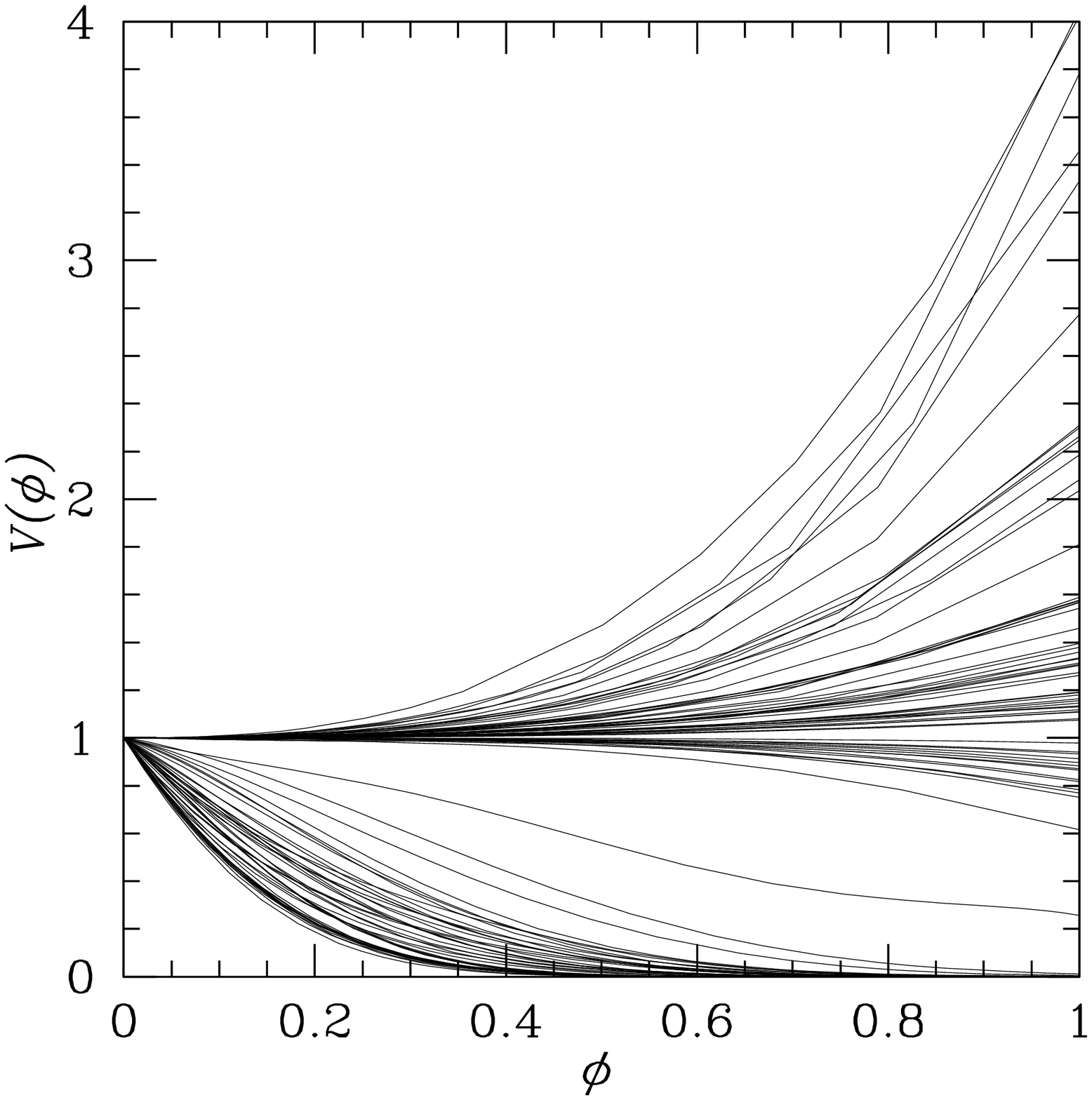}
\hspace*{24pt}
\includegraphics[width=2.7in]{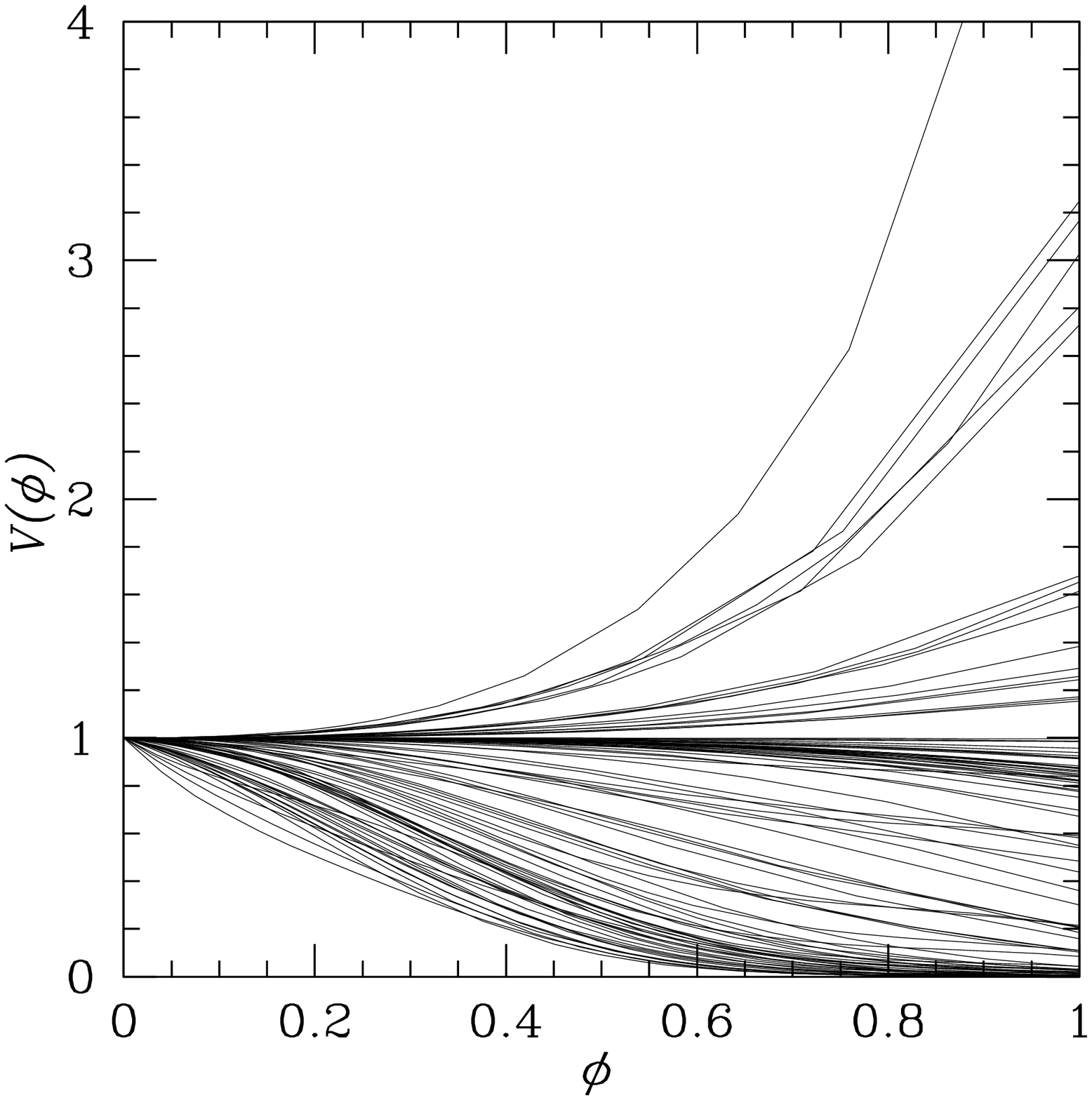}}
\caption{\label{fig_xsigma_rescaled}
Same as Fig.\ \ref{fig_xsigma}, but with the potentials rescaled to all have
the same height and width.}
\end{figure}

\begin{figure}
\hspace*{1.0in} {\bf {\large WMAP only\ \ \ \ }} \hspace*{1.0in} 
{\bf {\large WMAP plus seven other experiments}}
\centerline{\includegraphics[width=2.7in]{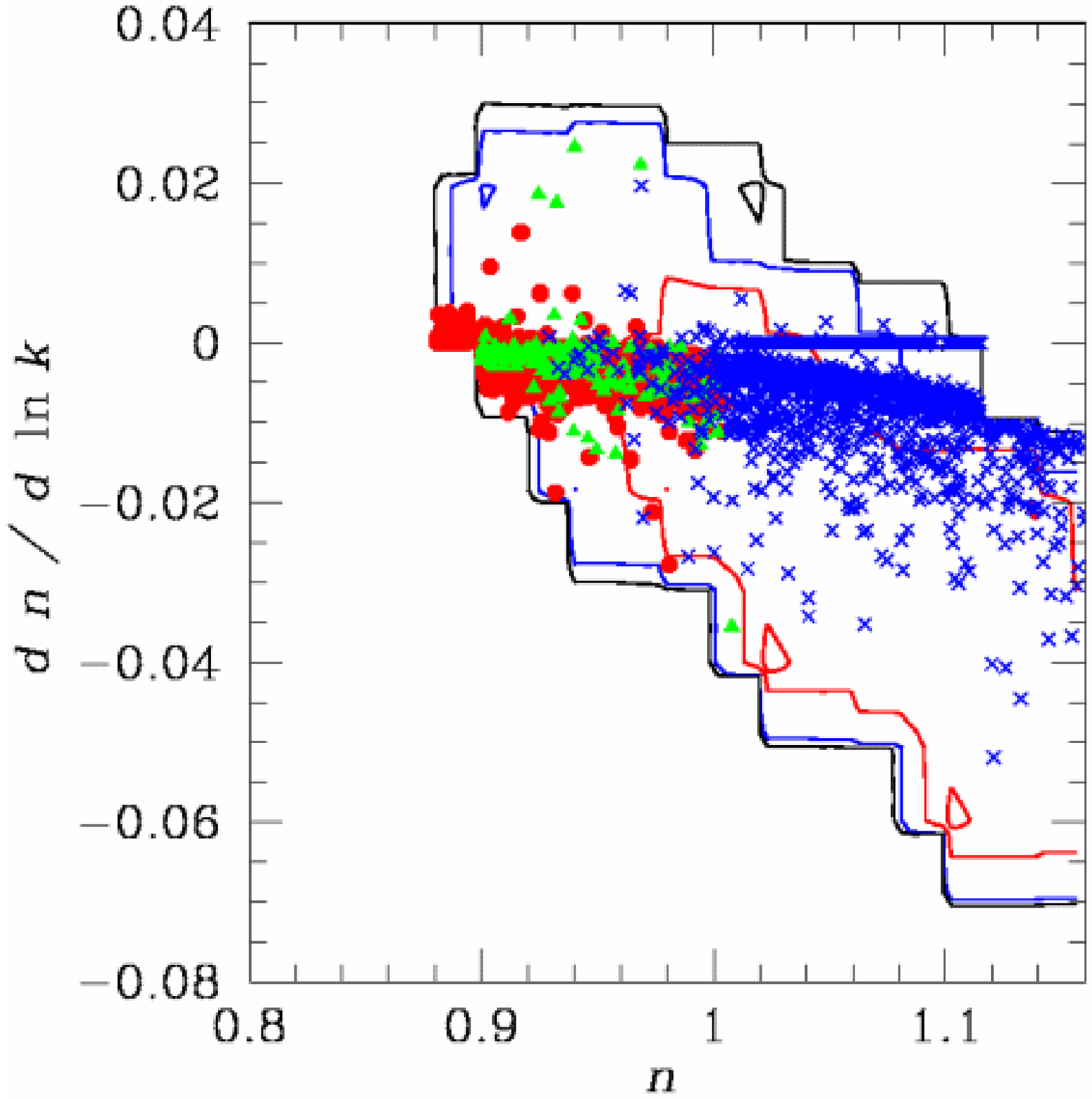}
\hspace*{24pt}
\includegraphics[width=2.7in]{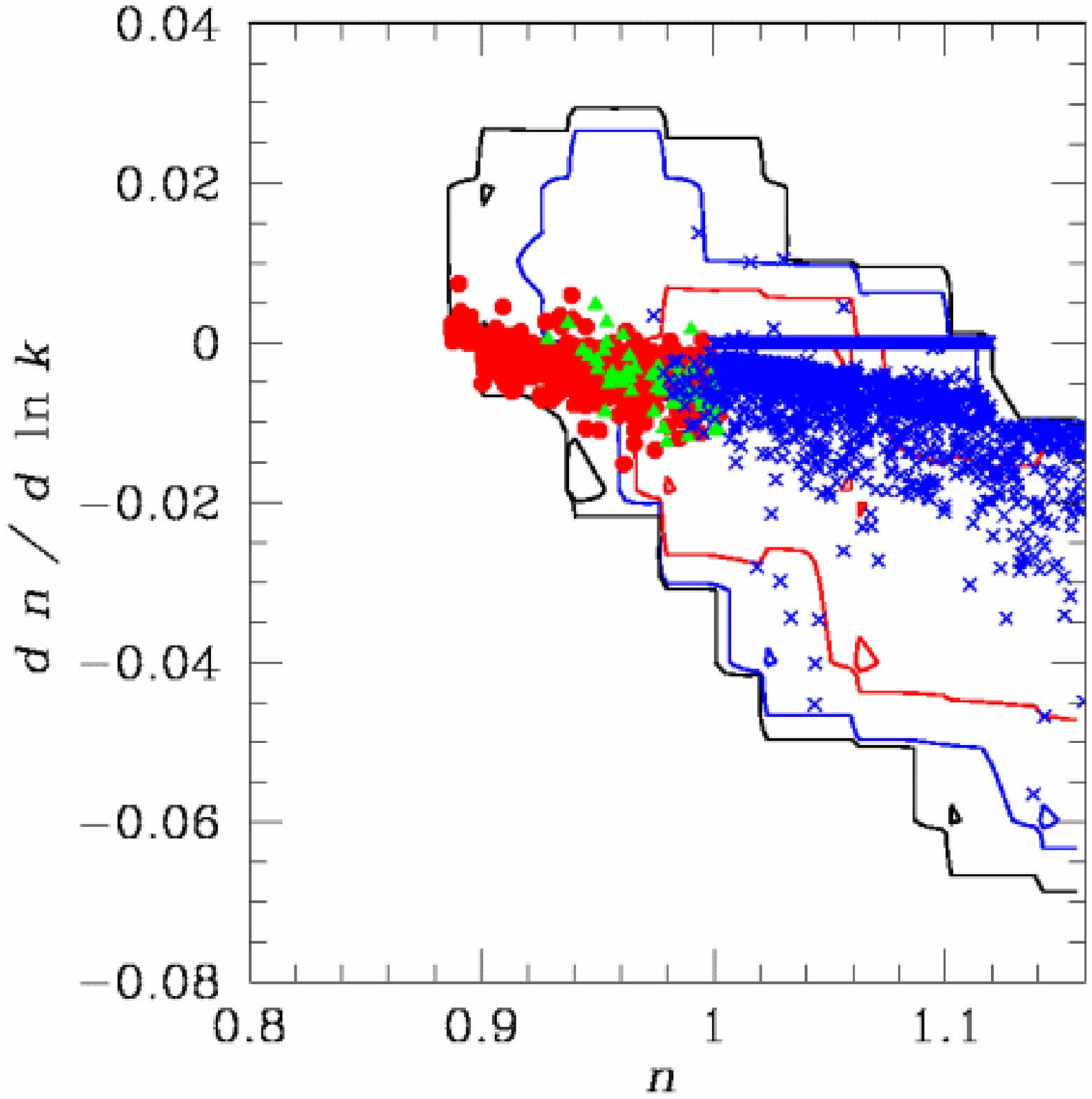}}
\centerline{\includegraphics[width=2.7in]{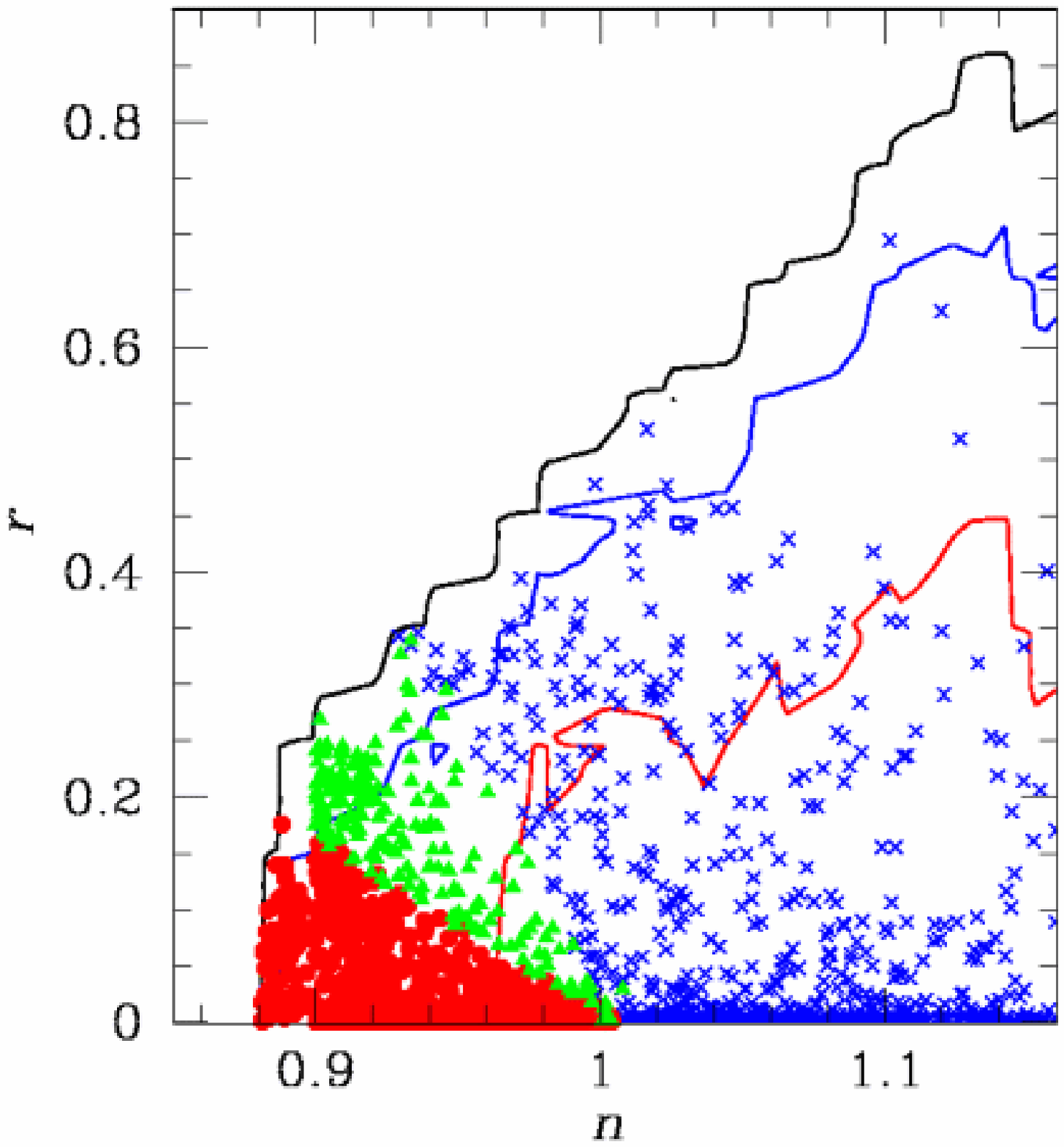}
\hspace*{24pt}
\includegraphics[width=2.7in]{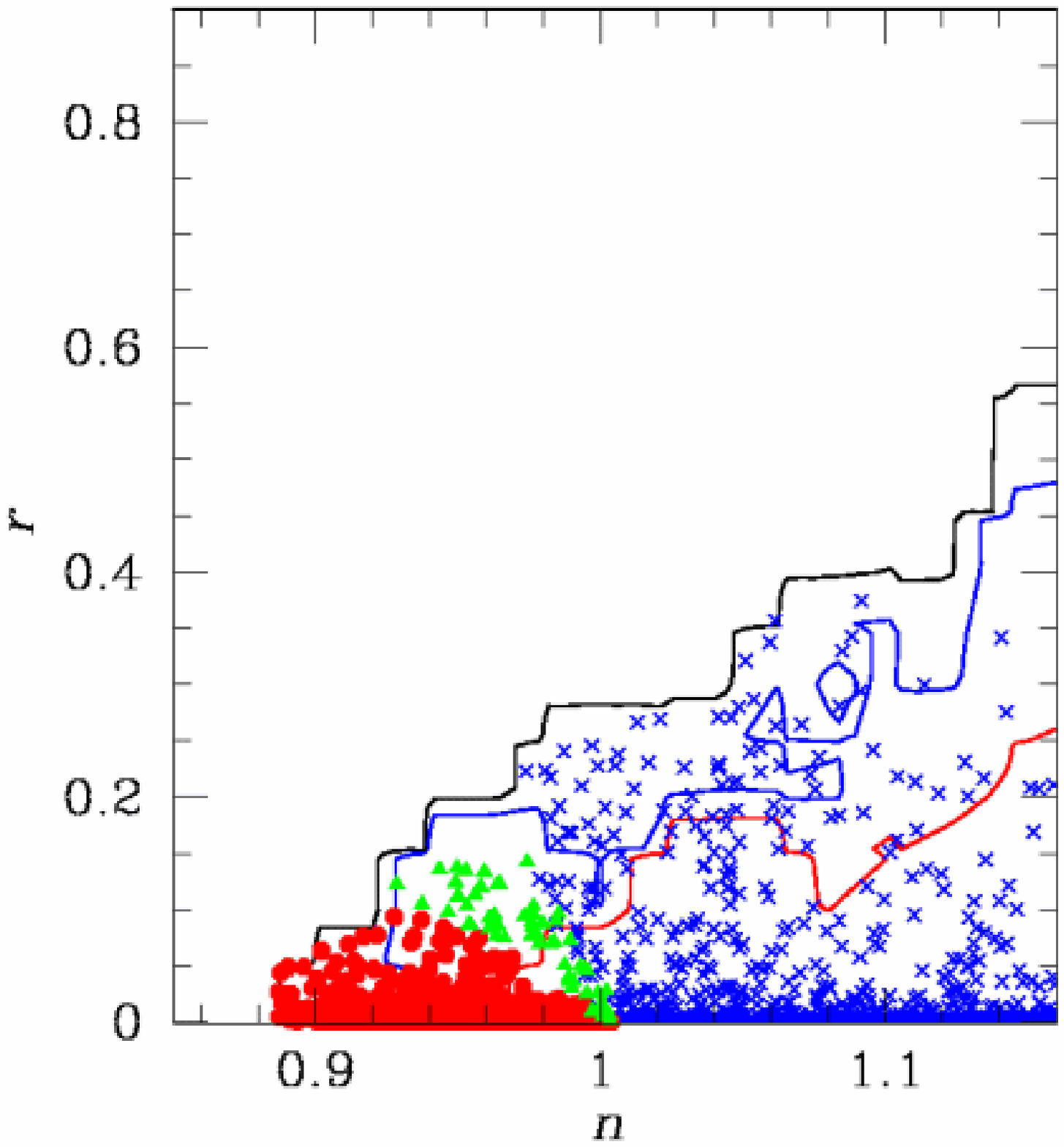}}
\centerline{\includegraphics[width=2.7in]{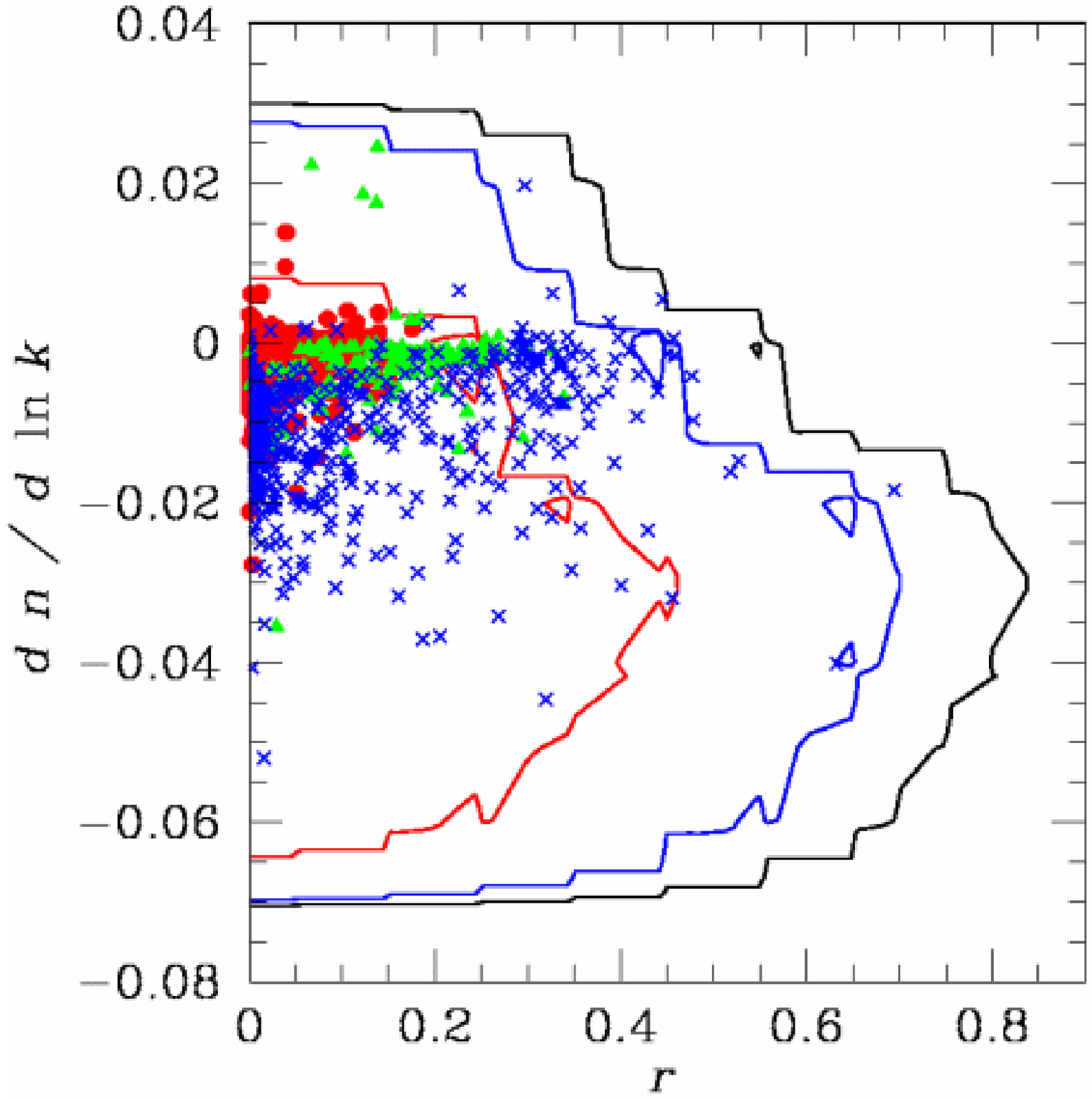}
\hspace*{24pt}
\includegraphics[width=2.7in]{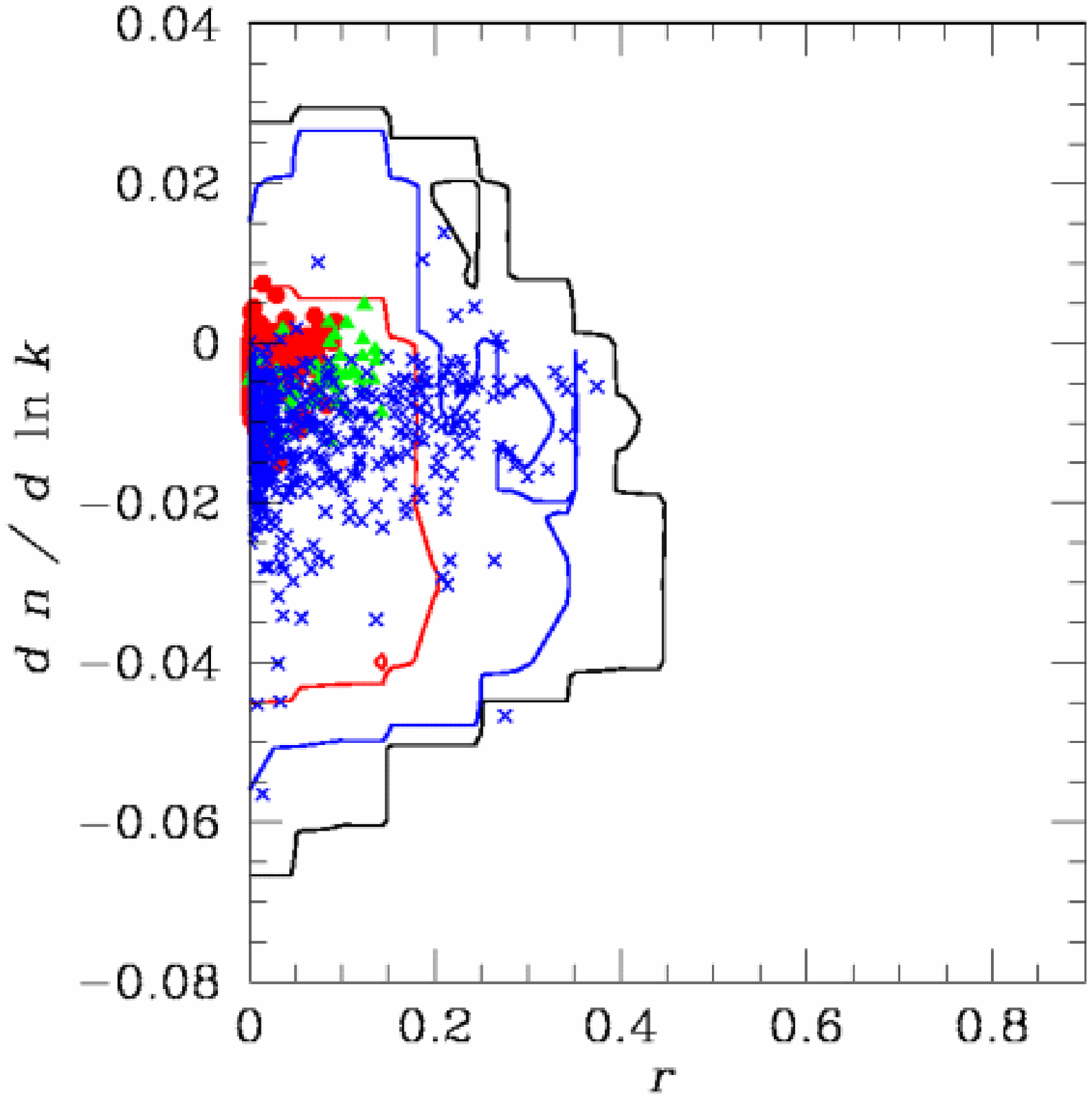}}
\caption{\label{fig_like_zoo}
Likelihood contours plotted in the ($n,dn/d\ln{k}$) plane (top), the ($n,r$)
plane (center), and the ($r,dn/d\ln{k}$) plane (bottom).  The points represent
the results of the Monte Carlo sampling, color coded by model type: small-field
(red, dots), large-field (green, triangles), and hybrid (blue, crosses).  The
left column is WMAP only while the right column is WMAP plus seven other
experiments}
\end{figure}

\begin{figure}
\hspace*{1.0in} {\bf {\large WMAP only\ \ \ \ }} \hspace*{1.0in} 
{\bf {\large WMAP plus seven other experiments}}
\centerline{\includegraphics[width=3.0in]{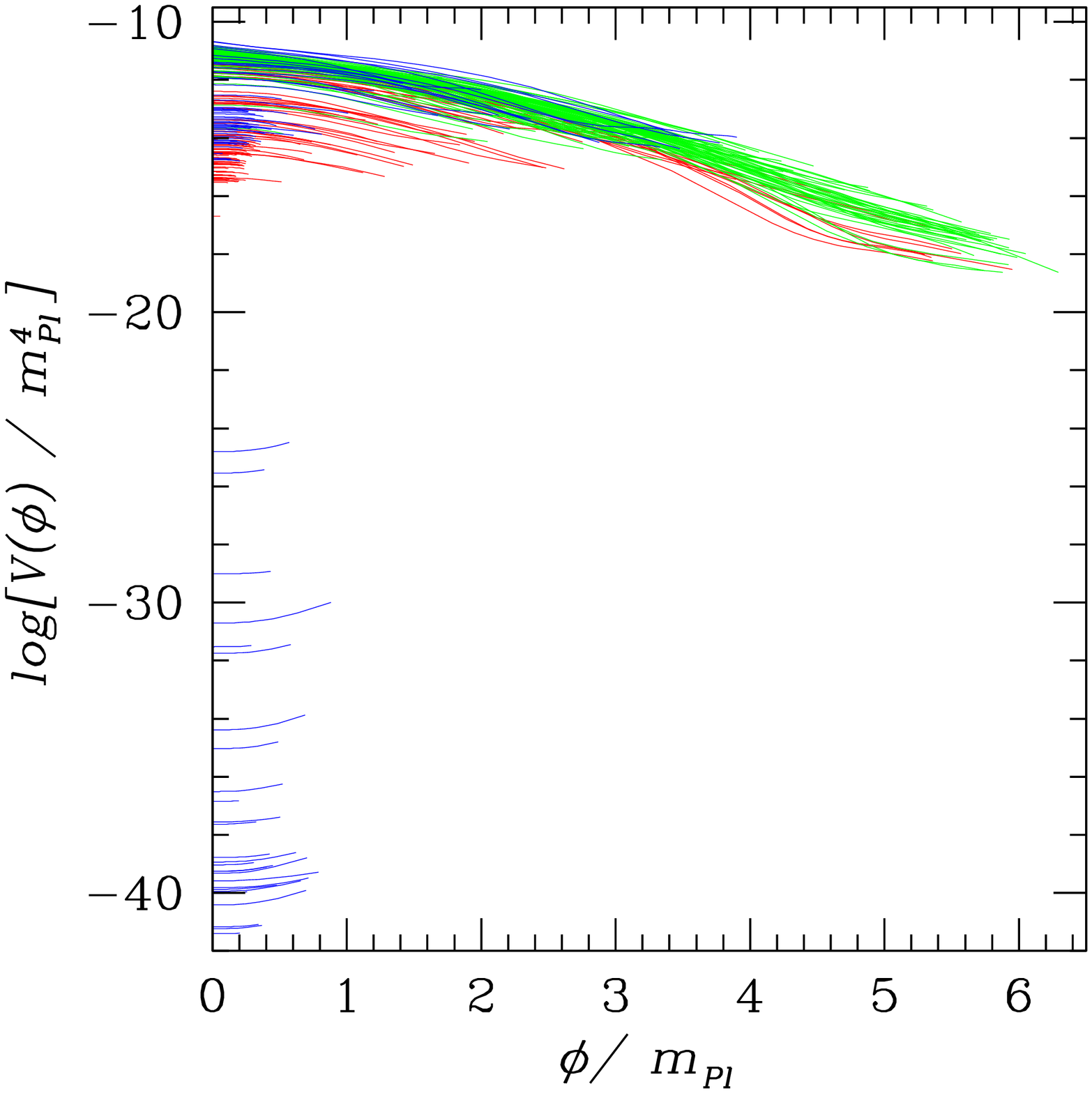}
\hspace*{24pt}
\includegraphics[width=3.0in]{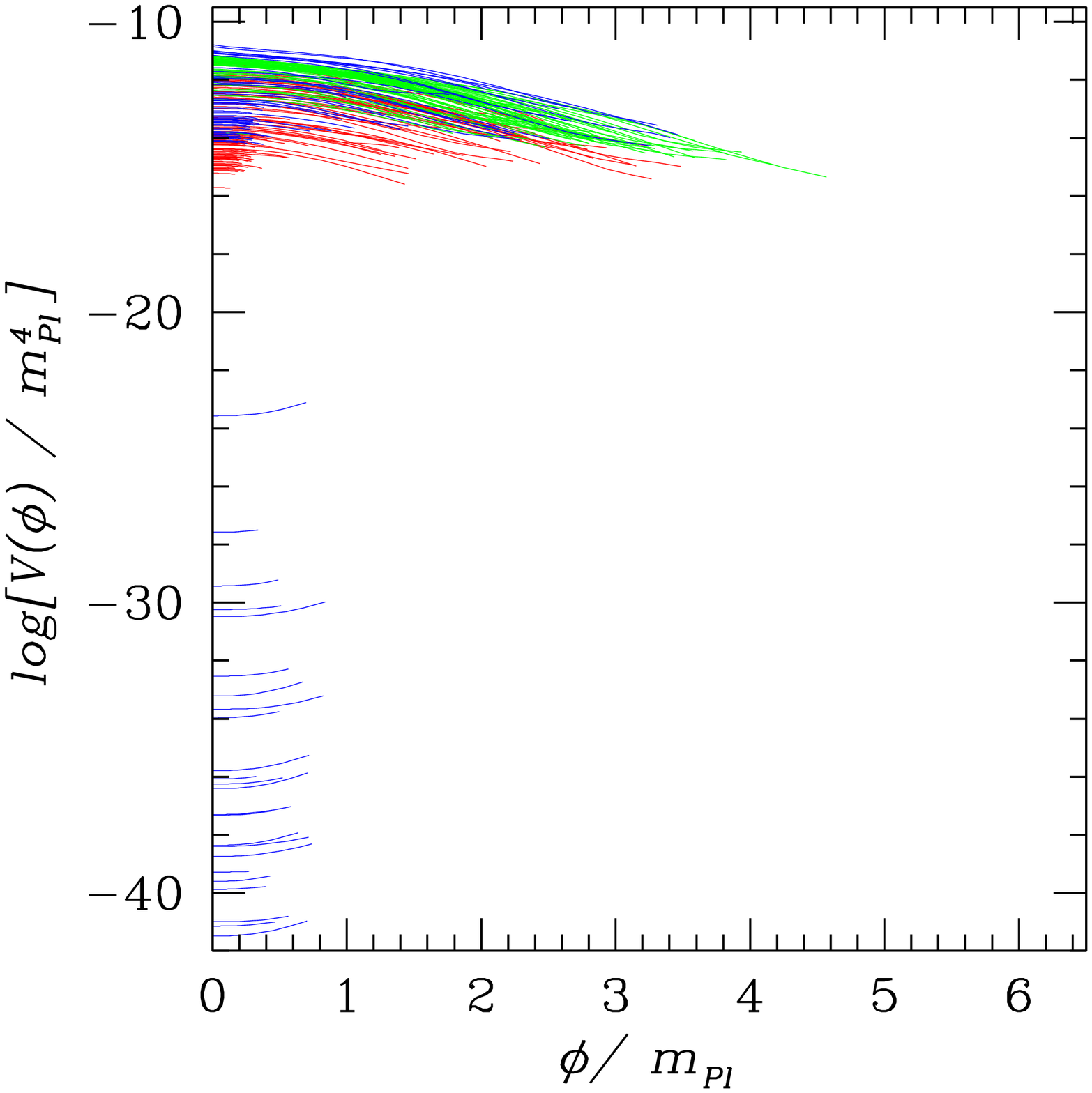}}
\centerline{\includegraphics[width=3.0in]{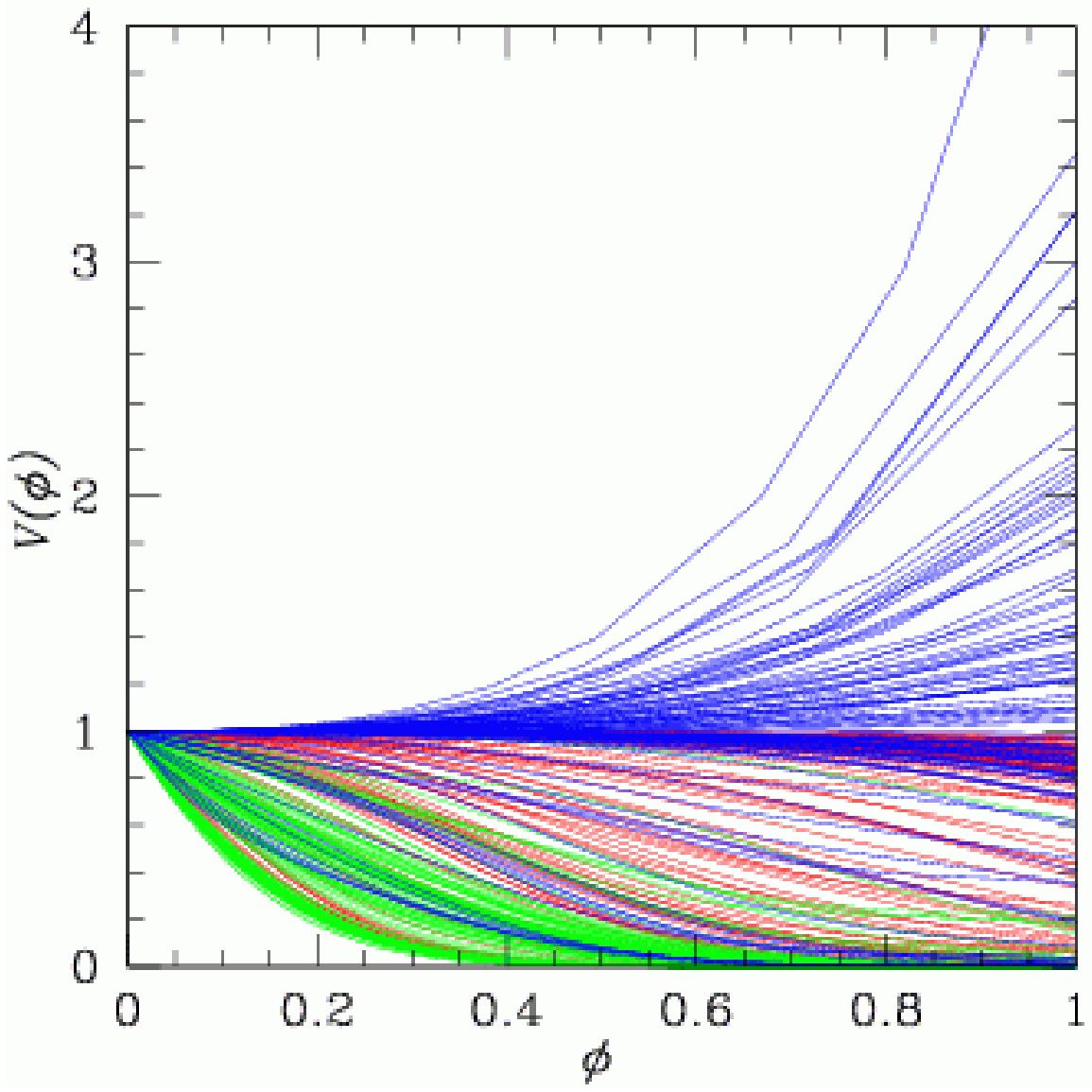}
\hspace*{24pt}
\includegraphics[width=3.0in]{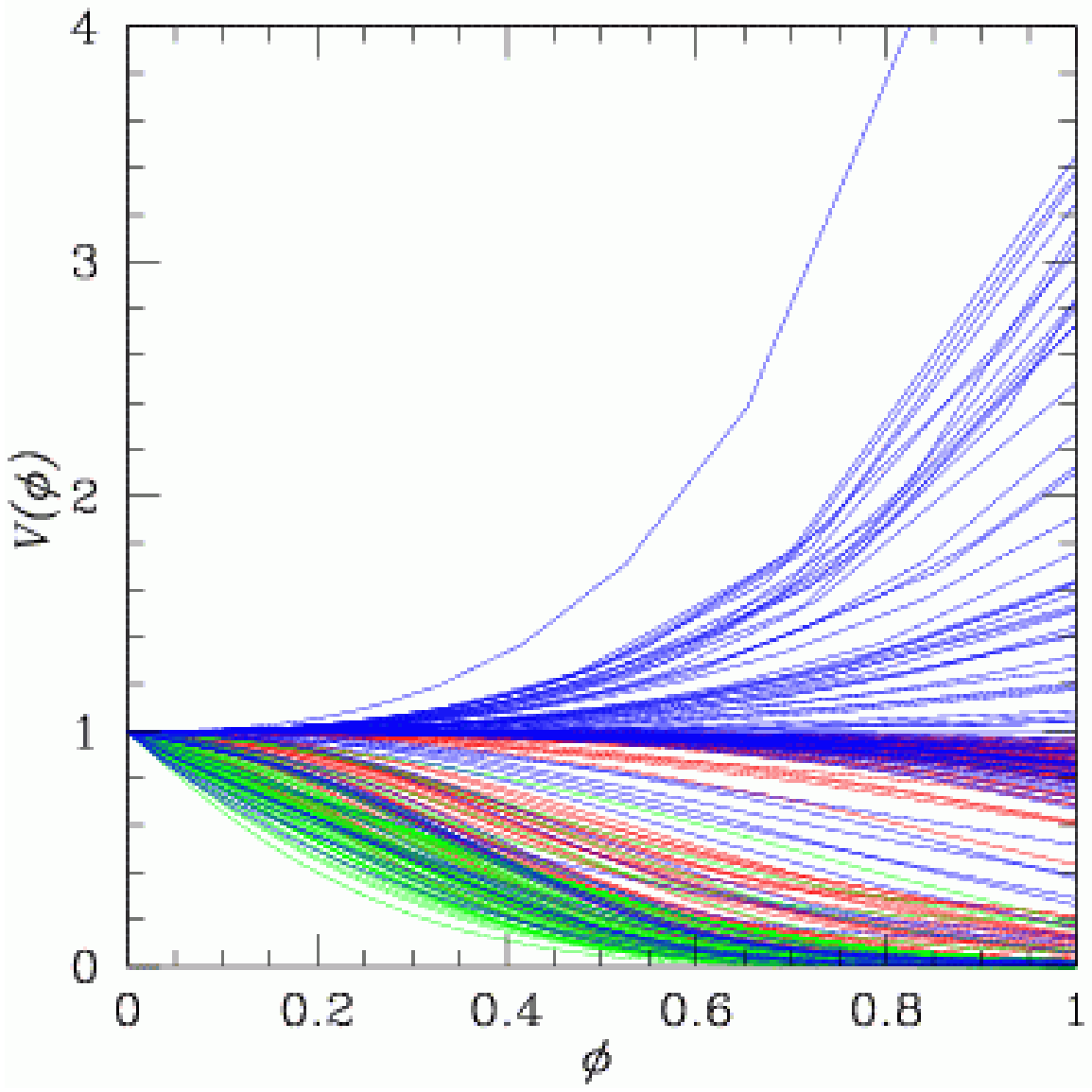}}
\caption{\label{fig_V_zoo}
Three hundred reconstructed potentials chosen from the sampling shown in Fig.\
\ref{fig_like} for WMAP (left column) and WMAP plus seven other experiments 
(right column). The potentials are color-coded according to model type:
small-field (red), large-field (green), and hybrid (blue).  The top figure
shows the potentials with height and width plotted in units of $m_{\rm Pl}$,
and the bottom figure shows the same potentials rescaled to all have the same
height and width.}
\end{figure}

\begin{figure}
\hspace*{1.0in} {\bf {\large WMAP only\ \ \ \ }} \hspace*{1.0in} 
{\bf {\large WMAP plus seven other experiments}}
\centerline{\includegraphics[width=2.7in]{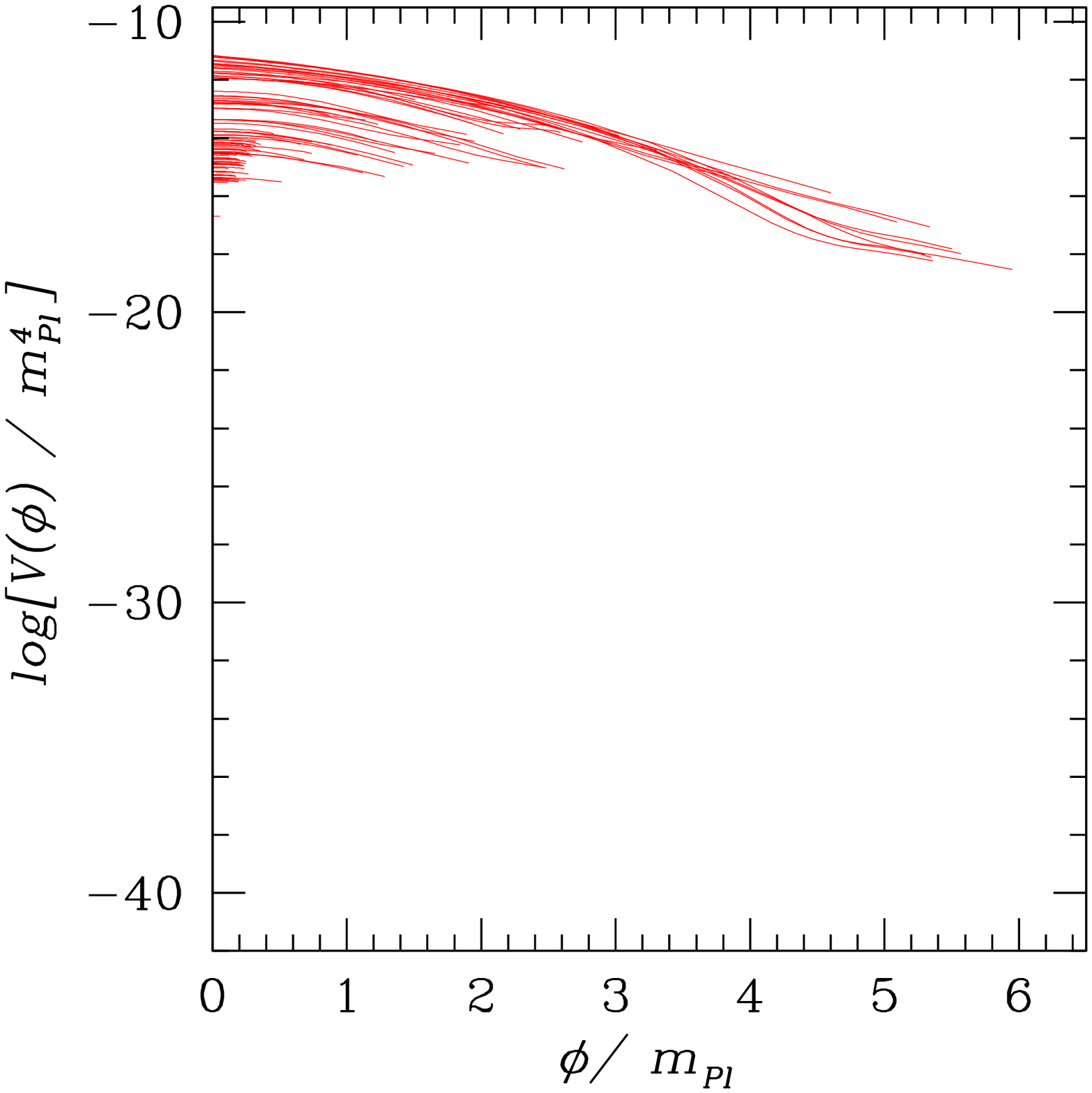}
\hspace*{24pt}
\includegraphics[width=2.7in]{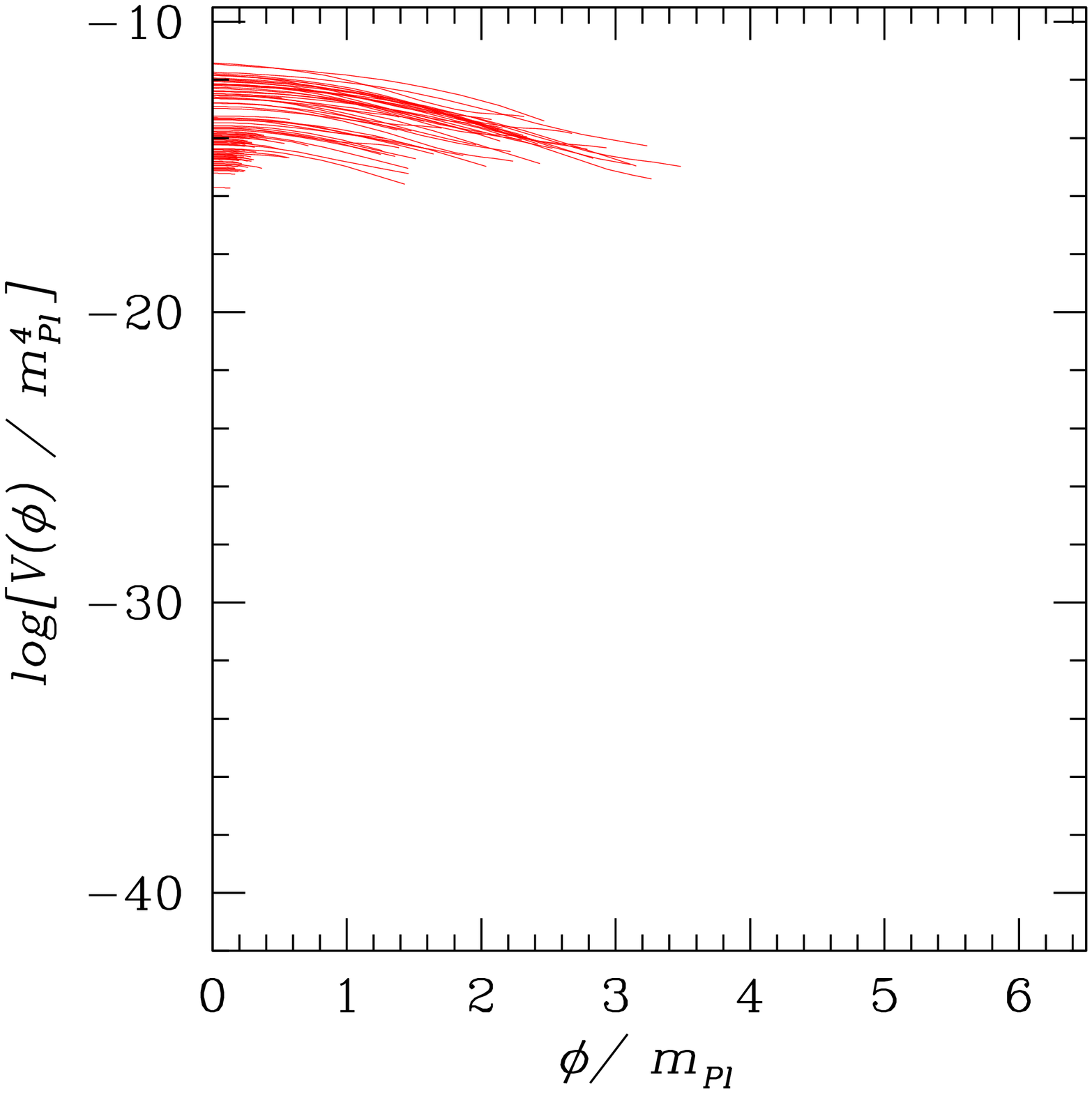}}
\centerline{\includegraphics[width=2.7in]{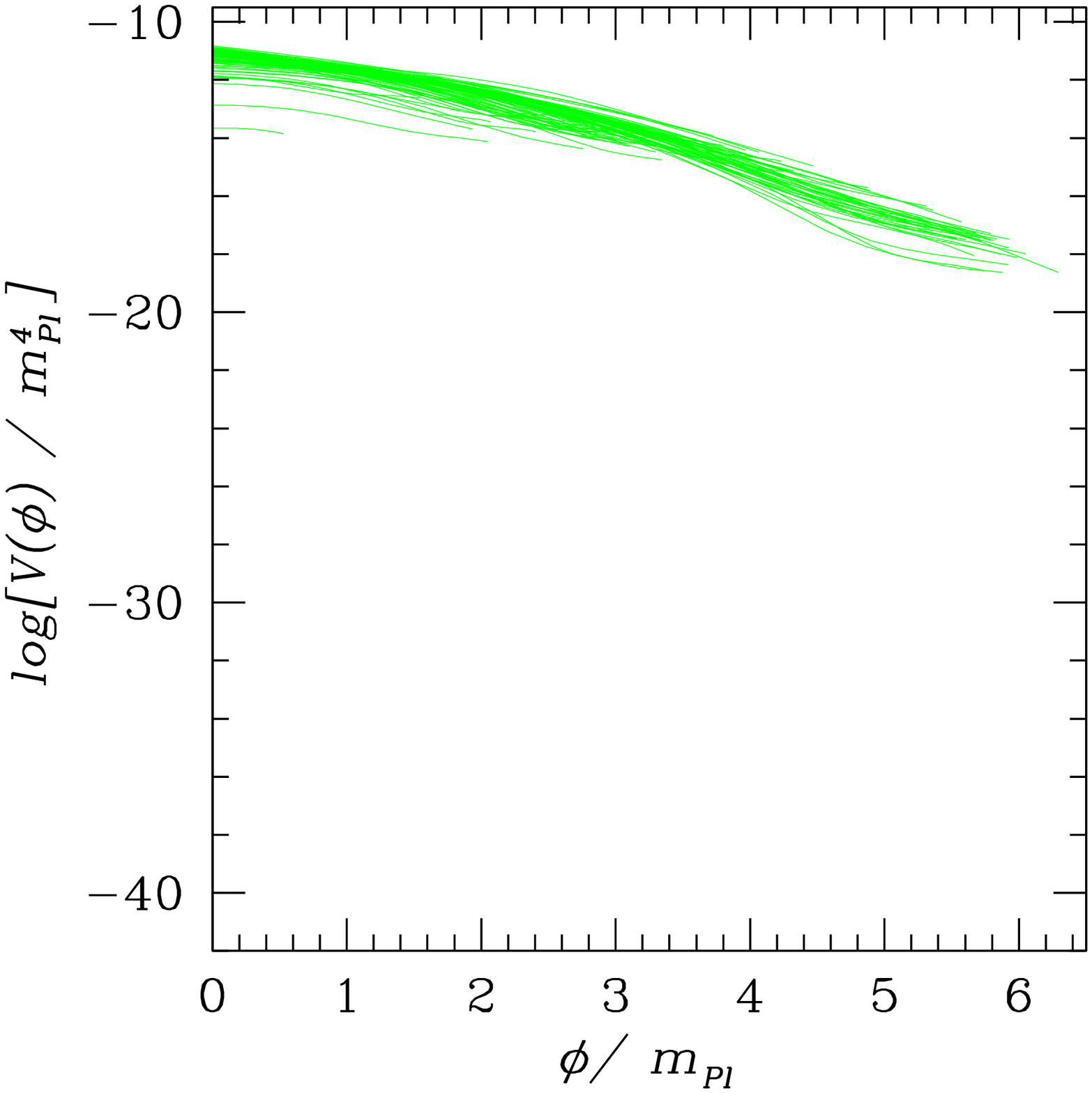}
\hspace*{24pt}
\includegraphics[width=2.7in]{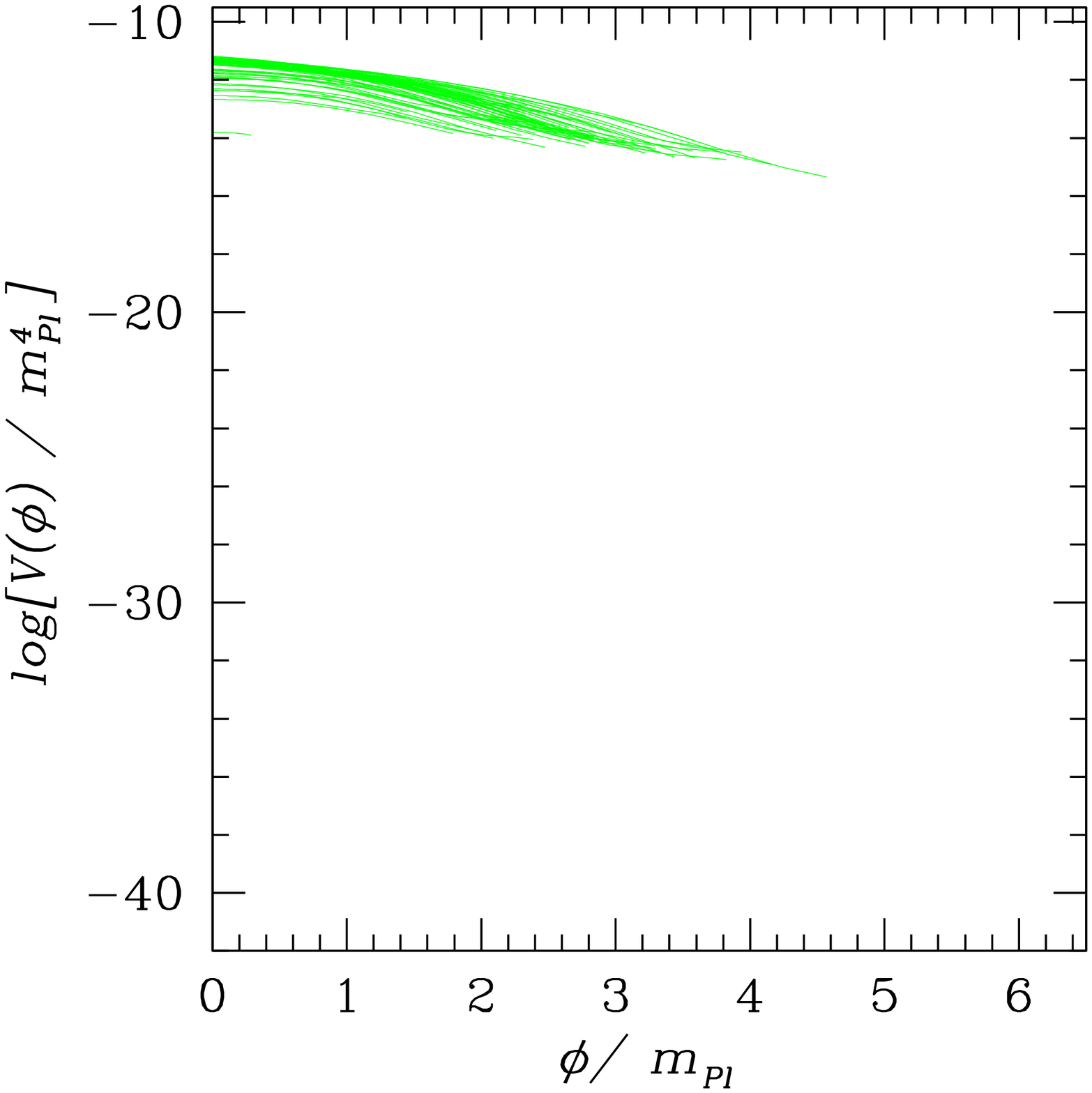}}
\centerline{\includegraphics[width=2.7in]{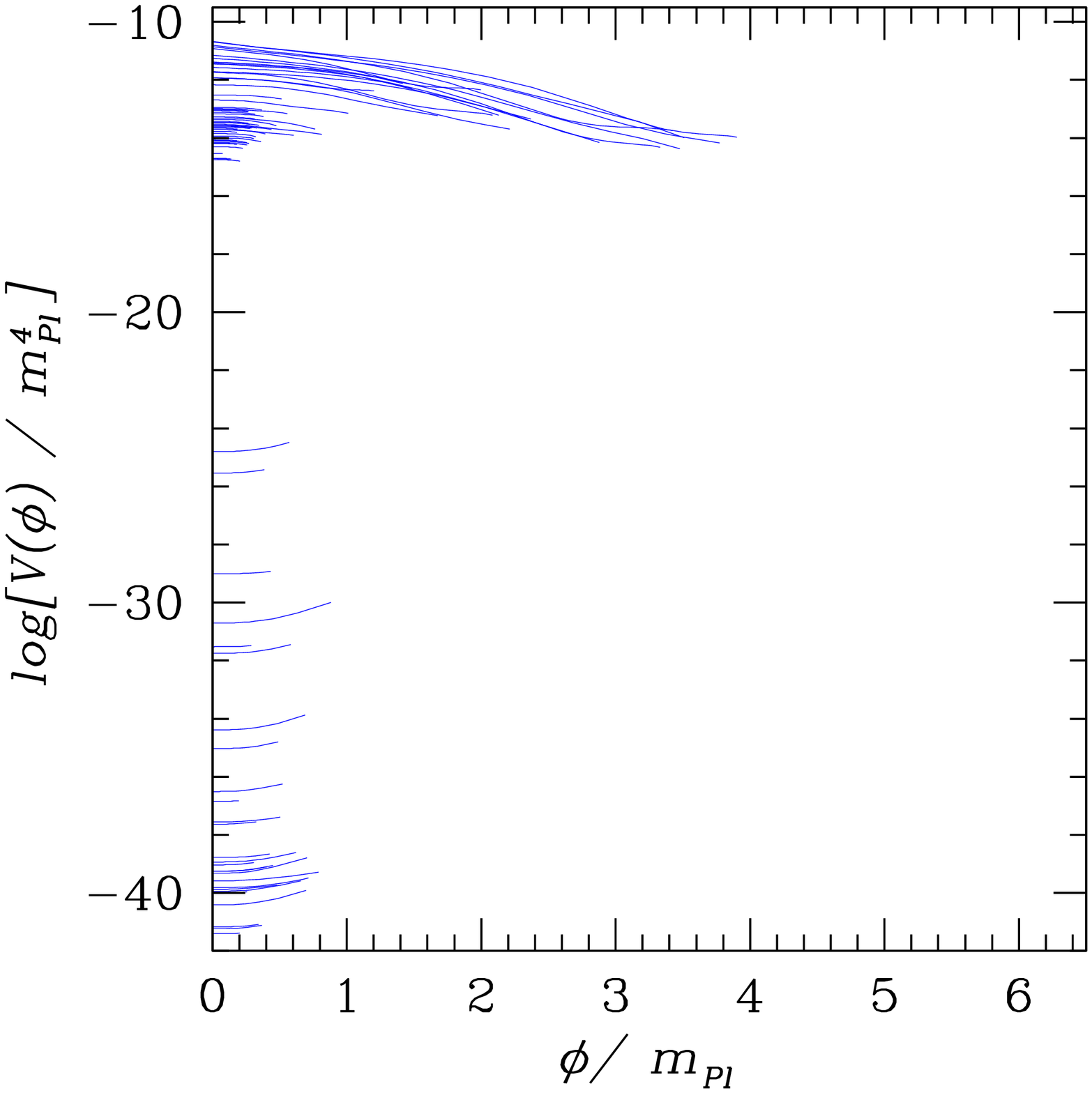}
\hspace*{24pt}
\includegraphics[width=2.7in]{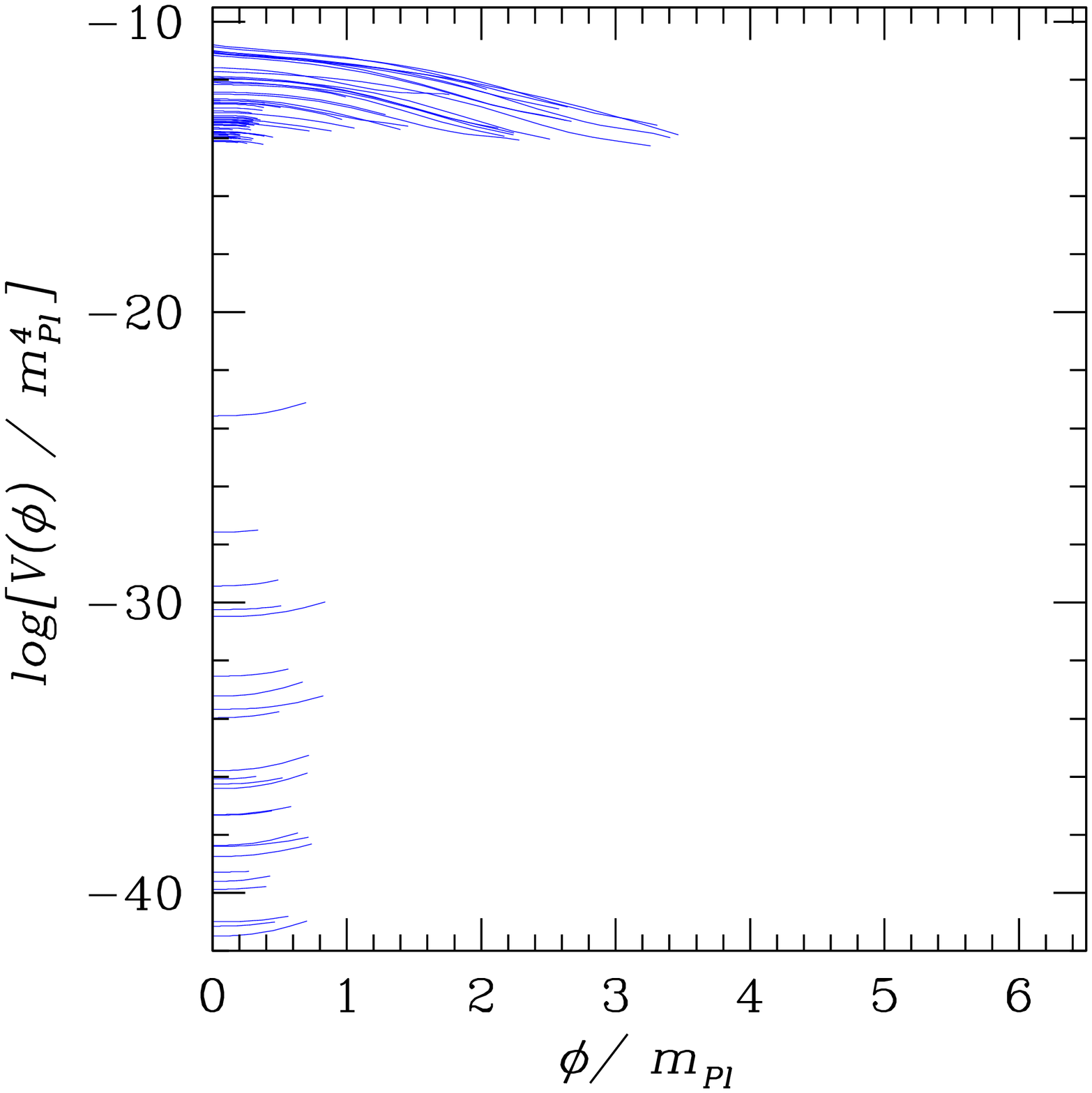}}
\caption{\label{fig_V_zoo_type}
Reconstructed potentials from Fig.\ \ref{fig_V_zoo} plotted separately by model
type: small-field (red, top), large-field (green, center), and hybrid (blue,
bottom). The WMAP results are shown in the left column while WMAP plus seven
other experiments are shown onthe right column.}
\end{figure}

\begin{figure}
\hspace*{1.0in} {\bf {\large WMAP only\ \ \ \ }} \hspace*{1.0in} 
{\bf {\large WMAP plus seven other experiments}}
\centerline{\includegraphics[width=2.7in]{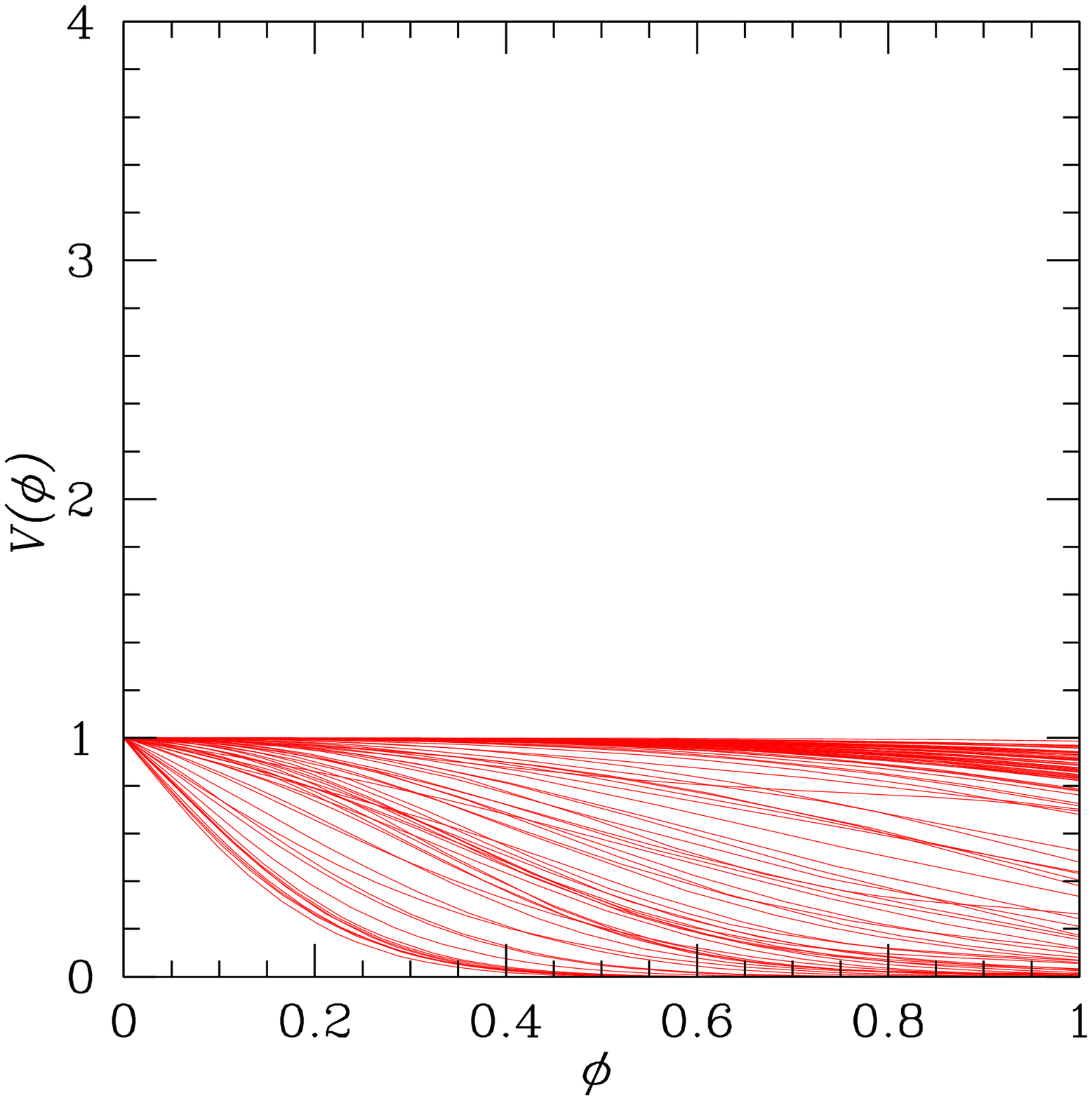}
\hspace*{24pt}
\includegraphics[width=2.7in]{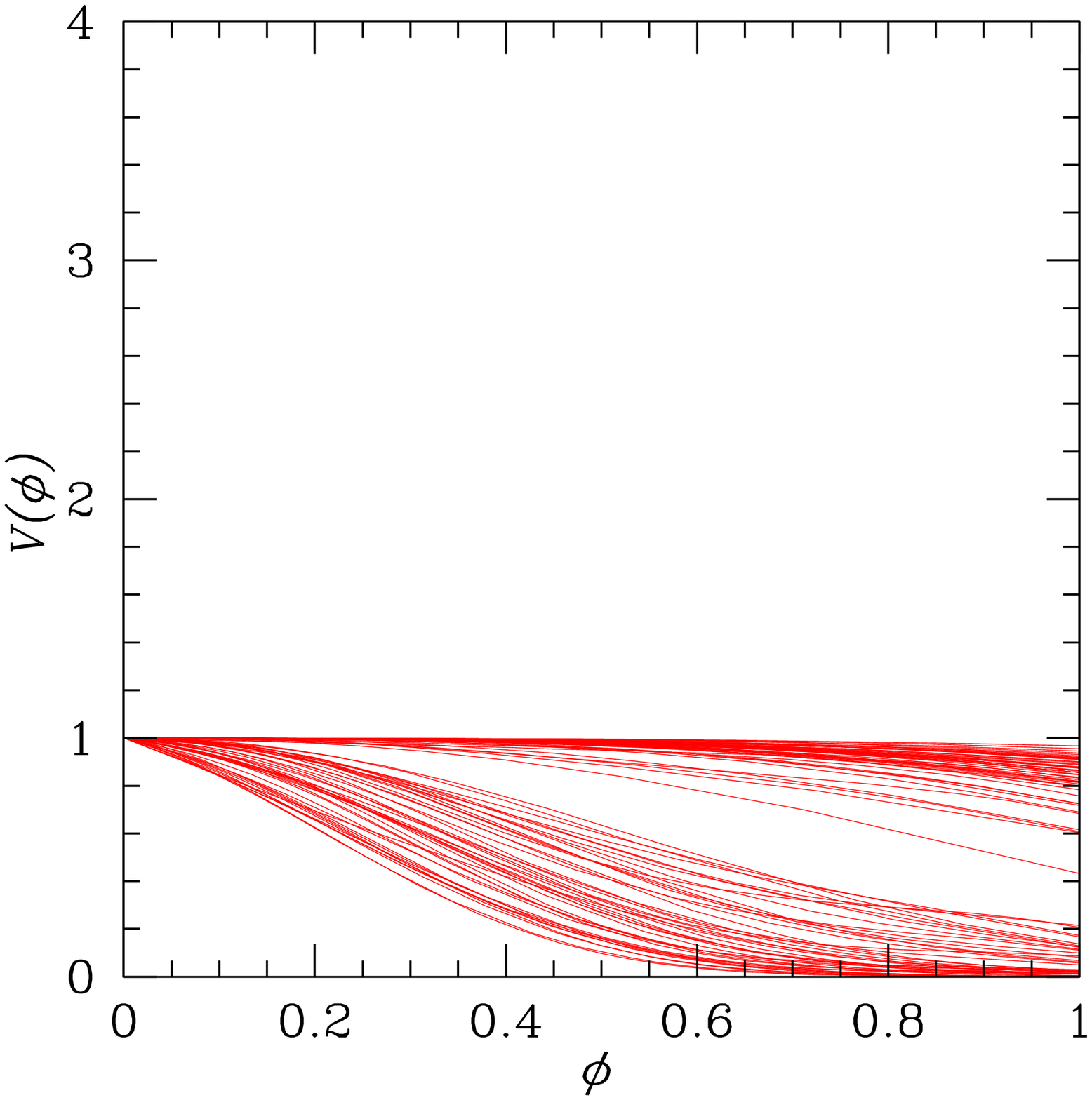}}
\centerline{\includegraphics[width=2.7in]{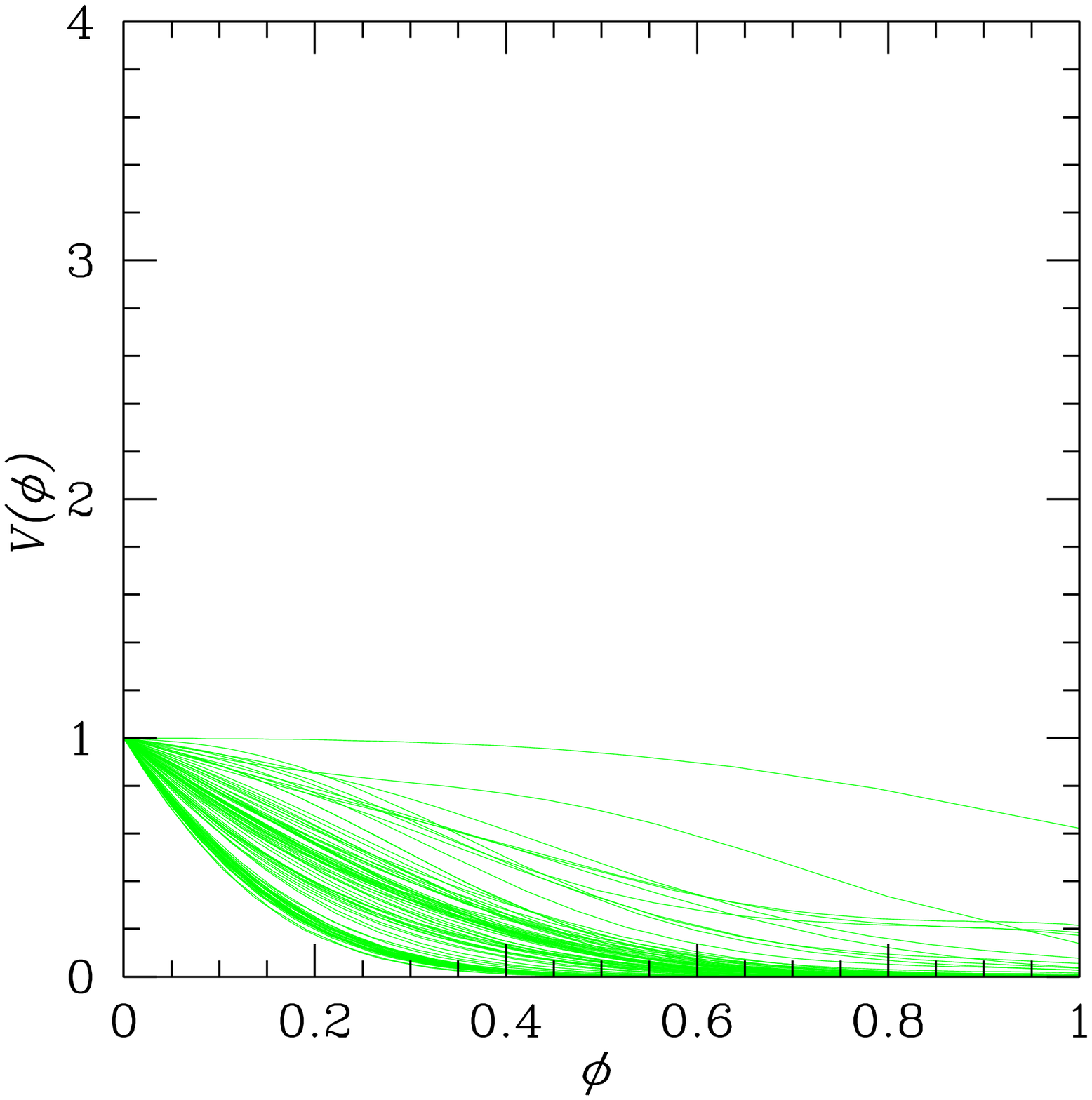}
\hspace*{24pt}
\includegraphics[width=2.7in]{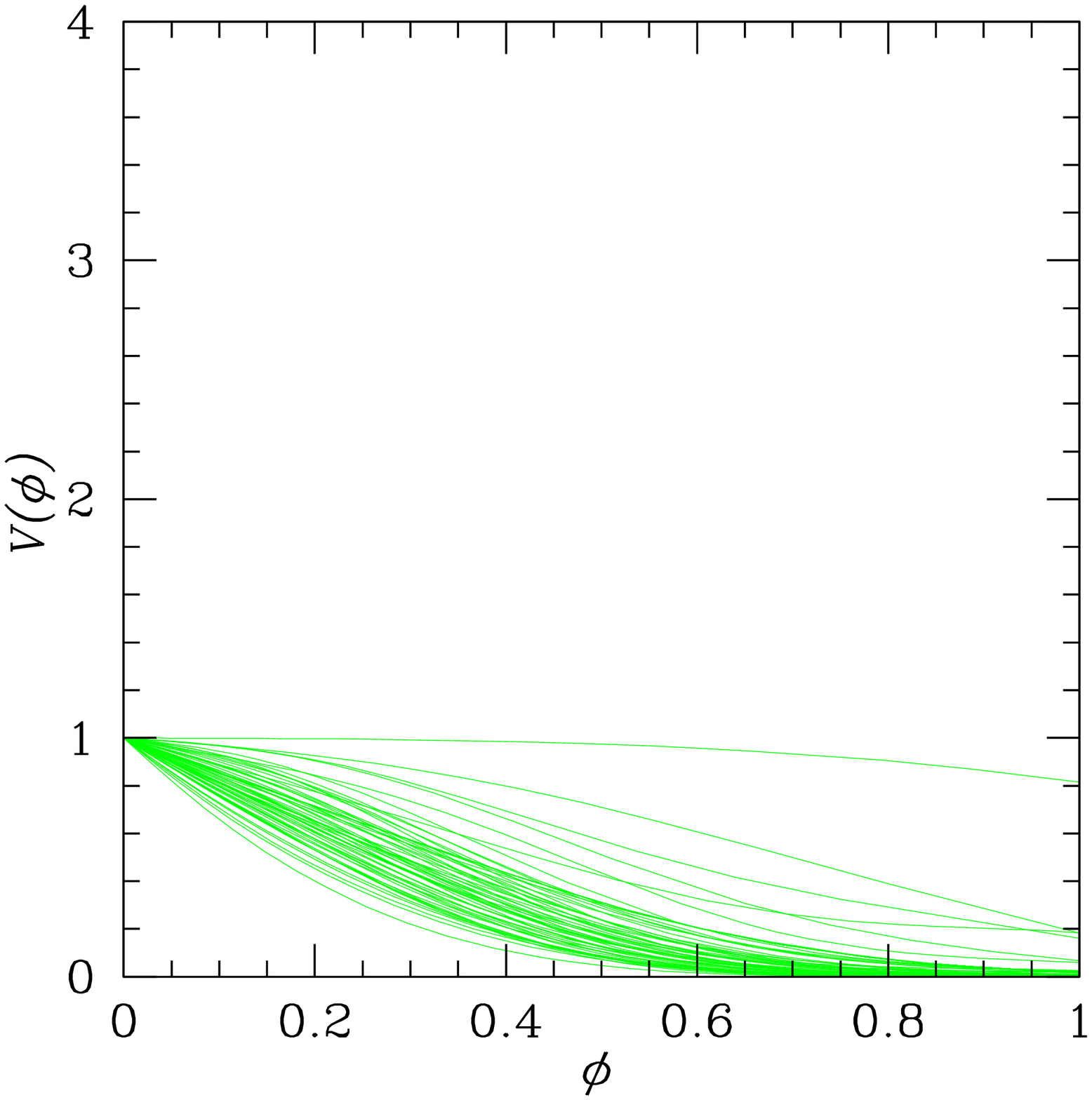}}
\centerline{\includegraphics[width=2.7in]{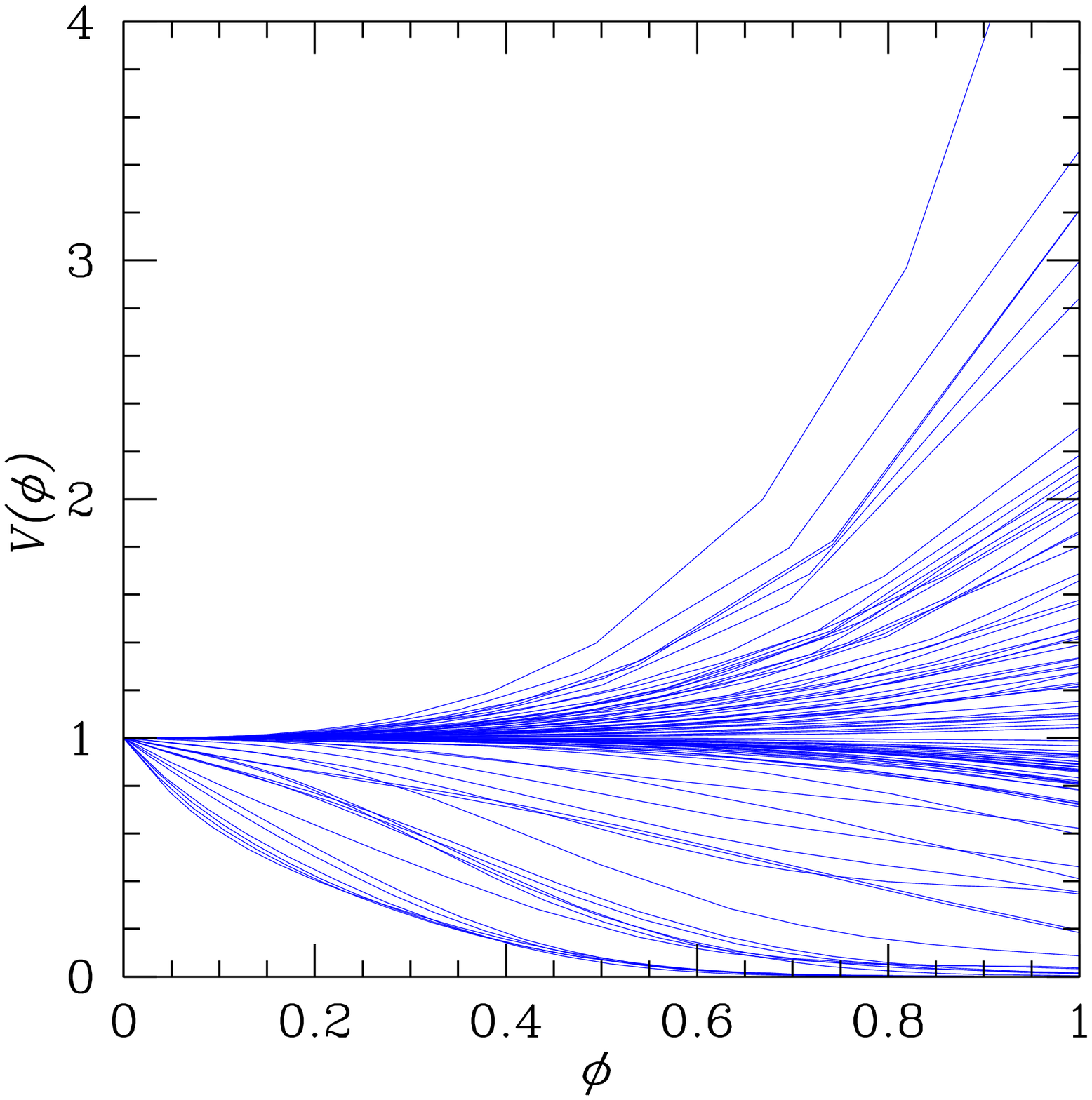}
\hspace*{24pt}
\includegraphics[width=2.7in]{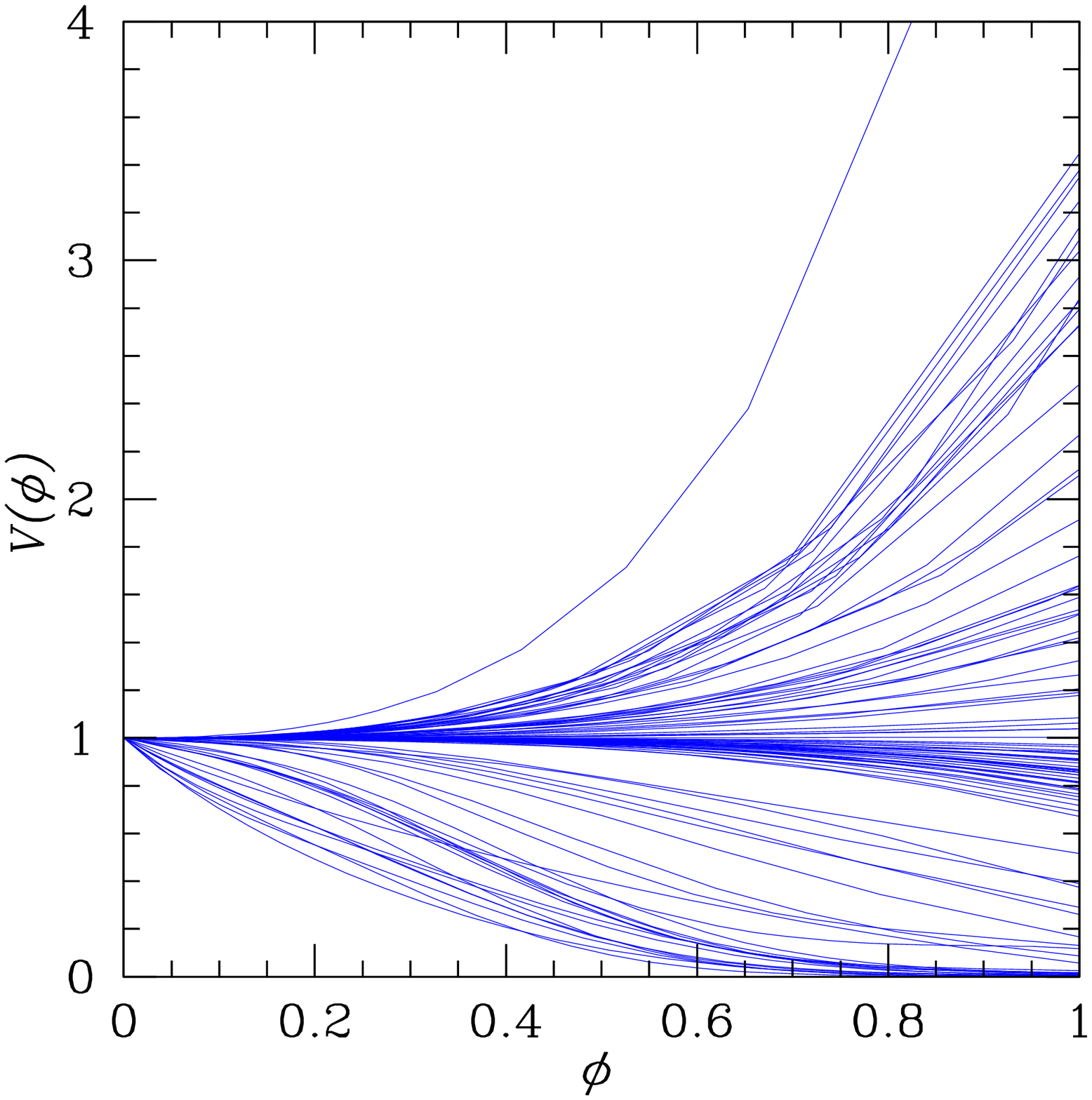}}
\caption{\label{fig_V_zoo_type_rescaled}
Same as Fig.\ \ref{fig_V_zoo_type}, but with the potentials rescaled to all
have the same height and width.}
\end{figure}

\end{widetext}


\begin{thebibliography}{99}
\frenchspacing

\bibitem{guth81} A. Guth, Phys. Rev. D {\bf 23}, 347 (1981)

\bibitem{lrreview} 
D. H. Lyth and A. Riotto, Phys. Rept. {\bf 314} 1 (1999); A. Riotto, 
hep-ph/0210162; W. H. Kinney, astro-ph/0301448.

\bibitem{muk81} 
V. F. Mukhanov and G. V. Chibisov, JETP Lett. {\bf 33}, 532 (1981).

\bibitem{hawking82} 
S. W. Hawking, Phys. Lett. {\bf 115B}, 295 (1982).

\bibitem{starobinsky82} 
A. Starobinsky, Phys. Lett. {\bf 117B}, 175 (1982).

\bibitem{guth82} 
A. Guth and S. Y. Pi, Phys. Rev. Lett. {\bf 49}, 1110 (1982).

\bibitem{bardeen83} 
J. M. Bardeen, P. J. Steinhardt, and M. S. Turner, Phys. Rev. D {\bf 28}, 679
(1983).

\bibitem{smoot92} 
G. F. Smoot {\it et al.} Astrophys. J. {\bf 396}, L1 (1992).

\bibitem{bennett96} 
C. L. Bennett {\it et al.} Astrophys. J. {\bf 464}, L1 (1996).

\bibitem{gorski96} 
K. M. Gorski {\it et al.} Astrophys. J. {\bf 464}, L11 (1996).

\bibitem{cp} 
See, for example, N.~Bahcall, J.~P.~Ostriker, S.~Perlmutter and
P.~J.~Steinhardt, Science {\bf 284}, 1481 (1999).

\bibitem{toco}
E.~Torbet {\it et al.}, Astrophys.\ J.\ {\bf 521} (1999) L79,
astro-ph/9905100; A.~D.~Miller {\it et al.}, Astrophys.\ J.\ {\bf 524}
(1999) L1, astro-ph/9906421.

\bibitem{b97}
P.~D.~Mauskopf {\it et al.}  [Boomerang Collaboration], Astrophys.\ J.\ {\bf
536} (2000) L59, astro-ph/9911444; A.~Melchiorri {\it et al.}
[Boomerang Collaboration], Astrophys.\ J.\ {\bf 536} (2000) L63
astro-ph/9911445.

\bibitem{Netterfield} 
C.~B.~Netterfield {\it et al.} [Boomerang Collaboration],
astro-ph/0104460.

\bibitem{halverson} 
N.~W.~Halverson {\it et al.}, astro-ph/0104489.

\bibitem{vsa} 
P. F. Scott {\it et al.}, astro-ph/0205380.

\bibitem{benoit} 
A. Benoit, astro-ph/0210305.

\bibitem{cbi} 
T. J. Pearson {\it et al.}, astro-ph/0205388. 

\bibitem{debe2001} 
P.~de Bernardis {\it et al.}, [Boomerang Collaboration],
astro-ph/0105296.

\bibitem{pryke} 
C.~Pryke, N.~W.~Halverson, E.~M.~Leitch, J.~Kovac, J.~E.~Carlstrom,
W.~L.~Holzapfel and M.~Dragovan, astro-ph/0104490.

\bibitem{stompor} 
R.~Stompor {\it et al.}, Astrophys.\ J.\ {\bf 561} (2001) L7,
astro-ph/0105062.

\bibitem{wang}
X. Wang, M. Tegmark, M. Zaldarriaga, astro-ph/0105091.

\bibitem{cbit} 
J. L. Sievers {\it et al.}, astro-ph/0205387.

\bibitem{vsat} 
J. A. Rubino-Martin {\it et al.}, astro-ph/0205367.

\bibitem{saralewis} 
A.~Lewis and S.~Bridle, astro-ph/0205436.

\bibitem{mesilk} 
A. Melchiorri and J. Silk, astro-ph/0203200;
S. H. Hansen, G. Mangano, A. Melchiorri, G. Miele and O. Pisanti,
Phys.\ Rev.\ D {\bf 65}, 023511 (2002), astro-ph/0105385;
A. Melchiorri, L. Mersini, C. J. Odman and M. Trodden,
astro-ph/0211522.

\bibitem{caro} 
C. J. Odman, A. Melchiorri, M. P. Hobson and A. N. Lasenby,
astro-ph/0207286.

\bibitem{dodelson97} 
S.~Dodelson, W.~H.~Kinney, and E.~W.~Kolb, Phys. Rev. D {\bf 56}, 3207 (1997), 
astro-ph/9702166.

\bibitem{kinney98a} 
W.~H.~Kinney, Phys. Rev. D {\bf 58}, 123506 (1998).

\bibitem{probes} 
W.~H.~Kinney, A.~Melchiorri, A.~Riotto, Phys. Rev. D {\bf 63}, 023505 (2001),
astro-ph/0007375; S.~Hannestad, S.~H.~Hansen, and F.~L.~Villante,
Astropart. Phys. {\bf 17} 375 (2002); S.~Hannestad, S.~H.~Hansen and
F.~L.~Villante, Astropart.\ Phys.\ {\bf 16}, 137 (2001); D.~J.~Schwarz,
C.~A.~Terrero-Escalante, and A.~A.~Garcia, Phys. Lett. B {\bf 517}, 243 (2001);
X.~Wang, M.~Tegmark, B.~Jain and M.~Zaldarriaga, astro-ph/0212417.

\bibitem{wmap1} 
C.~L.~Bennett {\it et al.}, astro-ph/0302207.

\bibitem{kogutt} 
Kogut {\it et al}, astro-ph/0302213.

\bibitem{wmapinf} 
H.~V.~Peiris {\it et al.}, astro-ph/0302225.

\bibitem{nong} 
V.~Acquaviva, N.~Bartolo, S.~Matarrese and A.~Riotto, astro-ph/0209156;
J.~Maldacena, astro-ph/0210603.

\bibitem{wmapnong} 
E.~Komatsu {\it et al.}, astro-ph/0302223.

\bibitem{kinney02} 
W.~H.~Kinney, Phys. Rev. D {\bf 66}, 083508 (2002), astro-ph/0206032.

\bibitem{easther02} 
R.~Easther and W.~H.~Kinney, Phys. Rev. D {\bf 67}, 043511 (2003), 
astro-ph/0210345.

\bibitem{barger} 
V.~Barger, H.~S.~Lee and D.~Marfatia, hep-ph/0302150.

\bibitem{acbar} 
C.L. Kuo {\it et al.}, astro-ph/0212289.

\bibitem{2df} 
W.J. Percival {\it et al.}, astro-ph/0105252.

\bibitem{lyman} 
R.A. Croft {\it et al.}, Astrophys. J. {\bf 581}, 20 (2002); N.Y. Gnedin and
A.J. Hamilton, astro-ph/0111194.

\bibitem{hst} 
W.L. Freedman {\it et al.}, Astrophys. J. {\bf 553}, 47 (2001).

\bibitem{ruhl} 
J. E. Ruhl {\it et al.} [Boomerang Collaboration], astro-ph/0212229.

\bibitem{lee} 
A. T. Lee {\it et al.}, Astrophys. J.  {\bf 561}, L1 (2001),
astro-ph/0104459.

\bibitem{grainge} K. Grainge et al., astro-ph/0212495.

\bibitem{linde83} 
A. D. Linde, Phys. Lett. {\bf 129B}, 177 (1983).

\bibitem{freese90} 
K. Freese, J. Frieman, and A. Olinto, Phys. Rev. Lett {\bf 65}, 3233 (1990).

\bibitem{linde91} 
A. D. Linde, Phys. Lett. {\bf 259B}, 38 (1991).

\bibitem{linde94} 
A. D. Linde, Phys. Rev. D {\bf 49}, 748 (1994).

\bibitem{copeland94} 
E. J. Copeland, A. R. Liddle, D. H. Lyth, E. D. Stewart, and D. Wands,
Phys. Rev. D {\bf 49}, 6410 (1994).

\bibitem{starobinsky80} 
A. A. Starobinsky, Phys. Lett. {\bf 91B}, 99 (1980).

\bibitem{grishchuk88} 
L. P. Grishchuk and Yu. V. Sidorav, in {\it Fourth Seminar on Quantum Gravity},
eds M. A. Markov, V. A. Berezin and V. P. Frolov (World Scientific, Singapore,
1988).

\bibitem{muslimov90} 
A. G. Muslimov, Class. Quant. Grav. {\bf 7}, 231 (1990).

\bibitem{salopek90} 
D. S. Salopek and J. R. Bond, Phys. Rev. D {\bf 42}, 3936 (1990).

\bibitem{lidsey95} 
J. E. Lidsey {\it et al.}, Rev. Mod. Phys. {\bf 69}, 373 (1997),
astro-ph/9508078.

\bibitem{lidsey97} 
J. E. Lidsey, A. R. Liddle, E. W. Kolb, E. J. Copeland, T. Barriero, and
M. Abney, Rev. Mod. Phys. {\bf 69}, 373 (1997).

\bibitem{dodelson03} S.~Dodelson and L.~Hui, astro-ph/0305113.

\bibitem{linde82} 
A.~D.~Linde, Phys.\ Lett.\  {\bf B108} 389, 1982.

\bibitem{albrecht82} 
A. Albrecht and P. J. Steinhardt, Phys. Rev. Lett {\bf48}, 1220 (1982).

\bibitem{kinney97a}
W.~H.~Kinney, Phys. Rev. D {\bf 56} 2002 (1997), hep-ph/9702427.

\bibitem{mukhanov85} 
V. F. Mukhanov, JETP Lett. {\bf 41}, 493 (1985).

\bibitem{mukhanov88} 
V. F. Mukhanov, Sov. Phys. JETP {\bf 67}, 1297 (1988).

\bibitem{mukhanov92} 
V. F. Muknanov, H. A. Feldman, and R. H. Brandenberger, Phys. Rep. {\bf 215},
203 (1992).

\bibitem{stewart93} 
E. D. Stewart and D. H. Lyth, Phys. Lett. {\bf 302B}, 171 (1993).

\bibitem{stewart97} 
E. D. Stewart, Phys. Lett. {\bf B391}, 34 (1997), hep-ph/9606241.

\bibitem{linde97}
A. D. Linde and A. Riotto, Phys. Rev. D {\bf 56}, 1841 (1997), 
hep-ph/9703209.

\bibitem{stewart97a} E. D. Stewart, Phys. Rev. D {\bf 56}, 2019 (1997),
hep-ph/9703232.

\bibitem{copeland97} 
E. J. Copeland, I. J. Grivell, and A. R. Liddle, astro-ph/9712028.

\bibitem{kinney97} 
W. H. Kinney and A. Riotto, Astropart. Phys. {\bf 10}, 387 (1999),
hep-ph/9704388.

\bibitem{covi98} 
L. Covi, D. H. Lyth and L. Roszkowski, Phys. Rev. D {\bf 60}, 023509 (1999),
hep-ph/9809310.

\bibitem{kinney98} 
W. H. Kinney and A. Riotto, Phys. Lett.{\bf 435B}, 272 (1998), hep-ph/9802443.

\bibitem{covi99} 
L. Covi and D. H. Lyth, Phys. Rev. D {\bf 59} (1999) 063515, hep-ph/9809562.

\bibitem{covi00} 
D. H. Lyth and L. Covi, astro-ph/0002397.

\bibitem{featuresinpowerspectrum} 
A.~A.~Starobinski, Grav. Cosmol. {\bf 4}, 88 (1998), astro-ph/9811360.
D.~J.~Chung, E.~W.~Kolb, A.~Riotto and I.~I.~Tkachev, hep-ph/9910437.

\bibitem{starobinsky79} 
A. A. Starobinsky, JETP Lett. {\bf 30}, 682 (1979).

\bibitem{rubakov82}
V. Rubakov, M. Sazhin, and A. Veryaskin, Phys. Lett. {\bf 115B}, 189 (1982).

\bibitem{fabbri83} 
R. Fabbri and M. Pollock, Phys. Lett. {\bf 125B}, 445 (1983).

\bibitem{abbot84} 
L.F. Abbott and M.B. Wise, Nucl. Phys. {\bf B244}, 541 (1984).

\bibitem{starobinsky85} 
A .A. Starobinsky, Sov. Astron.Lett. {\bf 11}, 133 (1985).

\bibitem{bunn96} 
E. F. Bunn and M. White, Astrophys. J. {\bf 480}, 6 (1997).

\bibitem{lyth96}
D. H. Lyth, hep-ph/9609431.

\bibitem{polarski95} 
D. Polarski and A. A. Starobinsky, Phys. Lett. {\bf 356B}, 196 (1995).

\bibitem{bellido95} 
J. Garc\'\i a-Bellido and D. Wands, Phys. Rev. D {\bf 52}, 6739 (1995).

\bibitem{sasaki96} 
M. Sasaki and E. D. Stewart, Prog. Theor. Phys. {\bf 95}, 71 (1996).

\bibitem{bellido98} 
J. Garc\'\i a-Bellido, Phys. Lett. {\bf 418B}, 252 (1998).

\bibitem{barrow93} 
J. D. Barrow and A. R. Liddle, Phys. Rev. D {\bf 47}, R5219 (1993).

\bibitem{liddle94} 
A.~R.~Liddle, P. Parsons, and J. D. Barrow, Phys. Rev. D {\bf 50}, 7222 (1994),
astro-ph/9408015.

\bibitem{hoffman00}
M.~B.~Hoffman and M.~S.~Turner, Phys. Rev. D {\bf 64}, 023506 (2001),
astro-ph/0006321.

\bibitem{schwarz01} D.~J.~Schwarz, C.~A.~Terrero-Escalante, and A.~.A.~Garcia,
Phys. Lett. {\bf B517}, 243 (2001), astro-ph/0106020.

\bibitem{liddle95} 
A.~R.~Liddle. and M.~S.~Turner, Phys. Rev. D {\bf 50}, 758 (1994). 

\bibitem{hodges90} 
H.~M.~Hodges and G.~R.~Blumenthal, Phys. Rev. D {\bf 42}, 3329 (1990).

\bibitem{copeland93} 
E.~J.~Copeland et al. Phys. Rev. Lett. {\bf 71}, 219 (1993).

\bibitem{beato00} 
E. Ayon-Beato, A. Garcia, R. Mansilla, and C. A. Terrero-Escalante,
Phys. Rev. D {\bf 62}, 103513 (2000).

\bibitem{bunn94} 
E.~F.~Bunn, D.~Scott, and M.~White, Astrophys. J. Lett. {\bf 441}, L9 (1995),
astro-ph/9409003.

\bibitem{stompor95} 
R.~Stompor, K.~M.~Gorski and A.~J.~Banday, Mon. Not. R.  Astron. Soc. {\bf
277}, 1225 (1995), astro-ph/9502035.

\bibitem{sz} 
M.~Zaldarriaga and U.~Seljak, Astrophys. J. {\bf 469}, 437 (1996),
astro-ph/9603033.

\bibitem{burles}
S.~Burles, K.~M.~Nollett and M.~S.~Turner, Astrophys.\ J.\ {\bf 552}, L1
(2001), astro-ph/0010171.

\bibitem{cyburt}
R.~H.~Cyburt, B.~D.~Fields and K.~A.~Olive, Compilation and Big Bang
Nucleosynthesis,''\ New Astron.\ {\bf 6} (1996) 215, astro-ph/0102179.

\bibitem{super1}  
P.M. Garnavich et al, Ap.J. Letters \textbf{493}, L53-57 (1998); S. Perlmutter
et al, Ap. J. \textbf{483}, 565 (1997); S.  Perlmutter et al (The Supernova
Cosmology Project), Nature \textbf{391} 51 (1998); A.G. Riess et al,
Ap. J. \textbf{116}, 1009 (1998);

\bibitem{thx}
M. Tegmark, A. J. S. Hamilton, Y. Xu, astro-ph/0111575.

\bibitem{ciardi}
B. Ciardi, A. Ferrara and S. D. M. White,
astro-ph/0302451.

\bibitem{Verde:2003ey}
L.~Verde {\it et al.}, astro-ph/0302218.
\bibitem{BJK} 
Bond, J.R., Jaffe, A.H., \& Knox, L.E. 2000, \apj, 533, 19

\bibitem{bridle} 
S.~L.~Bridle, R.~Crittenden, A.~Melchiorri, M.~P.~Hobson, R.~Kneissl and
A.~N.~Lasenby, astro-ph/0112114.

\bibitem{leach03} S.~M.~Leach and A.~R.~Liddle, astro-ph/0306305.


\end{thebibliography}
\end{document}